\documentclass[amsmath,amssymb,aps]{revtex4-2}

\usepackage{hyperref}
\usepackage{graphicx,epstopdf,pgfplots,caption,subcaption}

\usepackage{mathtools}
\mathtoolsset{showonlyrefs=false}

\usepackage{enumitem}

\newcommand{\p}[0]{\partial}

\newcommand{\bmth}[1]{\mbox{\boldmath $#1$}}
\newcommand{\grad}{\bmth{\nabla}}

\begin{document}
	
\title{\Large Viscous fingering patterns for Hele--Shaw flow \\ in a doubly connected geometry \\ driven by a pressure differential or rotation}

\author{\large Liam C. Morrow,$^1$ Nicolas De Cock,$^2$ and Scott W. McCue$^{3,*}$}
\affiliation{\vspace{2ex}$^1$Department of Engineering Science, University of Oxford, Oxford OX13PJ, United Kingdom\\
$^2$TERRA Research and Teaching Centre, Gembloux Agro Bio-Tech, University of Liege, Gembloux 5030, Belgium\\
$^3$School of Mathematical Sciences, Queensland University of Technology, Brisbane QLD 4001, Australia
\vspace{2ex}
}

\email[]{scott.mccue@qut.edu.au}

\date{\today}
	
\begin{abstract}
	Traditional mathematical models of Hele--Shaw flow consider the injection (or withdrawal) of an air bubble into (or from) an infinite body of viscous fluid.  The most commonly studied feature of such a model is how the Saffman-Taylor instability drives viscous fingering patterns at the fluid/air interface.  Here we consider a more realistic model, which assumes the viscous fluid is finite, covering a doubly connected two-dimensional region bounded by two fluid/air interfaces.  For the case in which the flow is driven by a prescribed pressure difference across the two interfaces, we explore this model numerically, highlighting the development of viscous fingering patterns on the interface with the higher pressure.  Our numerical scheme is based on the level set method, where each interface is represented as a separate level set function.  We show that the scheme is able to reproduce the characteristic finger patterns observed experimentally up to the point at which one of the interfaces bursts through the other.  The simulations are shown to compare well with experimental results.  Further, we consider a model for the problem in which an annular body of fluid is evolving in a rotating Hele--Shaw cell.  In this case, our simulations explore how either one or both of the interfaces can be unstable and develop fingering patterns, depending on the rotation rate and the volume of fluid present.
\end{abstract}
	
\date{\today}
	
\maketitle

\section{Introduction}

Viscous fingering pattern formation that develops in a Hele--Shaw flow is one of the most well studied phenomena in interfacial fluid dynamics.  These visually striking patterns, which are due to the Saffman-Taylor instability that applies when a more viscous fluid is displaced by a less viscous fluid \citep{Saffman1958}, are characterised by their tip-splitting and branching morphology. More broadly, the Hele--Shaw model has become a paradigm for studying interfacial instabilities occurring in other related moving boundary problem ranging from porous media flow \citep{Homsy1987} to dendrite solidification \citep{Ben1990}.

\begin{figure}
	\centering
	\includegraphics[width=0.4\linewidth]{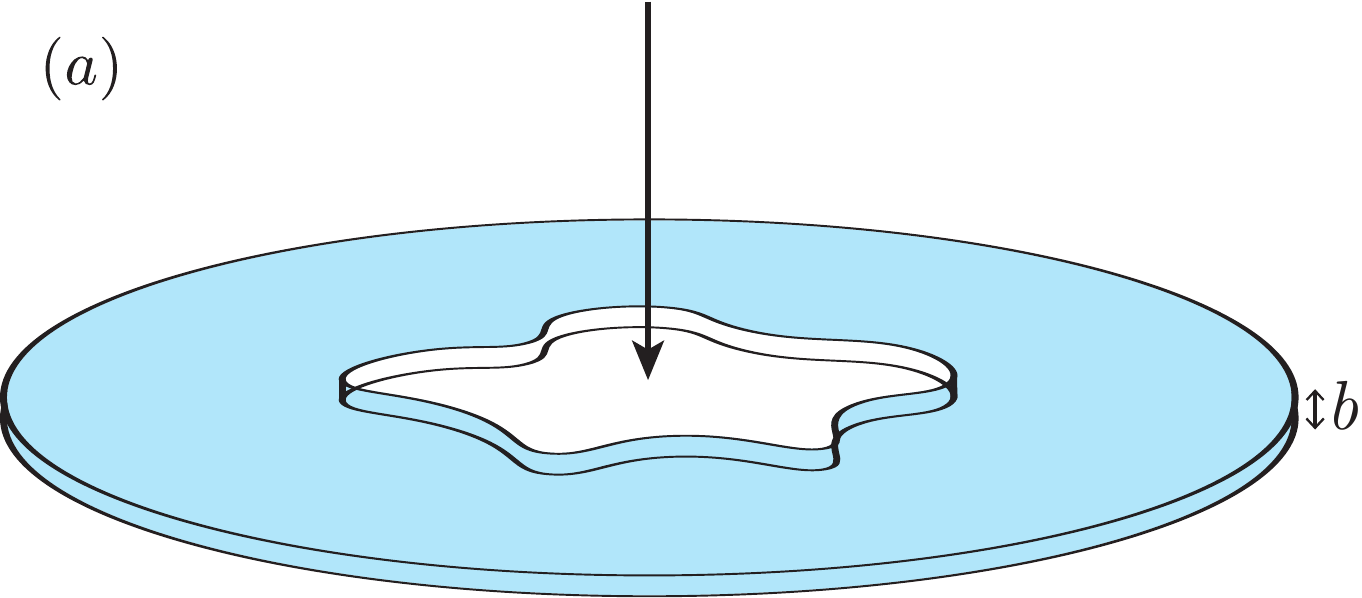}
	\includegraphics[width=0.4\linewidth]{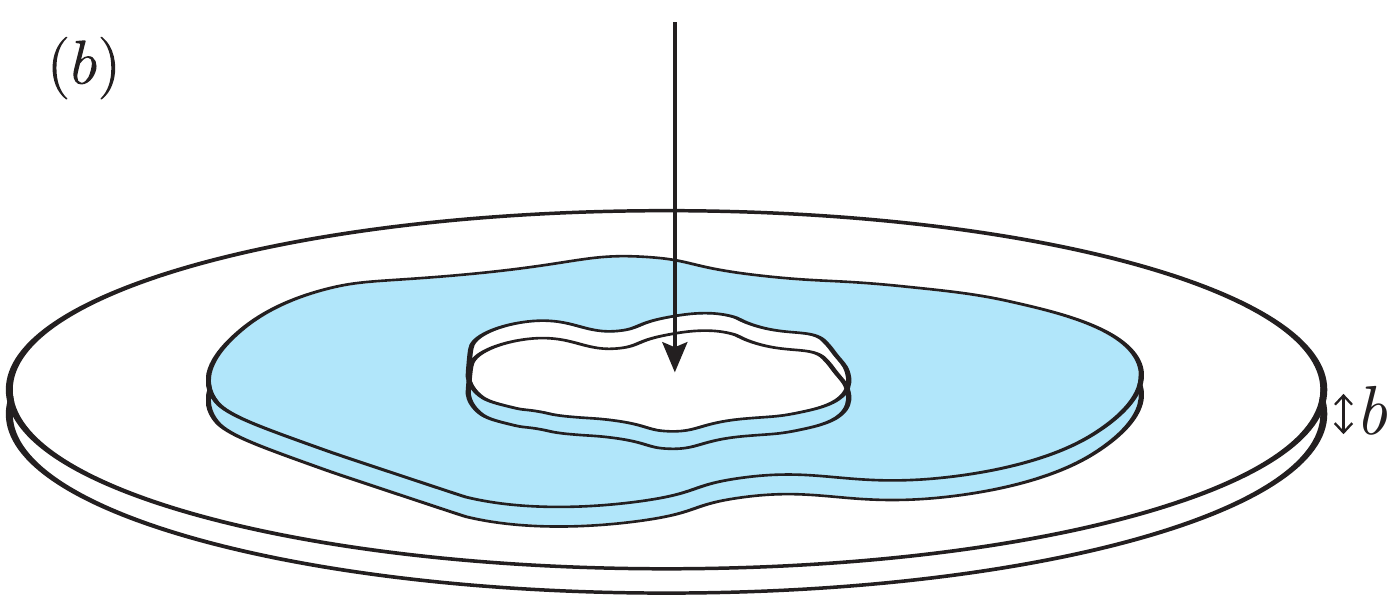}
	\caption{\label{fig:Figure1} $(a)$ An illustration of the classical model for an inviscid bubble (white) expanding or contracting in a Hele--Shaw cell, where the two parallel plates are separated by a small gap, $b$, and the viscous fluid (blue) is supposed to extend out forever.  If the inviscid bubble is injected into the viscous fluid, the interface between the two fluids develops viscous fingering patterns due to the Saffman-Taylor instability. $(b)$ A more realistic Hele--Shaw model involving a doubly connected geometry, where now the inviscid bubble is surrounded by a finite amount of viscous fluid, which in turn results in two interfaces. In this case the flow is driven by a pressure differential between the inner and outer regions.}
\end{figure}

The most common mathematical model used to study viscous fingering in Hele--Shaw flow, illustrated in Fig.~\ref{fig:Figure1}$(a)$, involves a fluid of negligible viscosity, air for example, being injected into or withdrawn from an infinite body of viscous fluid.  Under the injection scenario, linear and weakly nonlinear stability analysis shows that as the bubble of inviscid fluid expands, successive modes of perturbation become unstable \citep{Miranda1998,Paterson1981}, which in turn drives the viscous fingering patterns that are observed experimentally \citep{Chen1987,Chen1989,Paterson1981,Thome1989}. Recently, very many theoretical and numerical studies of variations of this type of injection problem with a single interface have been undertaken to study the effects of injection rate, fluid properties, miscibility, suspended particles, electric fields and geometrical alterations have on the viscous fingering structures \citep{Anjos2017b,Anjos2018,Anjos2021,Anjos2022,Arun2020,Li2021,Luo2018,Morrow2019,Morrow2021,Sharma2020,Tsuzuki2019,Vaquero2019}, some of which are supported by experimental results.  If, on the other hand, the inviscid fluid is withdrawn from the Hele--Shaw cell, the shape of the bubble boundary can be shown to be stable and, for example, if it is initially convex, the model predicts it will contract to a point \citep{Dallaston2013,Entov1991}.  For the complementary problem, where a viscous `blob' of fluid is completely surrounded by an inviscid fluid, stability analysis and numerical simulations predict the interface between the two fluids will be unstable when the blob is withdrawn from a point \citep{Chen2014,Ceniceros1999,Kelly1997,Paterson1981}. Experimental and numerically studies of this problem indicate that fingers develop inward, and appear to `race' each other toward the point of withdrawal \citep{Chen2014,Morrow2021,Thome1989}.

We are concerned here with the more realistic mathematical model for radial injection or withdrawal in which, instead of there being an infinite amount of viscous fluid, there is a finite amount of fluid in a doubly connected domain with two interfaces between the viscous fluid and the surrounding inviscid fluid, as indicated in Fig.~\ref{fig:Figure1}$(b)$.  In this configuration, when inviscid fluid is injected, the trailing (inner) interface will develop the traditional viscous fingering patterns, while the leading (outer) interface will be nominally stable.  On the other hand, when the inviscid fluid is being withdrawn, it is the outer interface that is unstable with inward fingering patterns developing.  While analytical studies of this doubly connected model are less prevalent in the literature than the single interface problem, a number of analytical and numerical studies of doubly connected or multiple-interface Hele--Shaw configurations have been undertaken, for example by applying linear or weakly nonlinear stability analysis \citep{Anjos2020,Gin2015a,Gin2019,Gin2021,Jackson2021} or using complex variable techniques for idealised scenarios without surface tension \citep{Crowdy2004,Dallaston2012}.  However, the only fully nonlinear simulations for this type of geometry have been conducted by \citet{Zhao2020}.  In Ref.~\cite{Zhao2020}, the authors concentrate on a geometry with three different fluids in three layers, with the middle fluid more viscous than the inner fluid, and the outer fluid more viscous than the middle fluid.  Simulations show unusual patterns on the outer interface.  Our own study is different as the model we consider has an inviscid fluid inside and outside the annular region of viscous fluid.  Therefore our fully nonlinear simulations complement those of Ref.~\cite{Zhao2020}.

\begin{figure}
	\centering
    \includegraphics[width=0.7\linewidth]{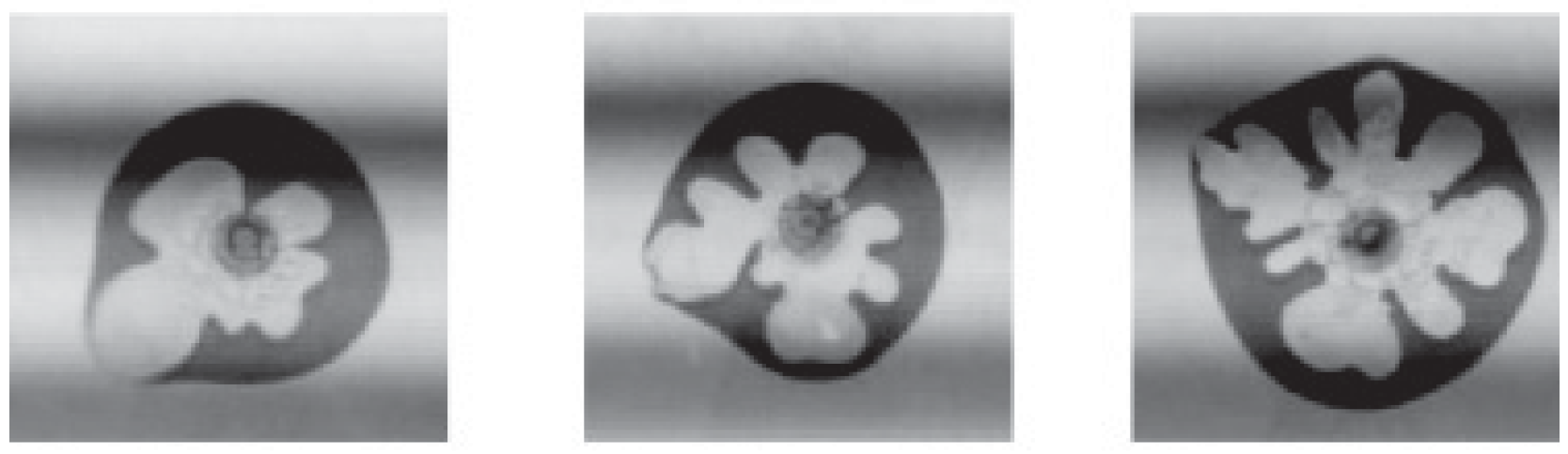}
	\caption{Experimental images from Ward \& White~\cite{Ward2011}, illustrating typical viscous fingering patterns in a doubly connected geometry at the time at which the air bubble bursts through the outer interface, with the viscous fluid (glycerol-water mixture) having, from left to right, kinematic viscosities of $4\times 10^{-6}$, $37\times 10^{-6}$ and $280\times 10^{-6}$ m$^2$s$^{-1}$.  These images are taken when the dimensional pressure differential was $\Delta \bar{p}=3.5$ kPa. Reproduced from Ref.~\cite{Ward2011} with permission from the American Physical Society.}
\label{fig:Figure2}
\end{figure}

Experiments that closely align with the doubly connected geometry illustrated in Fig.~\ref{fig:Figure1}$(b)$ have been conducted by Cardoso \& Woods~\cite{Cardoso1995} and Ward \& White~\cite{Ward2011}, for example.  A selection of images from Ref.~\cite{Ward2011} are shown in Fig.~\ref{fig:Figure2}.  These experiments were conducted with a glycerol-water mixture for the viscous fluid and air for the inviscid fluid, with a pressure differential of $3.5$ kPa between the inner and outer fluids (other pressure differentials were used but not shown here).  Other experiments in the radial Hele--Shaw configuration also appear to have effectively been performed with a doubly connected geometry, even if they were designed to concentrate on the inner interface only, simply due to the viscous fluid being necessarily finite in volume with a near circular outer boundary~\cite{Chen1989}.  We have conducted a small number of our own experiments in this doubly connected configuration (see Fig.~\ref{fig:FigureX}).

A variation on this doubly connected model that we are also concerned with here is where the inviscid fluid is neither injected nor withdrawn, but instead the entire Hele--Shaw cell is rotated.  In this case, the rotation of the experimental device propels the dense viscous fluid outward.  Mathematical models and experiments have previously been devoted to studying rotating Hele--Shaw scenarios with a focus on one interface~\cite{Alvarez2004,Alvarez2008,Anjos2017,Barua2022,Carrillo1996,Crowdy2002,Dias2011,Folch2009,Gadelha2004,Miranda2005,Morrow2019,Morrow2021,Paiva2019,Schwartz1989,Waters2005}. In the more complicated doubly connected case, experiments indicate that either one or both of the interfaces can be unstable \citep{Carrillo1999,Carrillo2000}.  These studies are supported by a comprehensive linear stability analysis in \citet{Carrillo2000}, who are able to track modes of perturbation, including a focus on the cases of a thick and thin annulus of viscous fluid.  Our contribution here is to report on fully nonlinear simulations of the rotating doubly connected geometry, thereby complementing previous experimental and analytical studies \citep{Carrillo1999,Carrillo2000}.  It is worth noting that, very recently, both linear and weakly nonlinear studies of Hele-Shaw flows have been studied in an annular geometry for ferrofluids \cite{Livera2021,Livera2022}, including quite complicated scenarios in which the external field induces a rotational motion.  Again, these interesting studies do not involve fully nonlinear numerical simulations.

In summary, we report on a numerical study of Hele--Shaw flow in two doubly connected geometries. Our scheme is based on the level set method, where we use separate level set functions to describe the evolution of each interface, and employ a modified finite difference stencil to solve for the pressure in the viscous fluid.  We consider two models.  In Sec.~\ref{sec:PressureDifferential}, we focus on the scenario in which inviscid fluid is injected or withdrawn subject to a prescribed pressure differential between the inner and outer boundaries (which could be positive or negative). We are able to perform simulations up to the point at which either one interface `bursts' through the other, or the interior bubble contracts to a point.   For the case of an expanding bubble, our simulations appear to compare well with experimental results (see Fig.~\ref{fig:FigureX}).  For the second model, which we treat in Sec.~\ref{sec:Rotating}, the fluid motion is driven by a centrifugal force that acts to propel the dense fluid outward. This scenario leads to a competition for stability on each interface between traditional Saffman-Taylor effects (which destabilise the interface between an inviscid fluid displacing a viscous fluid) and centrifugal effects (which have the opposite effect of destabilising the interface between a viscous fluid invading an inviscid fluid) \citep{Carrillo2000}.  Our simulations illustrate how either one or both of the interfaces can develop viscous fingering patterns, and the morphological features of these fingers are distinct for each interface.  Finally, we end the paper in Sec.~\ref{sec:Discussion} with a brief summary and a discussion of how our results
contribute to our understanding of how the Saffman-Taylor instability and its variants in Hele--Shaw and porous media flows applies in geometries with multiple interfaces.  Note that the numerical results in Sec.~\ref{sec:PressureDifferential} and \ref{sec:Rotating} are based on preliminary results reported in the PhD thesis~\cite{Morrow2020}.

\section{Flow driven by prescribed pressure differential} \label{sec:PressureDifferential}

\subsection{Summary of mathematical model}

In what follows, we use overbars to denote dimensional variables.  We consider an annular region of viscous fluid that occupies the domain $\bar{\bmth{x}} \in \bar{\Omega}(\bar{t})$ in a Hele--Shaw cell. The viscous fluid is doubly connected with inner and outer boundaries denoted by $\p \bar{\Omega}_i(\bar{t})$ and $\p \bar{\Omega}_o(\bar{t})$, respectively, as indicated by Fig.~\ref{fig:Figure1}$(b)$. The velocity of the viscous fluid is governed by Darcy's law
\begin{align} \label{eq:DarcysLaw}
	\bar{\bmth{v}}(\bar{\bmth{x}},\bar{t}) = -\frac{b^2}{12 \mu}  \bar{\grad} \bar{p},
\end{align}
where $\bar{\bmth{v}}$ is the velocity of the fluid, $\bar{p}$ is its pressure, $b$ is the gap between the plates, and $\mu$ is the viscosity of the fluid. The kinematic boundary conditions are given by
\begin{align}
	\bar{v}_n = \bar{\bmth{v}} \cdot \bar{\bmth{n}}, \hspace{2em} \bar{\bmth{x}} \in \p \bar{\Omega}_i,  \p \bar{\Omega}_o , \label{eq:DC2}
\end{align}
which relate the motion of the fluid to the velocity of each interface. We assume the pressure of the inviscid fluid is spatially independent, leading to the dynamic boundary conditions
\begin{align}
\bar{p} =
\begin{cases}
\displaystyle
\bar{p}_i(\bar{t}) - \gamma \left( \bar{\kappa} + \frac{2 \cos \theta_c}{b}  \right) \qquad  \bar{\bmth{x}} = \partial \bar{\Omega}_i \label{eq:DC3}
\\
\displaystyle		
\bar{p}_o(\bar{t}) - \gamma \left( \bar{\kappa} + \frac{2 \cos \theta_c}{b}  \right) \qquad \bar{\bmth{x}} = \partial \bar{\Omega}_o
	\end{cases},
\end{align}
where $\bar{p}_i$ and $\bar{p}_o$ are pressures within the inner and outer inviscid fluid regions, respectively, $\theta_c$ is the contact angle, $\bar{\kappa}$ denotes the signed curvature of each boundary (defined to be negative if the interface is locally convex from the viscous fluid side), and $\gamma$ is the surface tension parameter.

To nondimensionalise \eqref{eq:DarcysLaw}-\eqref{eq:DC3}, we scale space, time, pressure, and velocity according to
\begin{align}
\bar{\bmth{x}} = \bar{R}_i \bmth{x}, \qquad \bar{t} = \dfrac{12 \mu \bar{R}_i^3}{b^2 \gamma} t, \qquad  \bar{p} = \frac{\gamma}{\bar{R}_i} p, \qquad \bar{\bmth{v}} = \dfrac{b^2 \gamma}{12 \mu \bar{R}_i^2}\bmth{v},
\label{eq:Scaling}
\end{align}
where $\bar{R}_i$ is a length scale associated with the inner inviscid region.  For almost all of our calculations, we choose $\bar{R}_i$ to be the (dimensional) average radius of the inner bubble $\partial \bar{\Omega}_i$ at $\bar{t}=0$ so that the dimensionless average radius is unity.  Under this scaling, Darcy's law \eqref{eq:DarcysLaw} becomes $\bmth{v} = -\bmth{\nabla} p$. Assuming the viscous fluid is incompressible, then $\grad \cdot \bmth{v} = 0$;  thus, our model becomes
\begin{subequations}
\begin{alignat}{3}
\nabla^2 p &= 0  &\bmth{x} &\in \Omega,
\label{eq:Model1} \\
v_n &= -\grad p \cdot \bmth{n} & \bmth{x} &\in \partial \Omega_I, \partial \Omega_O, \label{eq:Model2} \\
p &= p_I - \kappa \qquad \qquad  &  \bmth{x} &\in \partial \Omega_I,
\label{eq:Model3}\\
p &= p_O - \kappa & \bmth{x} & \in \partial \Omega_O,
\label{eq:Model4}
	\end{alignat}
\end{subequations}
where $p_I$ and $p_O$ are the dimensionless quantities
\begin{align}
p_I = \frac{\bar{R}_i \bar{p}_i}{\gamma} - \frac{2 \cos \theta_c}{b/\bar{R}_i}
\qquad \textrm{and} \qquad
p_O = \frac{\bar{R}_i \bar{p}_o}{\gamma} - \frac{2 \cos \theta_c}{b/\bar{R}_i}.
\end{align}
There are two important parameters for this model.  First, we prescribe the dimensionless pressure differential between the two interfaces
\begin{align}
\Delta p = p_I - p_O = \dfrac{\bar{R}_i \left( \bar{p}_i - \bar{p}_o \right)} {\gamma},
\label{eq:deltap}
\end{align}
and assume that $\Delta p$ is a constant. Note that different values of advancing and receding contact angle simply lead to a change in $\Delta p$. Further, since (almost) all of our calculations are for initial conditions for which the interfaces are perturbations of circles, the second important parameter is the average initial radius of the outer interface,
\begin{align}
	R_O = \frac{\bar{R}_o}{\bar{R}_i}.
\label{eq:FreeParameter2}
\end{align}
Under this scaling, we expect that $\Delta \hat{p} > 0$ and $\Delta \hat{p} < 0$ results in an expanding and contracting interior bubble, respectively. Finally, the other parameters in the problem relate to the actual shape of the inner and outer interfaces, for example the details of the perturbations include the radius and modes of perturbation (remembering that the inner bubble is scaled so that initially its average radius is unity).

\subsection{Numerical scheme using level set method} \label{sec:NumericalScheme}

In this section, we describe our numerical scheme for solving \eqref{eq:Model1}-\eqref{eq:Model4}. This scheme is based on the numerical framework presented in Ref.~\citep{Morrow2019,Morrow2021}, where we presented a numerical scheme that employs the level set method for solving a generalised model of (simply connected) Hele--Shaw flow. It uses the concept of representing the interface between the viscous and inviscid fluids as the zero level set of a higher dimensional surface, $\phi(\bmth{x},t)$, that satisfies $\phi > 0$ if $\bmth{x} \in \Omega(t)$ and $\phi < 0$ if $\bmth{x} \in \mathbb{R}^2 \backslash \Omega(t)$. The level set function $\phi$, and in turn the interface, is evolved by solving the level set equation
\begin{align} \label{eq:LevelSetEqn}
	\frac{\partial \phi}{\partial t} + F |\grad \phi| = 0,
\end{align}
on the two-dimensional computational domain $D$, where $F = -\grad p \cdot \grad \phi/|\grad \phi|$. For all results, simulations are performed on the square domain $-L \le x \le L$ and $-L \le y \le L$ which is discretised into $n \times n$ equally spaced nodes. In this section, we summarise how the numerical scheme presented in \citep{Morrow2019,Morrow2021} can be adapted to solve \eqref{eq:Model1}-\eqref{eq:Model4}.

\begin{figure}
	\centering
	\includegraphics[width=0.3\linewidth]{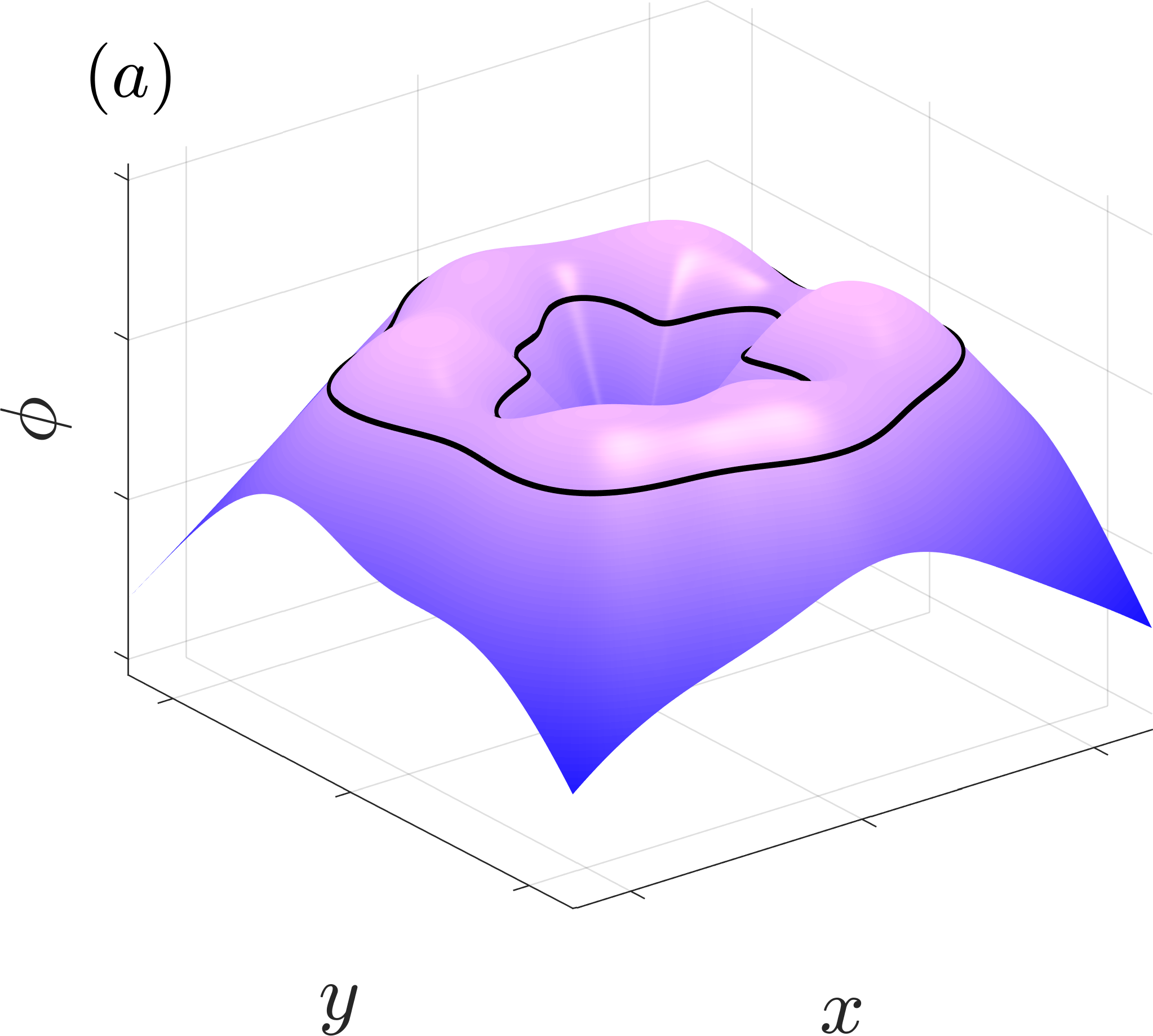}
	\includegraphics[width=0.3\linewidth]{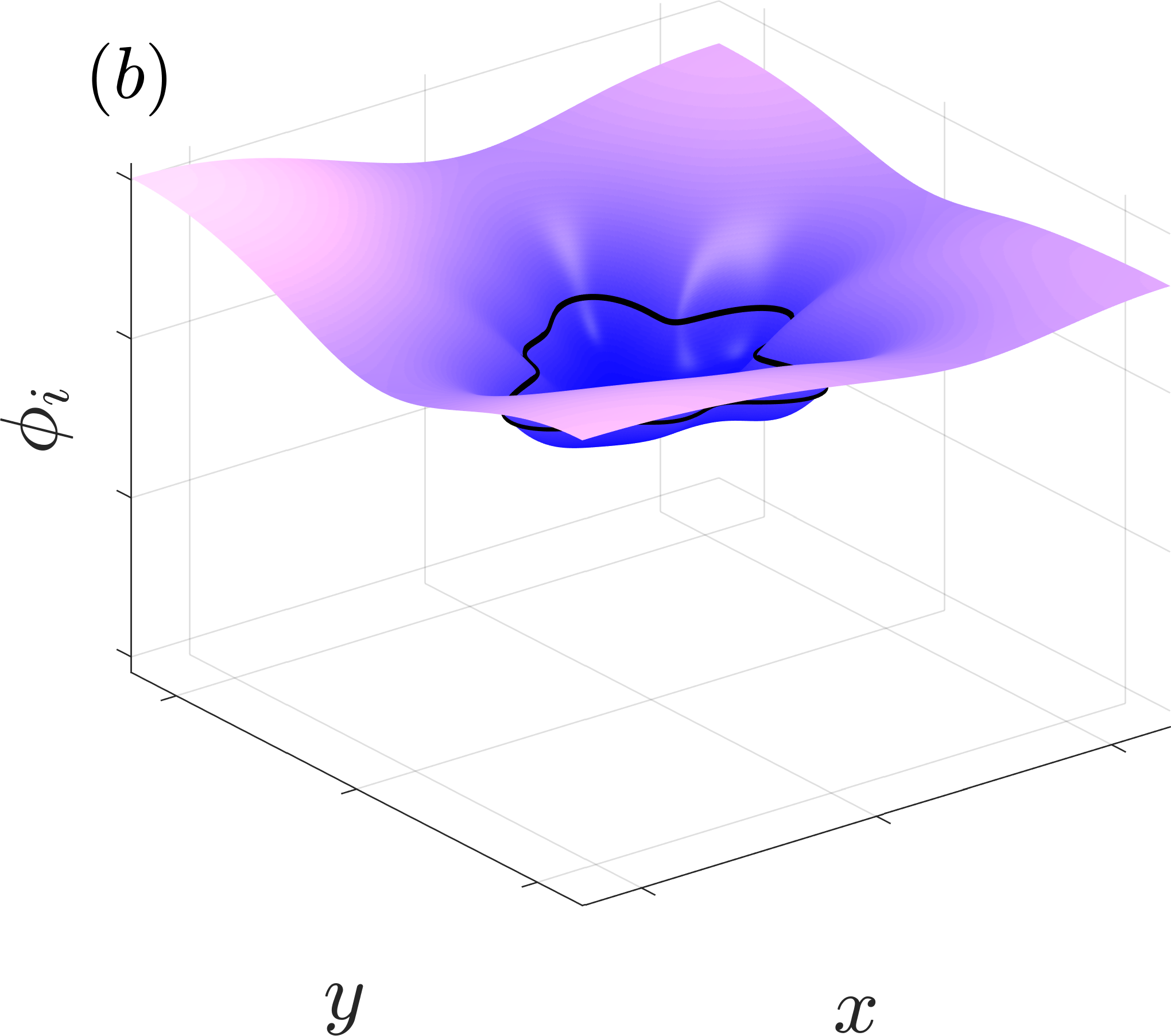}	
	\includegraphics[width=0.3\linewidth]{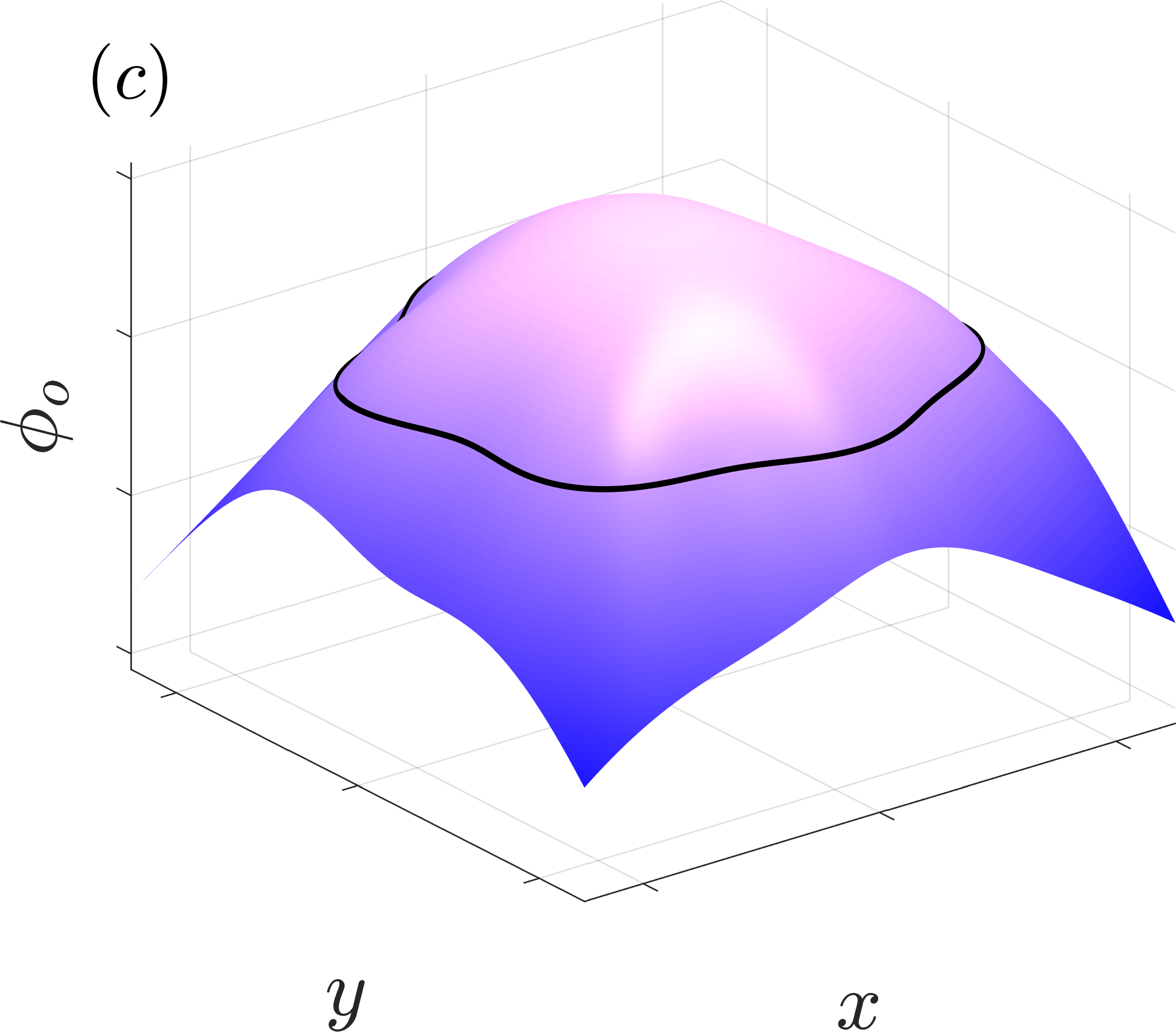}	
	\caption{$(a)$ An example level set function which represents two interfaces such that $\phi>0$ and $\phi<0$ correspond to the viscous and inviscid fluid regions, respectively. $(b)$ and $(c)$ show these two interfaces represented by separate level set functions such that now the viscous fluid region occupies the region where both $\phi_I > 0$ and $\phi_O > 0$.}
	\label{fig:LevelSet}
\end{figure}

It is straightforward to choose a level set function that satisfies $\phi > 0$ where $\bmth{x} \in \Omega$ and $\phi < 0$ otherwise; an example of which is given in Fig.~\ref{fig:LevelSet}$(a)$. As per usual, the location of the interfaces can be found by determining where $\phi$ changes sign. A limitation of this approach is that by representing multiple interfaces with a single level set function, it is not straightforward to determine which of the two interfaces has been found when determining where $\phi$ changes sign. To overcome this issue, we represent each of the interfaces between the inviscid and viscous fluids with a separate level set function, $\phi_I$ and $\phi_O$, as illustrated in Figs \ref{fig:LevelSet}$(b)$ and $(c)$. Thus the viscous fluid will occupy the region where both $\phi_I > 0$ and $\phi_O > 0$; otherwise, the region is filled with inviscid fluid. Both $\phi_I$ and $\phi_O$ are updated according to
\begin{align} \label{eq:DClevelseteqn}
	\frac{\partial \phi_I}{\partial t} + F_I |\grad \phi_I| = 0 \quad \textnormal{and} \quad \frac{\partial \phi_O}{\partial t} + F_O |\grad \phi_O| = 0,
\end{align}
where
\begin{align} \label{eq:SpeedFunction}
	F_I = -\frac{\grad p \cdot \grad \phi_I}{|\grad \phi_I|} \quad \textrm{and} \quad F_O = -\frac{\grad p \cdot \grad \phi_O}{|\grad \phi_O|}.
\end{align}
The concept of representing multiple interfaces with separate level set functions has previously been implemented to study multi-phase moving boundary problems \citep{Gunther2014,Zhao1996}.

\subsubsection{Velocity extension} \label{sec:VelocityExtension}

\begin{figure}
	\centering
	\includegraphics[width=0.35\linewidth]{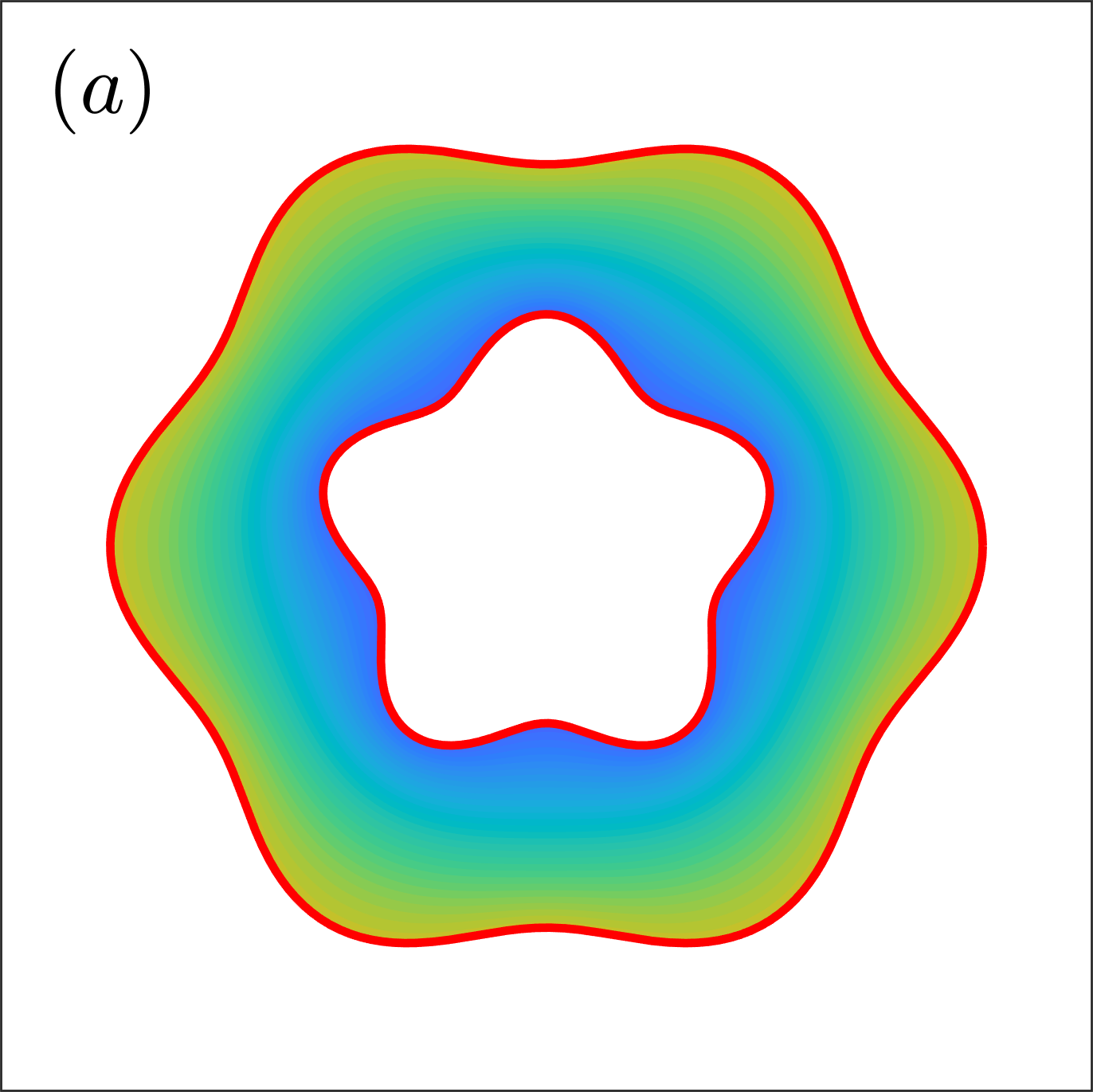}
	\hspace{0.75cm}
	\includegraphics[width=0.35\linewidth]{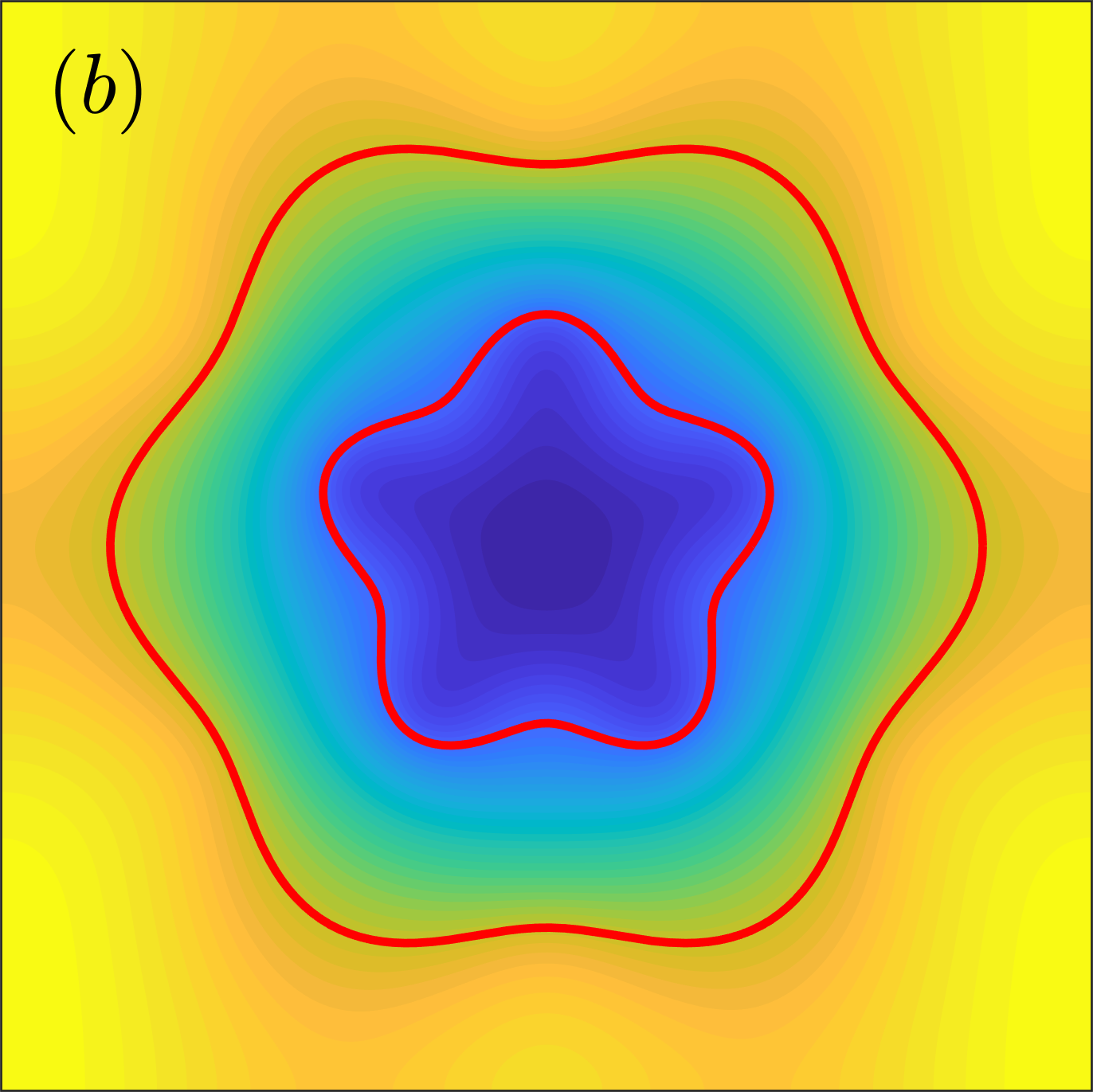}	
	\caption{An illustration of the velocity extension process used to extend $F_i$ and $F_o$ in a doubly connected domain. $(a)$ An example function that is defined in the region where $\boldsymbol{x} \in \Omega(t)$. $(b)$ The region $\boldsymbol{x} \not\in \Omega(t)$ has been `filled-in' by solving the biharmonic equation \eqref{eq:Biharmonic}, which gives a smooth, differentiable function over the entire computational domain.}
	\label{fig:VelocityExtension}
\end{figure}

From \eqref{eq:SpeedFunction}, we have continuous expressions for $F_I$ and $F_O$ in the viscous fluid region, $\bmth{x} \in \Omega$ which satisfy the kinematic boundary conditions \eqref{eq:Model2}. However, to solve the level set equations \eqref{eq:DClevelseteqn}, we require continuous expressions for $F_I$ and $F_O$ over the entire computational domain. As such, we need to extend $F_I$ and $F_O$ into $\bmth{x} \in \mathbb{R}^2 \backslash \Omega$. To achieve this, we solve the biharmonic equation
\begin{align} \label{eq:Biharmonic}
	\nabla^4 F_I = 0\quad \textnormal{and} \quad \nabla^4 F_O=0 \qquad \mbox {in} \quad \bmth{x} \in \mathbb{R} \backslash \Omega(t).
\end{align}
subject to $\nabla F_I = \nabla F_O = 0$ and $\nabla^2 F_I = \nabla^2 F_O = 0$ on $\partial D$. By doing so, we obtain a smooth continuous normal velocity over the entire domain. An advantage of this method, proposed by \citet{Moroney2017}, is that it does not require the location of either interface to be explicitly known. Instead, we form the biharmonic stencil which is modified such that the values where $F_I$ and $F_O$ are already known are not overwritten. The resulting system of linear equations is solved exactly using LU decomposition. This gives a continuous expression for $F_I$ and $F_O$ over the entire computational domain, and ensures that \eqref{eq:Model2} is still satisfied. We illustrate the velocity extension process in Fig.~\ref{fig:VelocityExtension}, where Fig.~\ref{fig:VelocityExtension}$(a)$ shows an example $F$ defined in the region where $\bmth{x} \in \Omega(t)$, while Fig.~\ref{fig:VelocityExtension}$(b)$ show $F$ after solving the biharmonic equation \eqref{eq:Biharmonic} in the region $\bmth{x} \not\in \Omega(t)$. We see that this process gives a smooth differentiable expression for $F$ over the entire computational domain. We have used this biharmonic extension process to solve a variety of different moving boundary problems \citep{Morrow2019,Morrow2019b,Morrow2019c,Morrow2021}, and note that multiply connected domains pose no additional difficulty.

\subsubsection{Solving for the pressure} \label{sec:SolvingVelocityPotential}

To solve for the governing equation for the pressure \eqref{eq:Model1}, we use a modified finite difference stencil to solve Laplace's equation in the region where $\bmth{x} \in \Omega$. For nodes not adjacent to either interface, $p$ is computed using a standard five point stencil such that
\begin{align} \label{eq:FDstabcil}
	\frac{p_{i-1,j} - 2p_{i,j} + p_{i+1,j}}{\Delta x^2} + \frac{p_{i,j-1} - 2p_{i,j} + p_{i,j+1}}{\Delta y^2}=0.
\end{align}
For nodes adjacent to one of the interfaces, where either $\phi_I$ or $\phi_O$ changes sign, we modify our stencil \eqref{eq:FDstabcil} by imposing a ghost node on the interface. For example, suppose that the location of the interior interface, $x_I$, falls between the two nodes $x_{i-1,j}$ and $x_{i,j}$. As $x_{i-1,j}$ is not in the domain $\bmth{x} \in \Omega$, we are unable to use $p_{i-1,j}$ in our finite difference stencil \eqref{eq:FDstabcil}. Instead, following \citet{Chen1997}, we impose a ghost node at $x = x_I$ whose value is $p_I$ such that
\begin{align}
	\frac{\partial^2 p}{\partial x^2} \to \frac{2}{h(\Delta x + h)} p_I -\frac{2}{h + \Delta x}p_{i,j} + \frac{2}{\Delta x (h + \Delta x)}p_{i+1,j},
\end{align}
where
\begin{align}
	h = \Delta x \left| \frac{\phi_{i,j}}{\phi_{i,j} - \phi_{i-1,j}} \right|.
\nonumber
\end{align}
The value of $p_I$ is determined from the dynamic boundary condition \eqref{eq:Model3}, or \eqref{eq:Model4} for the exterior interface.  The curvature term in \eqref{eq:Model3} is computed, as is standard, via $\kappa = \grad \cdot \left( \grad \phi / | \grad \phi | \right)$. Similar adjustments are made if an interface falls between two nodes in the $y$-direction. The resulting linear system of equations is solved exactly using LU decomposition.

The general algorithm for solving \eqref{eq:Model1}-\eqref{eq:Model4}, together with a discussion on numerical verification, is provided in the Appendix~\ref{sec:furthernumerical}.

\subsection{Linear stability analysis}\label{sec:lsa1}

Some insight into how the Saffman-Taylor instability applies on both inner and outer interfaces can be gleaned by applying a standard linear stability analysis.  This approach has been undertaken at length by Ref.~\cite{Anjos2020,Gin2015a,Gin2019,Gin2021} for a variation of the problem where there are three layers of viscous fluids, the lowest viscosity on the inner fluid and the highest viscosity on the outer fluid.  Here we are concentrating on the case in which the innermost and outermost fluids are inviscid, as this is the scenario that is commonplace in experimental work (typically with air injected into a finite region of viscous fluid, also surrounded by air).

We start by denoting the inner and outer interfaces by $r=s_I(\theta,t)$ and $r=s_O(\theta,t)$, respectively.  Leaving out the details, using polar coordinates $(r,\theta)$, by writing out
\begin{equation}
p=P_0(r,t)+\epsilon\, P_1(r,\theta,t)+\mathcal{O}(\epsilon^2),
\label{eq:pLSA}
\end{equation}
\begin{equation}
s_I(\theta,t)=s_{I0}(t)+\epsilon\, s_{I1}(\theta,t)+\mathcal{O}(\epsilon^2),
\quad
s_O(\theta,t)=s_{O0}(t)+\epsilon\, s_{O1}(\theta,t)+\mathcal{O}(\epsilon^2),
\label{eq:sLSA}
\end{equation}
then, to leading order, the location of the interfaces is governed by the nonlinear system of differential equations
\begin{align}
	\frac{\textrm{d} s_{I0}}{\textrm{d} t} = \frac{1}{s_{I0} \ln (s_{O0}/s_{I0})} \left( \Delta p - \left( \frac{1}{s_{I0}} + \frac{1}{s_{O0}} \right)  \right), \label{eq:Circle1} \\
	\frac{\textrm{d} s_{O0}}{\textrm{d} t} = \frac{1}{s_{O0} \ln (s_{O0}/s_{I0})} \left( \Delta p - \left( \frac{1}{s_{I0}} + \frac{1}{s_{O0}} \right)  \right), \label{eq:Circle2}
\end{align}
where $s_{I0}(0)=1$, $s_{O0}(0)=R_O$.  Solving (\ref{eq:Circle1})-(\ref{eq:Circle2}) is a numerical task (although in the zero-surface-tension case the equations can be integrated exactly to write the solution in terms of dilogarithms~\cite{Shargatov2018}), but there are a couple of immediate observations that can be made.  First, by dividing one equation by another and integrating, we arrive at
\begin{equation}
s_{O0}^2-s_{I0}^2=R_O^2-1,
\label{eq:mass}
\end{equation}
which is nothing more than conservation of mass (or area).  Second, we see that the sign of
$$
\Delta p - \left( \frac{1}{s_{I0}} + \frac{1}{s_{O0}} \right),
$$
dictates whether the (unperturbed) interface speeds are positive or negative.  Therefore, we find that if $\Delta p< 1+ R_0^{-1}$ then the system is contracting while if $\Delta p > 1+ R_0^{-1}$ the system is expanding. Note that it is actually possible for the system to contract even if $\Delta p$ is positive, albeit very small (that is, positive but less than $1+ R_0^{-1}$).  For example, looking ahead to our experimental results in Fig.~\ref{fig:FigureX}(a), where the interface expansion is driven by a dimensional pressure differential of $16.2$ kPa, for that particular initial condition, we would need to decrease the pressure differential to be below $0.013$ kPa in order for the system to contract instead of expand.  Despite this technical clarification, we continue to simply associate $\Delta p<0$ with a contracting system and $\Delta p>0$ with an expanding one.

Further detailed calculations show that if we write
\begin{equation}
P_1=(C_n(t)r^n+D_n(t)r^{-n})\cos n\theta,
\quad
s_{I1}=\delta_{In}(t)\cos n\theta,
\quad
s_{O1}=\delta_{On}(t)\cos n\theta,
\label{eq:P1andsI1andsO1}
\end{equation}
for $n\geq 2$, then $\delta_{In}$ and $\delta_{On}$ must satisfy the non-autonomous linear system
\begin{equation}
\frac{\textrm{d}\delta_{In}}{\textrm{d}t}=
\left[\frac{1}{s_{I0}}\left(\frac{n(s_{I0}^{2n}+s_{O0}^{2n})}{s_{O0}^{2n}-s_{I0}^{2n}}
-1\right)\frac{\mathrm{d}s_{I0}}{\mathrm{d}t}
-\frac{n(n^2-1)(s_{I0}^{2n}+s_{O0}^{2n})}{s_{I0}^3(s_{O0}^{2n}-s_{I0}^{2n})}
\right]\delta_{In}
-\frac{2ns_{I0}^ns_{O0}^n}{s_{O0}^{2n}-s_{I0}^{2n}}
\left(\frac{1}{s_{O0}}\frac{\mathrm{d}s_{I0}}{\mathrm{d}t}
+\frac{n^2-1}{s_{I0}s_{O0}^2}\right)\delta_{On},
\label{eq:gamman}
\end{equation}
\begin{equation}
\frac{\textrm{d}\delta_{On}}{\textrm{d}t}=
\frac{2ns_{I0}^ns_{O0}^n}{s_{O0}^{2n}-s_{I0}^{2n}}
\left(\frac{1}{s_{I0}}\frac{\mathrm{d}s_{O0}}{\mathrm{d}t}
-\frac{n^2-1}{s_{I0}^2s_{O0}}\right)\delta_{In}
+\left[-\frac{1}{s_{O0}}\left(\frac{n(s_{I0}^{2n}+s_{O0}^{2n})}{s_{O0}^{2n}-s_{I0}^{2n}}
+1\right)\frac{\mathrm{d}s_{O0}}{\mathrm{d}t}
-\frac{n(n^2-1)(s_{I0}^{2n}+s_{O0}^{2n})}{s_{O0}^3(s_{O0}^{2n}-s_{I0}^{2n})}
\right]\delta_{On}.
\label{eq:deltan}
\end{equation}
Even though (\ref{eq:gamman})-(\ref{eq:deltan}) is linear, it is rather difficult to understand the qualitative behaviour of this system since the coefficients are time-dependent and depend on the solution of the leading order problem (\ref{eq:Circle1})-(\ref{eq:Circle2}).  As such, we cannot simply compute eigenvalues of a constant matrix, for example, to determine stability.

However, for a start we note two limits of (\ref{eq:gamman})-(\ref{eq:deltan}).  In the limit $s_{O0}\rightarrow\infty$, which corresponds to a traditional model like that illustrated in Fig.~\ref{fig:Figure1}$(a)$, then
$$
\frac{\textrm{d}\delta_{In}}{\textrm{d}t}\sim \left(\frac{n-1}{s_{I0}}\frac{\mathrm{d}s_{I0}}{\mathrm{d}t}
-\frac{n(n^2-1)}{s_{I0}^3}\right)\delta_{In},
$$
which is the usual equation for an inviscid bubble inside an infinite body of viscous fluid.  Here the stability is clearer with a contracting bubble being stable to perturbations, while an expanding bubble being unstable to sufficiently small wavenumbers \citep{Dallaston2013,Miranda1998,Paterson1981}.  Further, in the limit $s_{I0}\rightarrow 0$, which corresponds to the so-called ``blob'' problem, we have
$$
\frac{\textrm{d}\delta_{On}}{\textrm{d}t}\sim
\left(-\frac{n+1}{s_{O0}}\frac{\mathrm{d}s_{O0}}{\mathrm{d}t}
-\frac{n(n^2-1)}{s_{O0}^3}\right)\delta_{On}.
$$
Again, in this single interface limit the stability is easier to interpret, with an expanding bubble clearly stable and a contracting bubble unstable to perturbations with sufficiently small wavenumbers \citep{Chen2014,Kelly1997,Paterson1981}.

Returning to a qualitative description of the full system (\ref{eq:gamman})-(\ref{eq:deltan}), we see the terms involving $n^2-1$ are due to surface tension (which has been scaled to unity in our formulation).  Thus, as we expect, surface tension has the effect of stabilising both interfaces (just as it does in the two single-interface limits above), especially for higher modes with large $n$ (and small wavelength).  On the other hand, the effect of the speeds $\mathrm{d}s_{I0}/\mathrm{d}t$ and $\mathrm{d}s_{O0}/\mathrm{d}t$ is more interesting.  First, as we expect, if both speeds are positive then that has the effect of destabilising the inner interface via the term $(\mathrm{d}s_{I0}/\mathrm{d}t)\delta_{In}$ in (\ref{eq:gamman}), which is what normally happens with the Saffman-Taylor instability.  That is, the inviscid fluid displacing the viscous fluid causes the interface to be unstable and leads to viscous fingering patterns in the usual way.  However, we see in this scenario that the term $(\mathrm{d}s_{O0}/\mathrm{d}t)\delta_{In}$ in (\ref{eq:deltan}) also acts to destabilise the outer interface.  This tendency for the outer interface to be unstable applies even in the more extreme case for which it is initially perfectly circular and so $\delta_{On}(0)=0$ for $n\geq 2$.  This result is not entirely expected since the outer interface would otherwise be stable since, in isolation, we have a viscous fluid displacing an inviscid fluid.  Therefore, according to this model, the instability on the inner interface appears to ``infect'' the outer interface by causing a (more mild) instability there too.  Similar arguments hold when both speeds $\mathrm{d}s_{I0}/\mathrm{d}t$ and $\mathrm{d}s_{O0}/\mathrm{d}t$ are negative, in which case the outer interface is unstable in the usual way, but the inner interface is also subject to a (more mild) instability.

\begin{figure}
	\centering
	\includegraphics[width=0.4\linewidth]{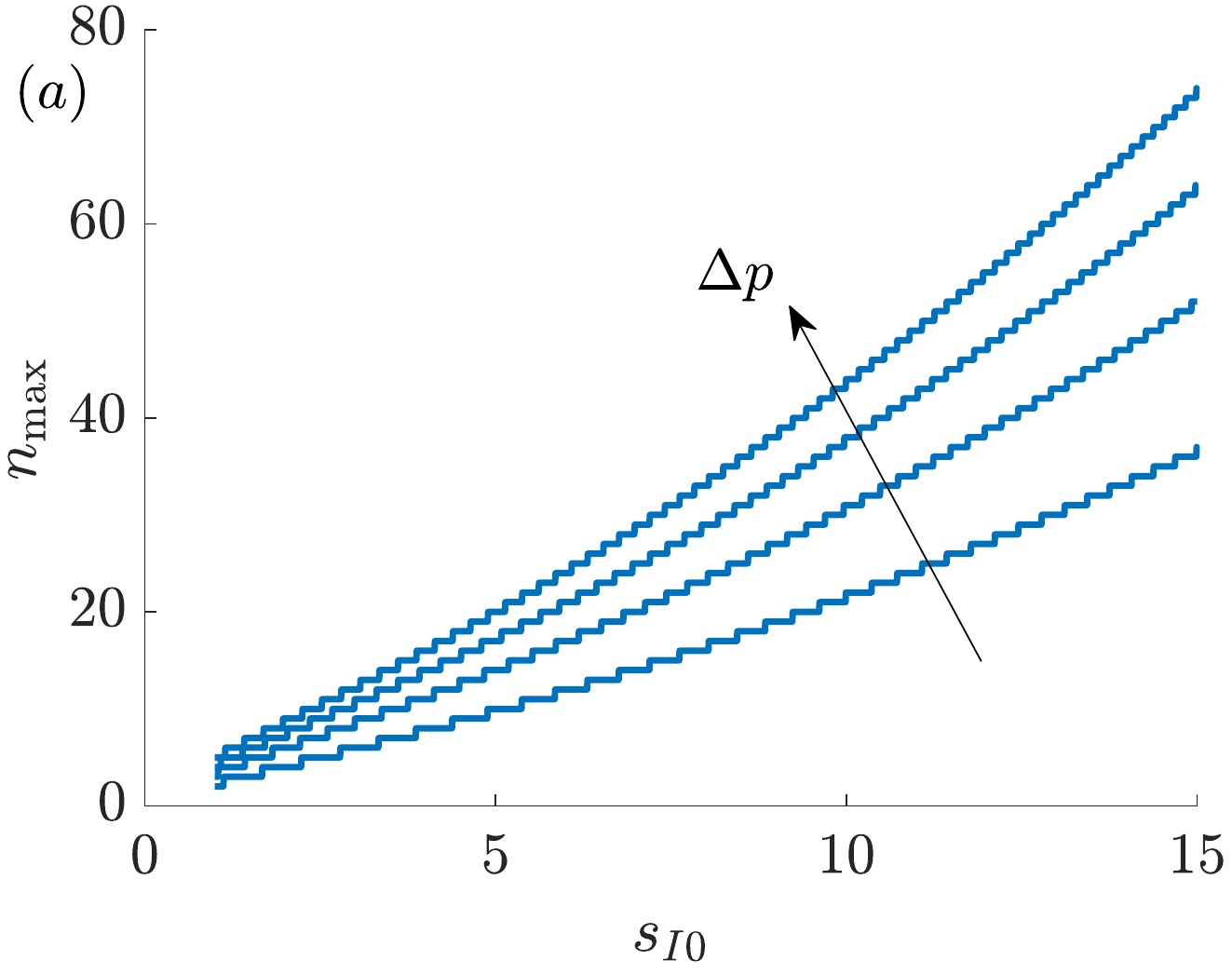}
	\includegraphics[width=0.4\linewidth]{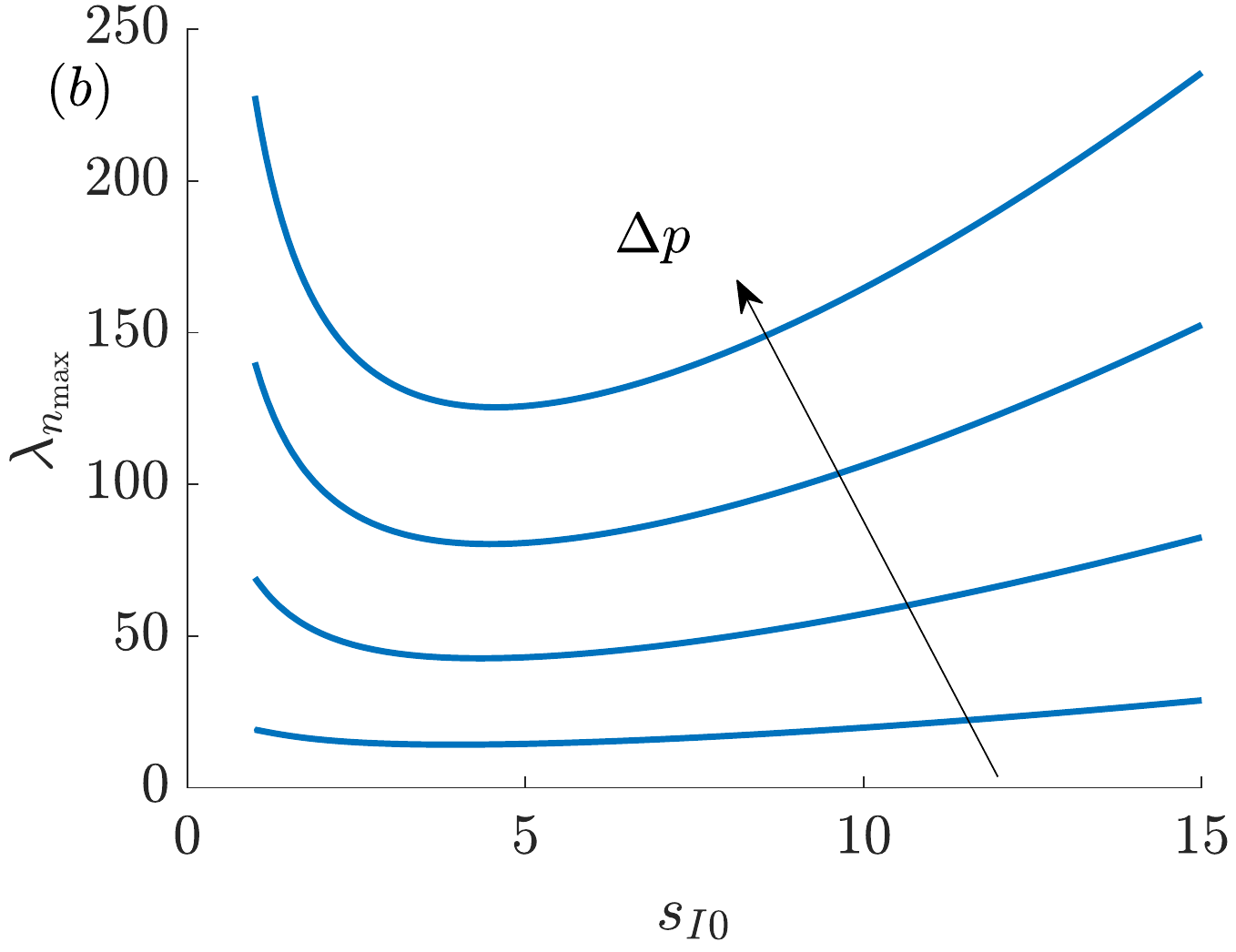}
	
	\includegraphics[width=0.4\linewidth]{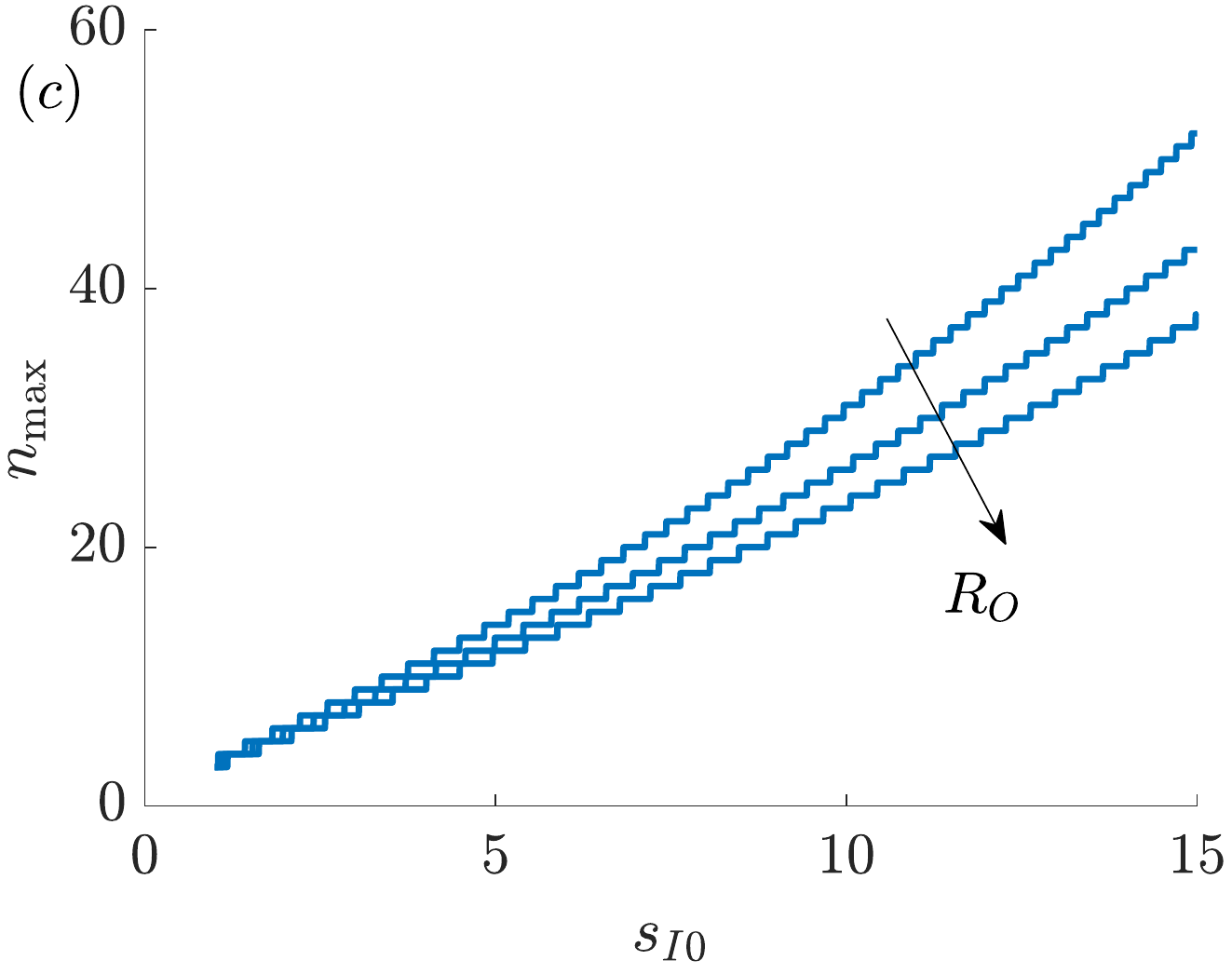}
	\includegraphics[width=0.4\linewidth]{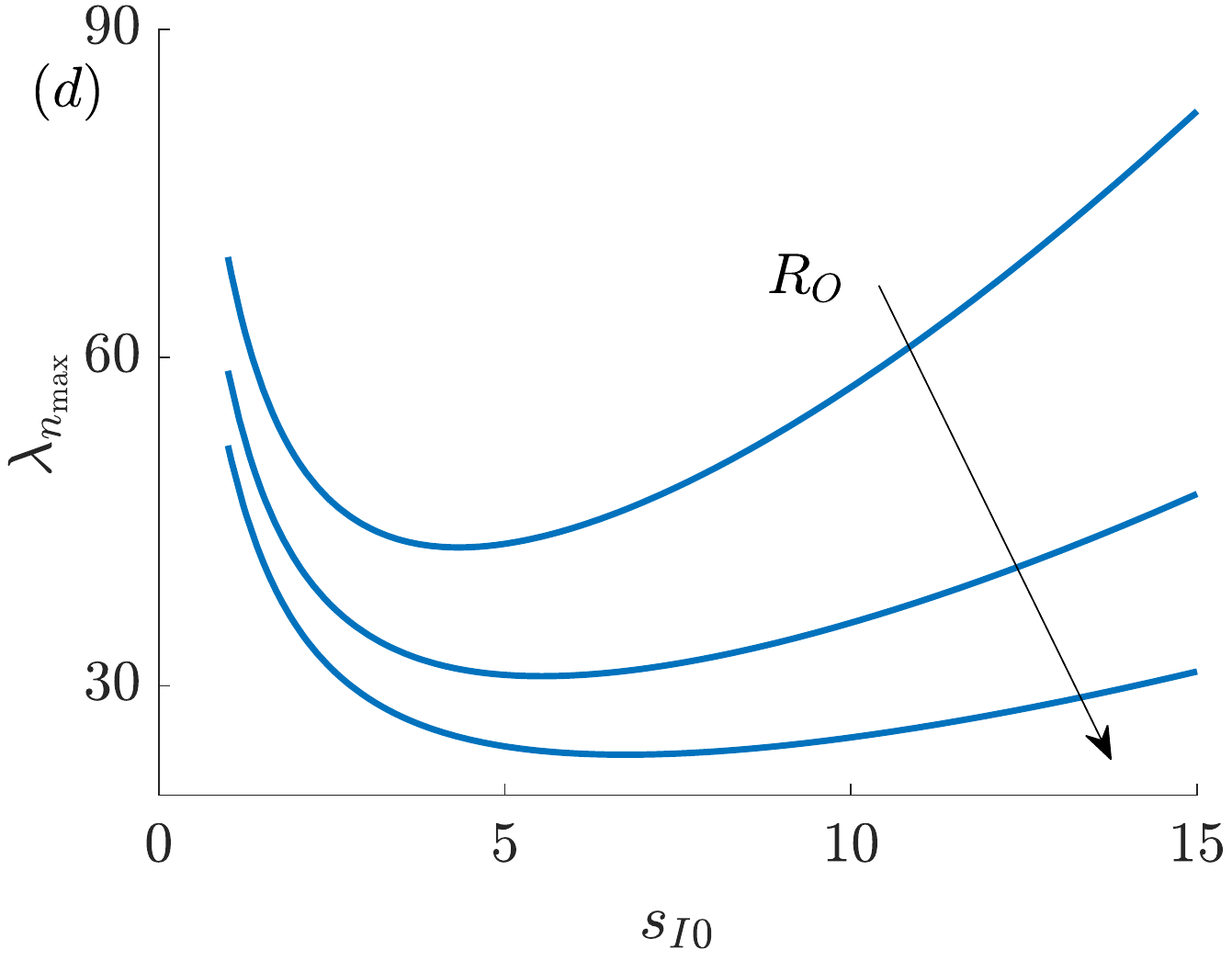}
	\caption{$(a)$ The most unstable mode of perturbation, $n_{\max}$, computed from \eqref{eq:gamman} and \eqref{eq:deltan} with $R_O = 10$ and (bottom to top) $\Delta p = 50$, 100, 150, and 200. $(b)$ Corresponding maximum eigenvalue at $n = n_{\max}$. $(c)$ $n_{\max}$ and $(d)$ $\lambda_{n_{\max}}$ with $\Delta p = 100$ and (top to bottom) $R_O = 10$, 12.5, and 15.}
	\label{fig:LinearStabiliityAnalysis}
\end{figure}

In an attempt to better understand the system (\ref{eq:gamman})-(\ref{eq:deltan}), we assume quasi-steady-state conditions, whereby modes of perturbation grow or decay on time-scales that are much faster than the base state evolves.  We can then interpret (\ref{eq:gamman})-(\ref{eq:deltan}) at each time as being a constant-coefficient system characterised by a $2\times 2$ matrix.  The largest eigenvalue of this matrix $\lambda_n$ is then a tentative measure of the growth rate of the $n$th mode at that time (assuming also that the perturbations remain small).  As we sweep across all $n$, we can compute what we may speculate is the most unstable mode $n_{\max}$ and corresponding growth rate $\lambda_{n_{\max}}$.  Provided $\lambda_{n_{\max}}>0$, the interpretation is that $n_{\max}$ is an estimate for the number of fingers that emerge at that particular time.  For example, in Fig.~\ref{fig:LinearStabiliityAnalysis}, we have plotted both $n_{\max}$ and $\lambda_{n_{\max}}$ as a function of $s_{I0}$ for a variety of combinations of $\Delta p$ and $R_O$.  We see in (a) and (c) that the most unstable mode $\lambda_{n_{\max}}$ increases with $s_{I0}$, which is consistent with observations of tip-splitting and increasing number of fingers on the inner interface as it expands.  Further, these plots show that, for a fixed inner radius $s_{I0}$, we expect more fingers for higher values of $\Delta p$ or lower values of $R_O$, which is consistent with the observation that a higher pressure difference or smaller distance between interfaces leads to higher speeds near finger tips compared to fjords, which leads to the classical Saffman-Taylor instability.  Interestingly, we see in  Fig.~\ref{fig:LinearStabiliityAnalysis} (b) and (d) that the growth rate of the most unstable mode decreases and then increases as the inner interface expands, although, again, these ideas extend beyond the limitations of linear stability analysis as perturbations themselves are no longer small.

\subsection{Results for expanding bubble} \label{sec:Expanding}

\subsubsection{Numerical simulations} \label{sec:NumericalSimulations}

In this section, we focus on the case in which $\Delta p > 1 + R_0^{-1}$, resulting in an expanding inner inviscid bubble with an unstable boundary.  While typical models for inviscid bubbles expanding into an infinite body of viscous fluid are set up to have a constant injection rate of inviscid fluid and therefore a constant rate of increase in bubble area \cite{Paterson1981}, our choice of constant pressure differential $\Delta p$ leads to a variable injection rate and non-constant rate of area increase that must be determined as part of the solution process.  We believe our choice of constant pressure difference $\Delta p > 0$ reflects natural experimental conditions for the doubly connected geometry in question, which requires a finite pressure on the outer interface (as opposed to models with an infinite body of viscous fluid, which have pressure increasing logarithmically in the far-field). Further, both our experiments and those of Ward \& White~\cite{Ward2011} involve a constant pressure differential, which provides further motivation for the boundary conditions in our model.

\begin{figure}
	\centering
	\begin{tabular}{c|ccc}
		\hline
		& $\Delta p = 50$ & $\Delta p = 100$ & $\Delta p = 300$ \\
		\hline
		\rotatebox{90}{\hspace{1.5cm}$R_O = 10$} & \includegraphics[width=0.25\linewidth]{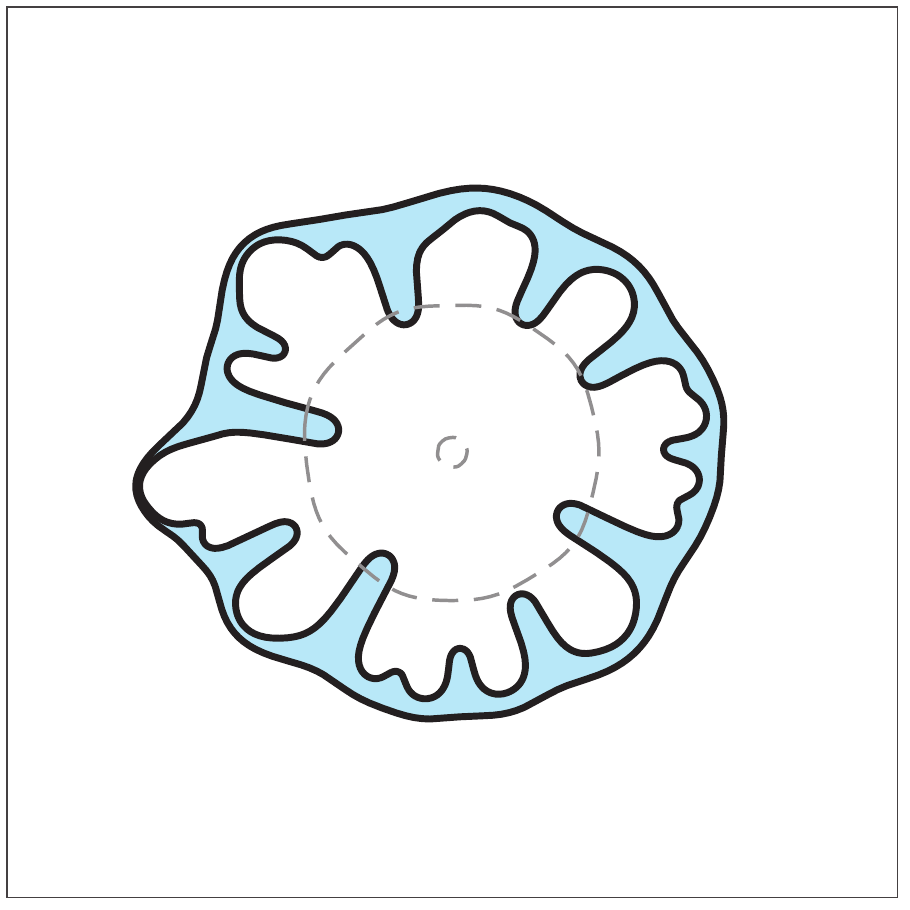} & \includegraphics[width=0.25\linewidth]{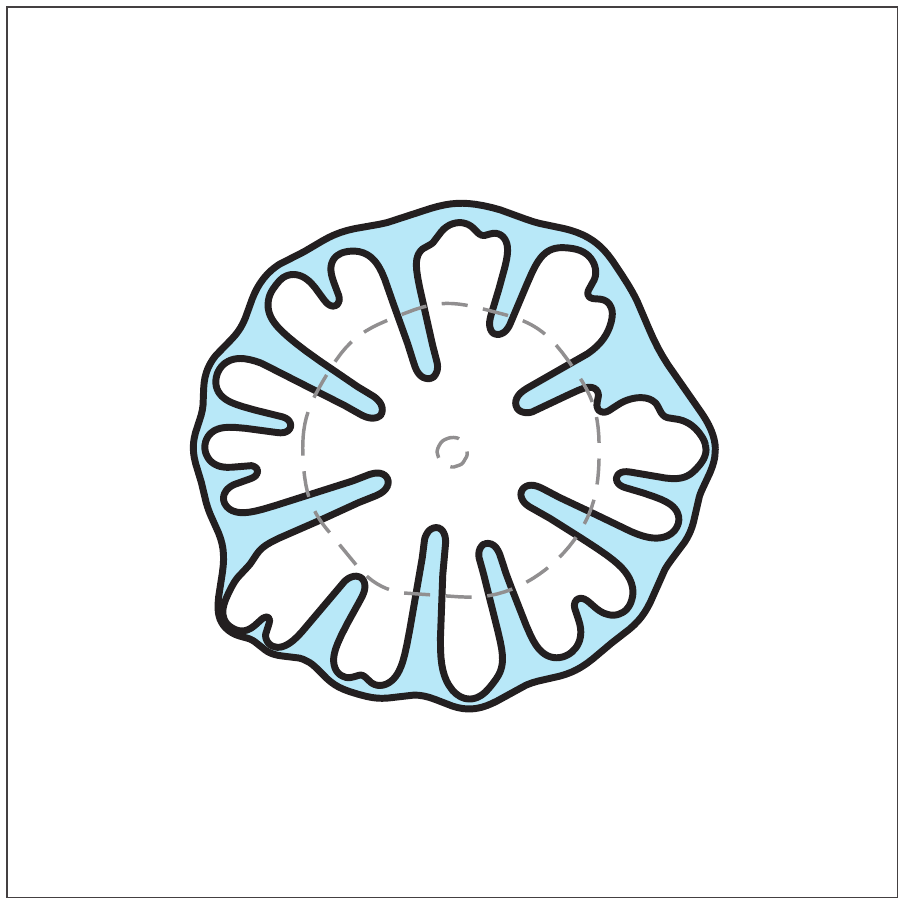}  & \includegraphics[width=0.25\linewidth]{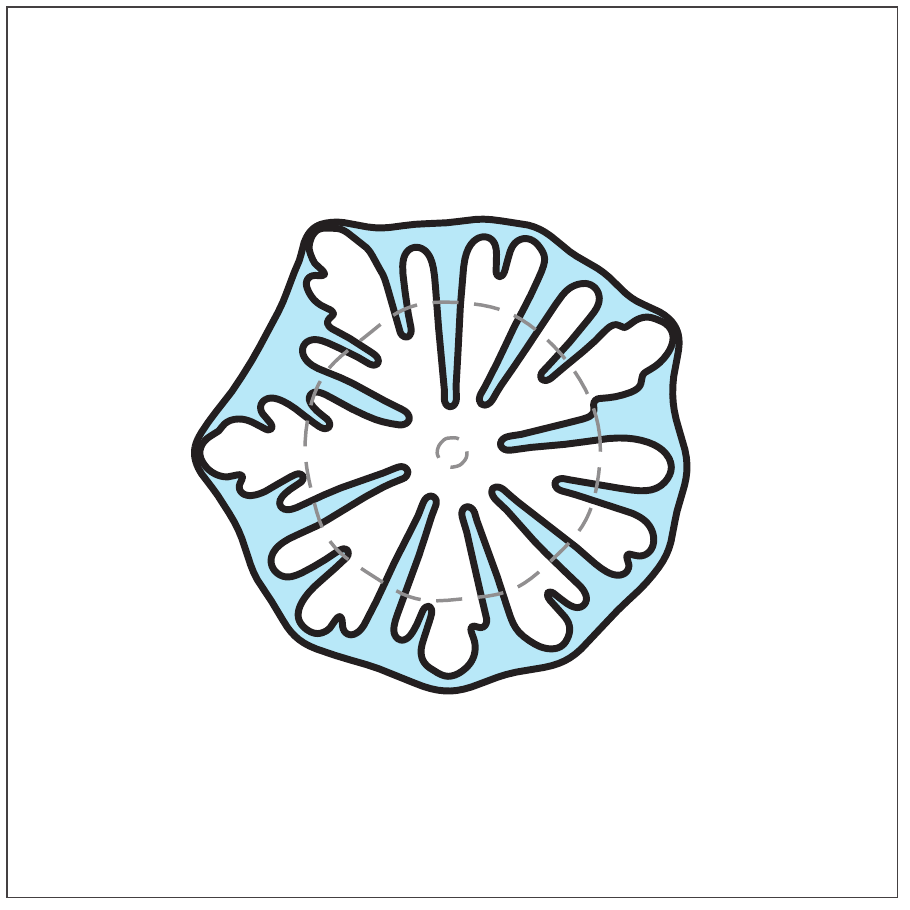}  \\
		\rotatebox{90}{\hspace{1.5cm}$R_O = 12.5$} & \includegraphics[width=0.25\linewidth]{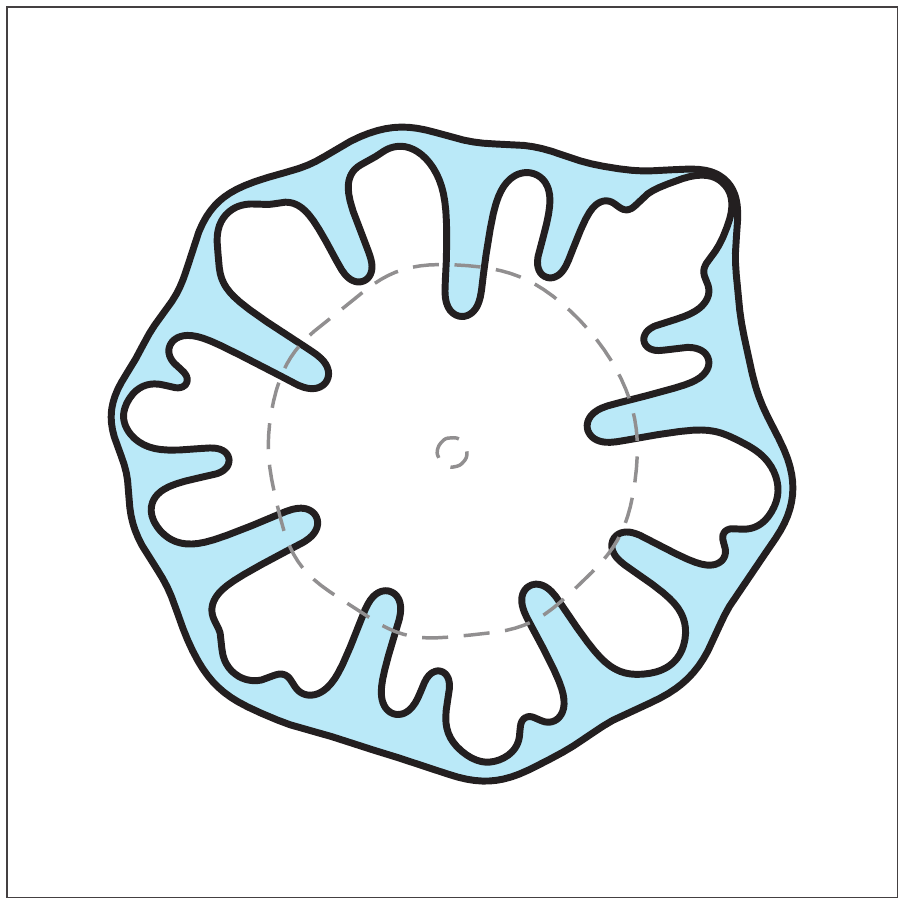} & \includegraphics[width=0.25\linewidth]{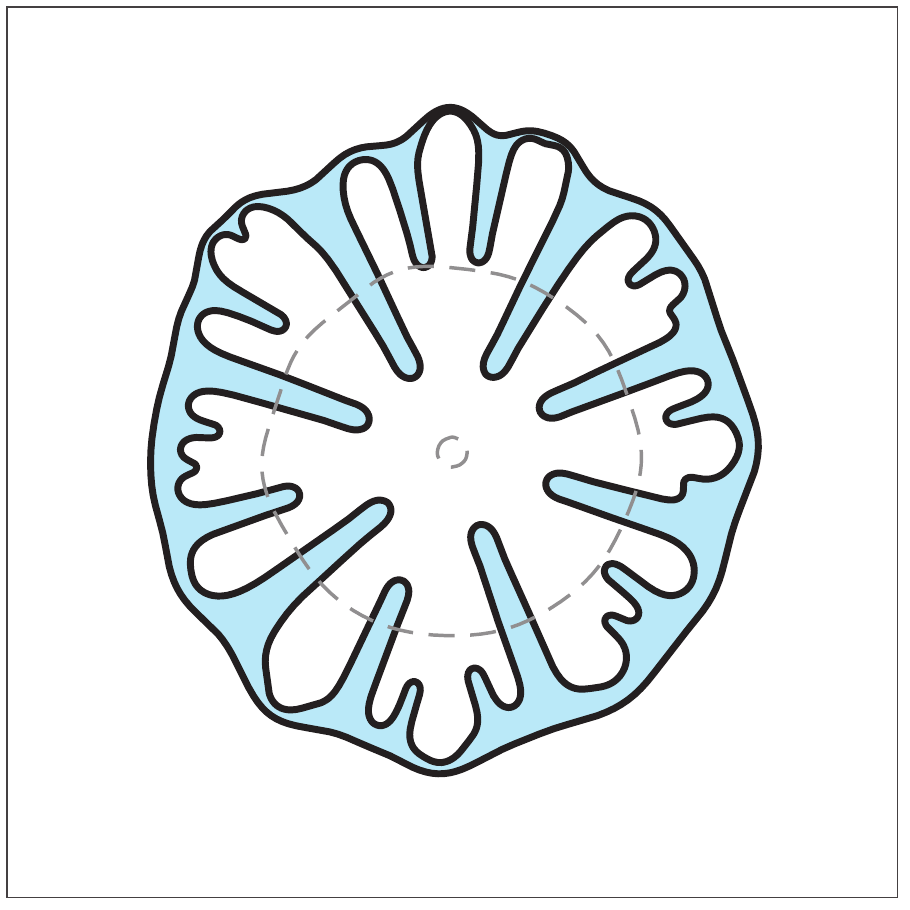}  & \includegraphics[width=0.25\linewidth]{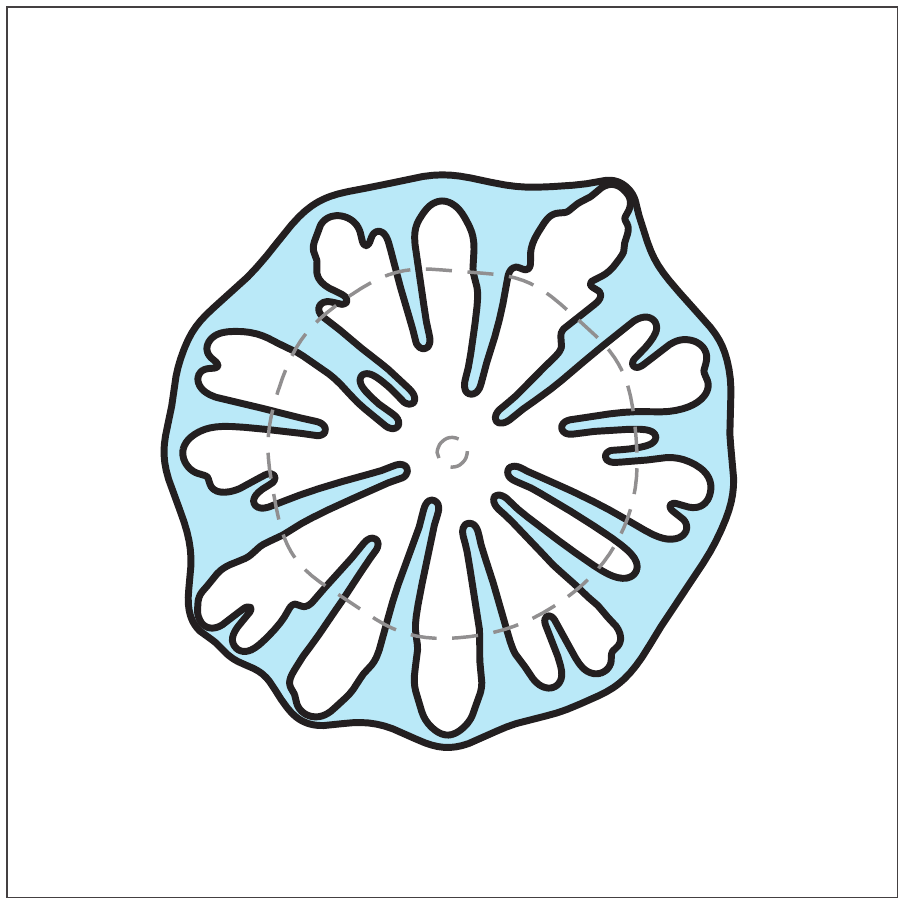}  \\
		\rotatebox{90}{\hspace{1.5cm}$R_O = 15$} & \includegraphics[width=0.25\linewidth]{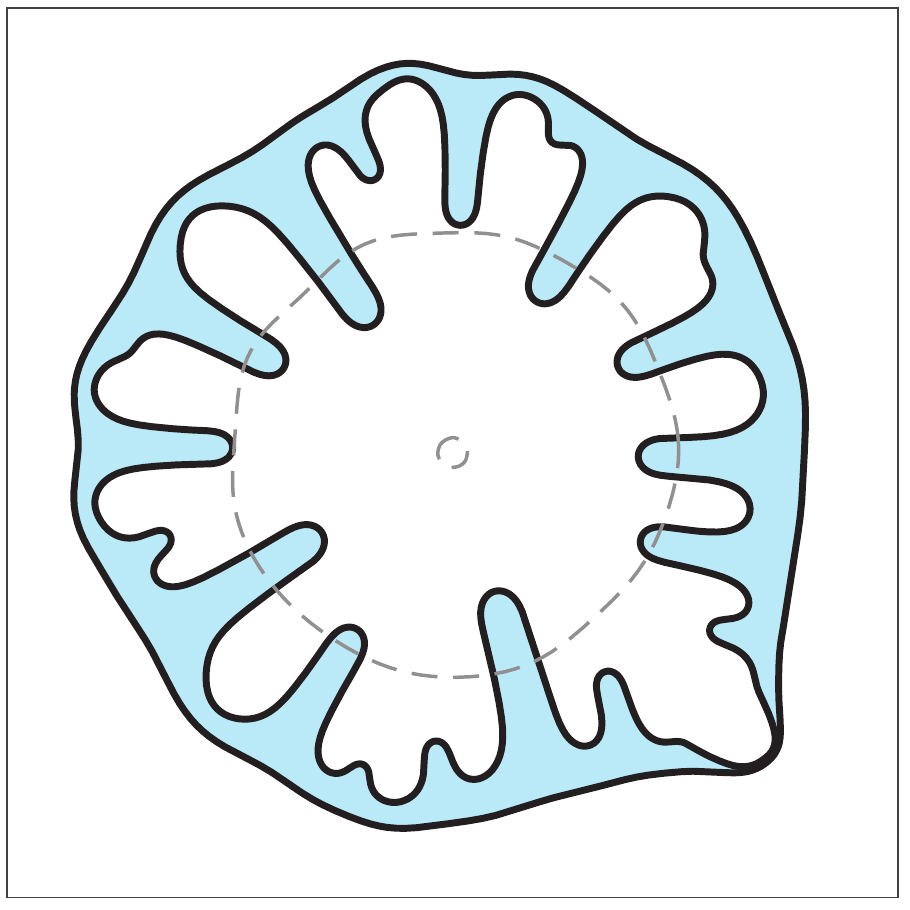} & \includegraphics[width=0.25\linewidth]{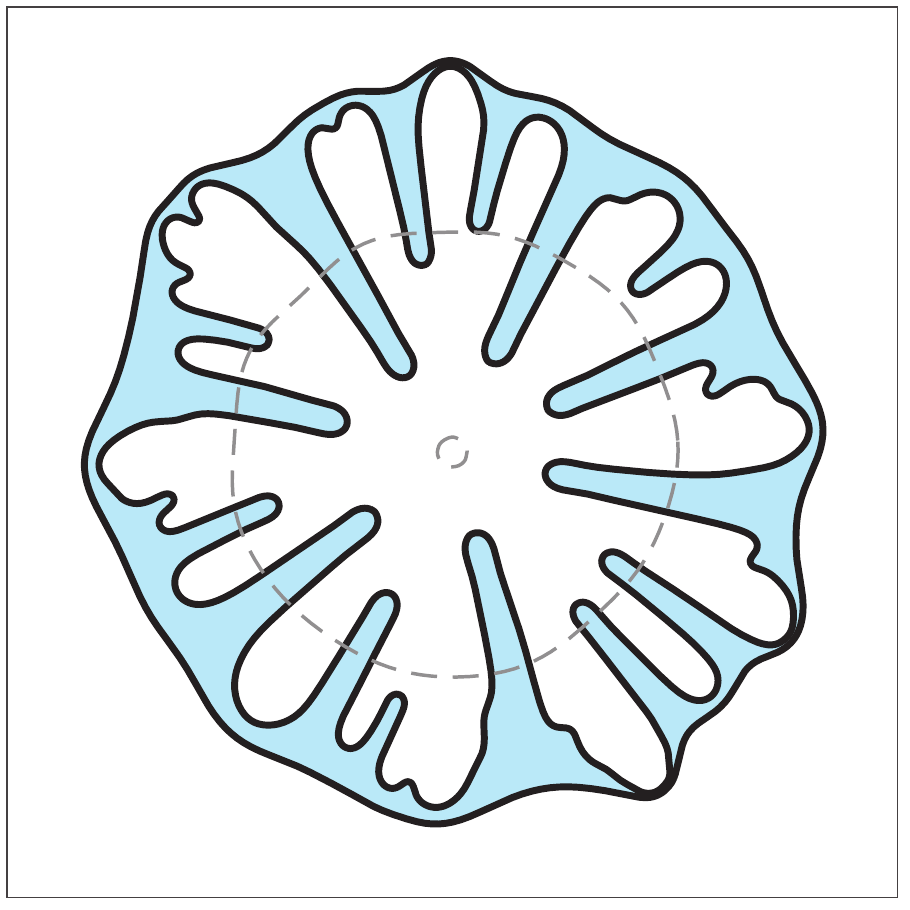}  & \includegraphics[width=0.25\linewidth]{Fig6h}  \\
	\end{tabular}
	\caption{Numerical solutions of \eqref{eq:Model1}-\eqref{eq:Model4} for different values of $\Delta p$ and $R_O$.
		Simulations are performed with the initial conditions \eqref{eq:RadialIC} where $N = 12$ and $\varepsilon = 5 \times 10^{-3}$. The dotted (grey) curves represent the initial condition for the inner and outer boundaries. Simulations are performed on the domain $-30 \le x \le 30$ and $-30 \le y \le 30$ using $900 \times 900$ equally spaced grid points.}
	\label{fig:DoublyConnected}
\end{figure}

In this section, we consider the initial conditions for the inner and outer boundaries
\begin{equation}
	s_I(\theta, 0) = 1 + \varepsilon \sum_{m = 2}^{N} \cos ( m (\theta - \theta_m) ),  \quad
	s_O(\theta, 0) = R_O + \varepsilon \sum_{n = 2}^{N} \cos ( n (\theta - \theta_n) ).  \label{eq:RadialIC}
\end{equation}
Here $\theta_m$ and $\theta_n$ are randomly chosen values between $0$ and $2\pi$, included to ensure the various sinusoidal modes are out of phase (as would be expected in a physical system).
Figure~\ref{fig:DoublyConnected} shows a selection of numerical solutions of \eqref{eq:Model1}-\eqref{eq:Model4} with (rows one to three) $R_O = 10$, 12.5, and 15, and (columns one to three) $\Delta p = 50$, $100$, and $300$. Solutions are shown at the time when simulations are stopped. For each parameter combination considered, the trailing interface destabilises while the leading interface remains near circular over the duration of a simulation. We find that both the number and length of fingers increases with $\Delta p$ (left to right), and tip-splitting behaviour is more pronounced. For a fixed value of $\Delta p$, we find that a larger number of fingers develop as $R_O$ is increased (top to bottom), however the length of these fingers relative to the size of the bubble does not appear to significantly vary with $R_O$. Note that the area enclosed by the inner bubble at the time simulations are halted increases noticeably with $R_O$. These observations are consistent with previous experimental results (see figure 2 in \citet{Ward2011} for example) and the linear stability analysis in the previous subsection. For the latter, we can use the results in Fig.~\ref{fig:LinearStabiliityAnalysis}(a) and (c) to tentatively predict the number of fingers in the simulation.  While counting fingers is difficult to nail down, as tip-splitting continues throughout the simulations and so some smaller protrusions may or may not be considered fingers, we see these predictions from linear theory appear to slightly over-estimate the number of fingers in the simulations.

\begin{figure}
	\centering
	\includegraphics[width=0.19\linewidth]{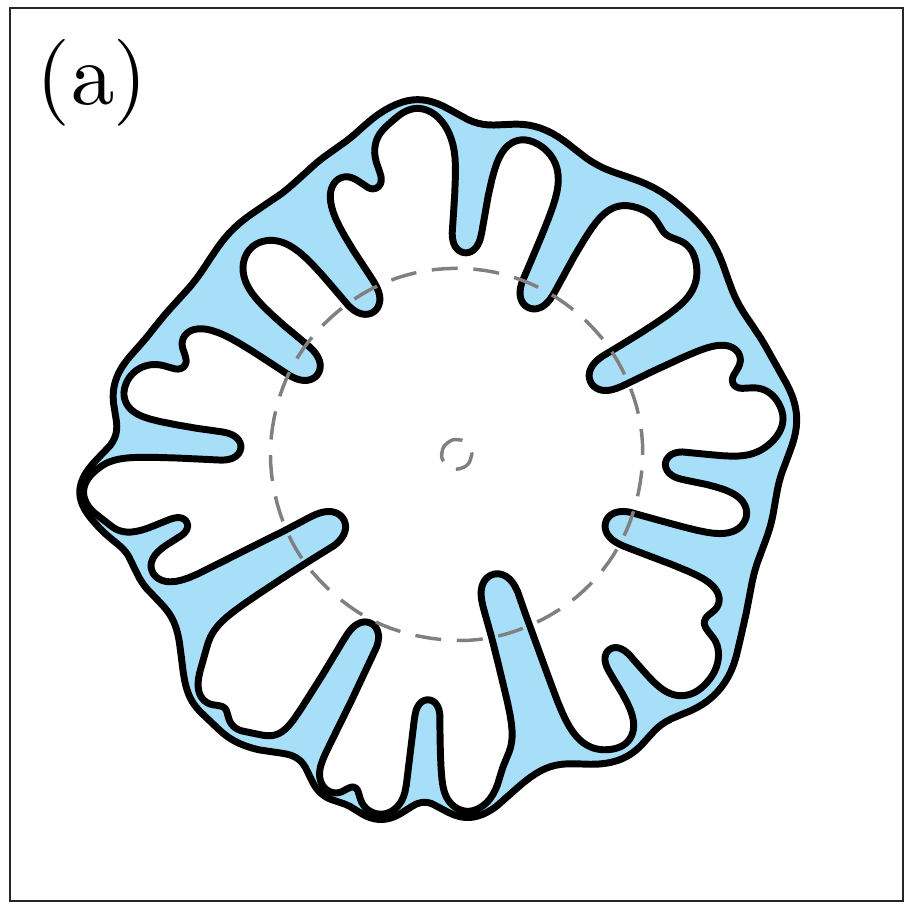}
	\includegraphics[width=0.19\linewidth]{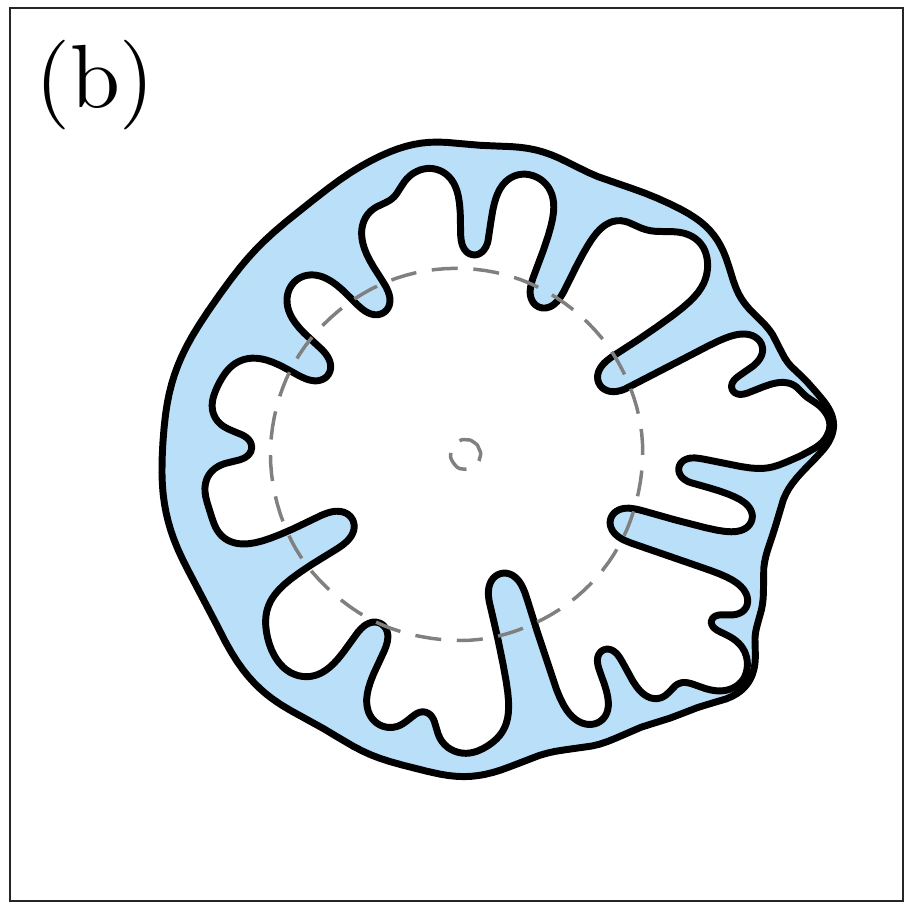}
	\includegraphics[width=0.19\linewidth]{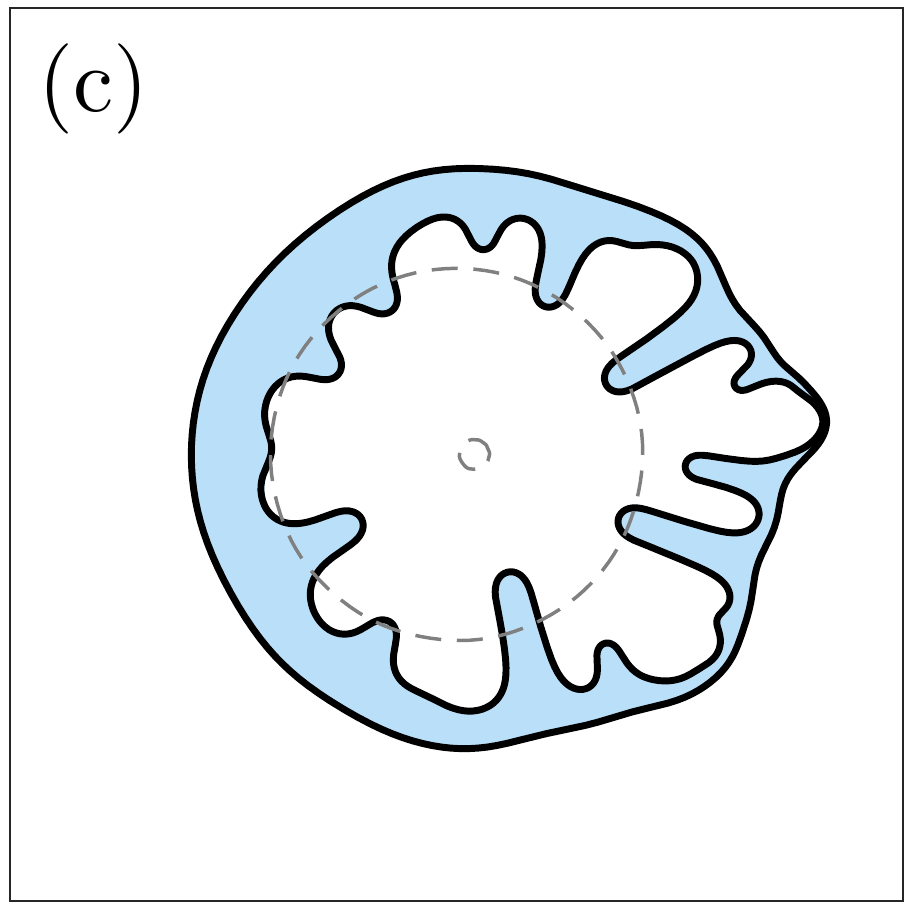}
	\includegraphics[width=0.19\linewidth]{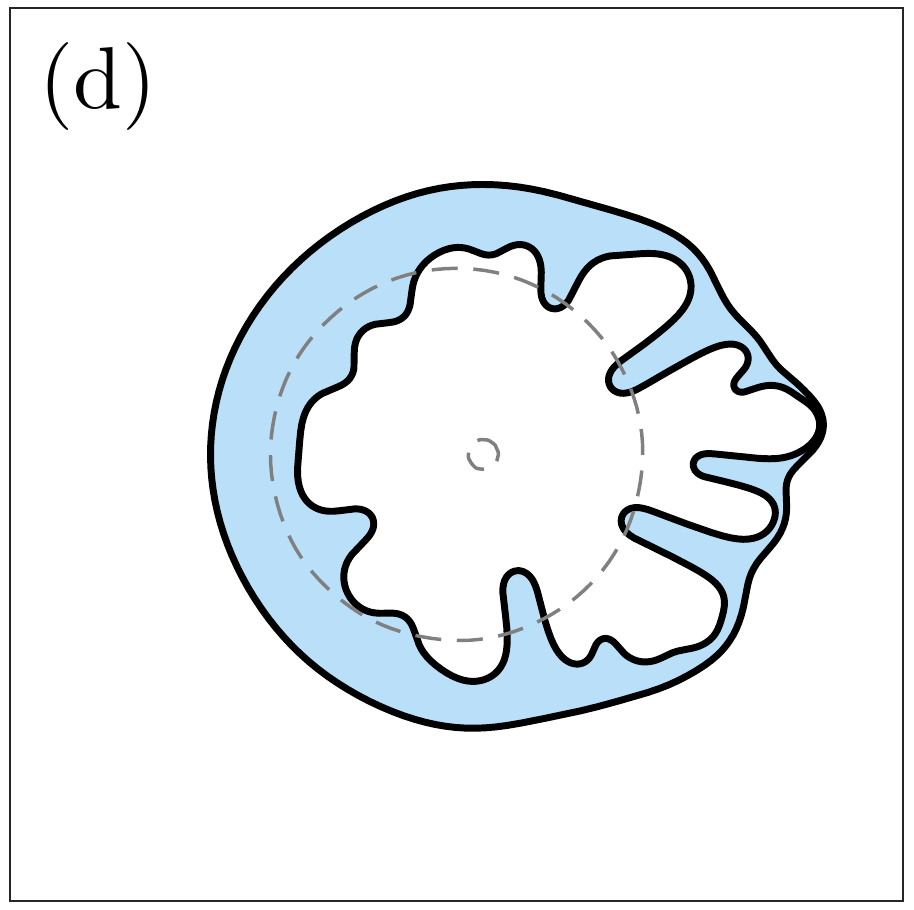}
	\includegraphics[width=0.19\linewidth]{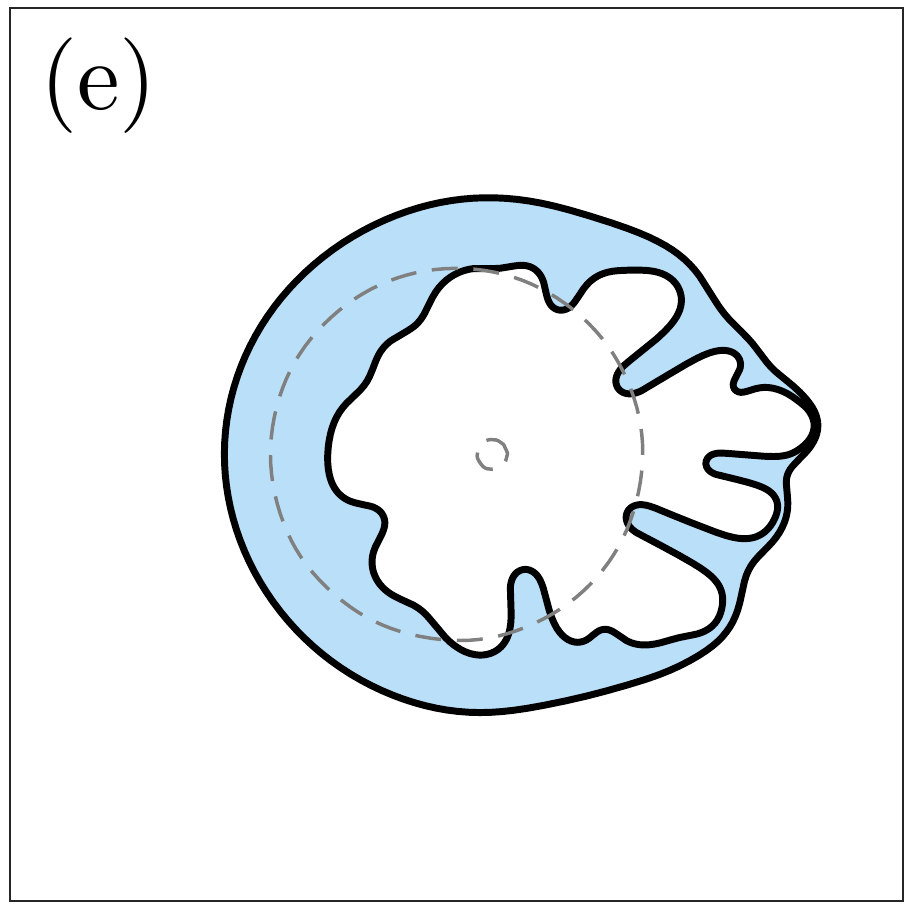}
	\caption{Numerical solutions of \eqref{eq:Model1}-\eqref{eq:Model4} for the value $\Delta p= 50$ and $R_O= 12.5$. Simulations are performed with the initial conditions \eqref{eq:RadialIC}, where $N = 12$ and $\varepsilon = 5 \times 10^{-3}$, except the centre of the inner boundary is shifted in the positive $x$-direction by $(a)$ 0 $(b)$ 0.6 $(c)$ 1.2 $(d)$ 1.8 and $(e)$ 2.4. The corresponding bursting times are $t_{\textrm{burst}} = 1.352$, 1.283, 1.18, 1.079, and 0.983.}
	\label{fig:offcentre}
\end{figure}

In practice, even if both the inner and outer interfaces are initially near circular as with the simulations in Fig.~\ref{fig:DoublyConnected}, the centres of the circles are not likely to perfectly align with each other.  In that case, this asymmetry will cause a preferential fingering pattern in the direction in which the two interfaces are closest.  For example, Fig.~\ref{fig:offcentre}(a) shows a simulation from an initial condition based on perturbed circular interfaces with a common centre, while the other four simulations in Fig.~\ref{fig:offcentre}(b)-(e) are for the same parameter values except that the centre of the inner base circle is shifted to the right.  It is clear that the prominent fingering in Fig.~\ref{fig:offcentre}(c)-(e) is towards the right and not at all evenly spaced in the radial direction.  This simple example shows how nonlinear competition between fingers favours those fingers with a very slight advantage in pressure gradient and, furthermore, demonstrates how sensitive the system is to this type of realistic symmetry breaking that is likely to occur in real experiments.

\subsubsection{Experimental results}

\begin{figure}
	\centering
	\includegraphics[width=0.15\linewidth]{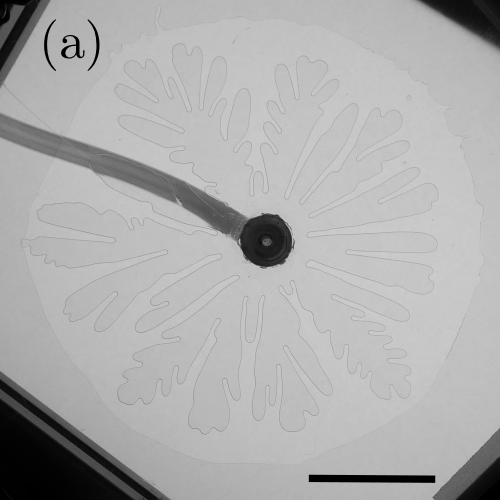}
	\includegraphics[width=0.15\linewidth]{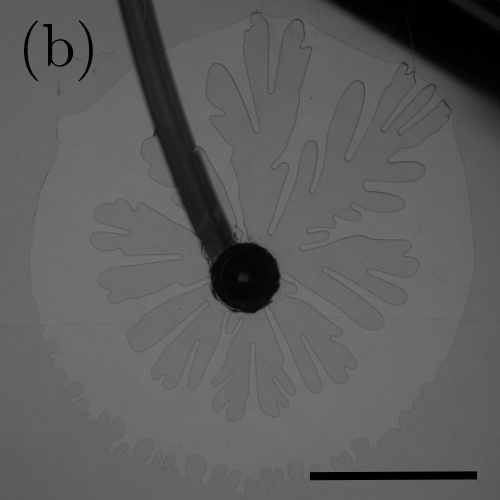}
	\includegraphics[width=0.15\linewidth]{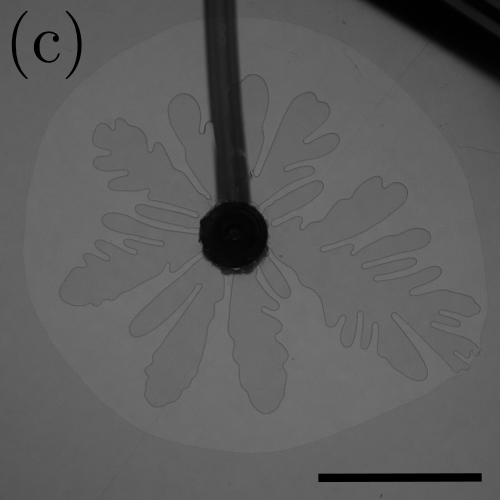}
	\includegraphics[width=0.15\linewidth]{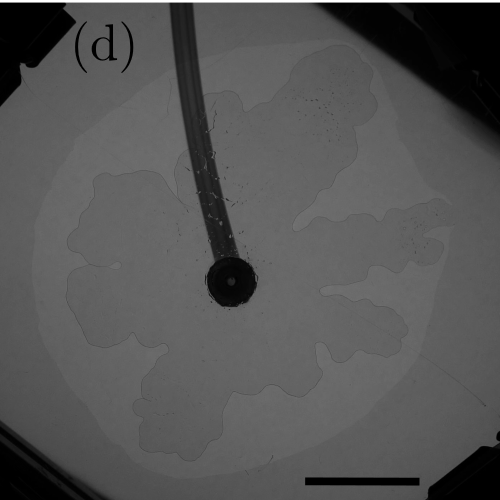}
	\includegraphics[width=0.15\linewidth]{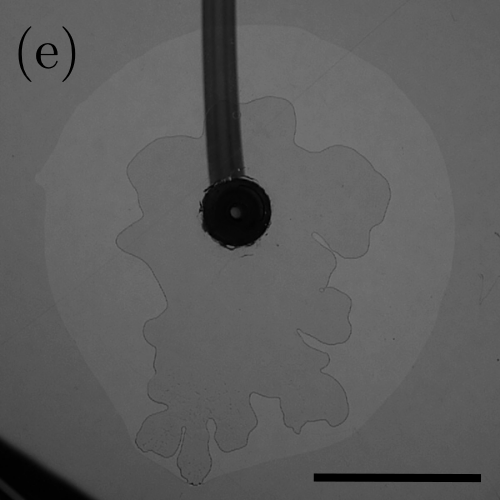}
	\includegraphics[width=0.15\linewidth]{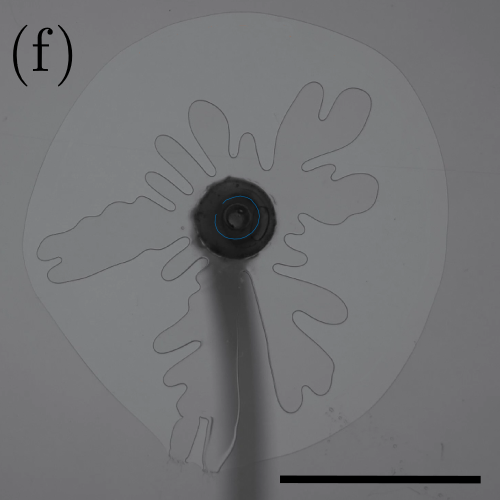}
	
	\includegraphics[width=0.15\linewidth]{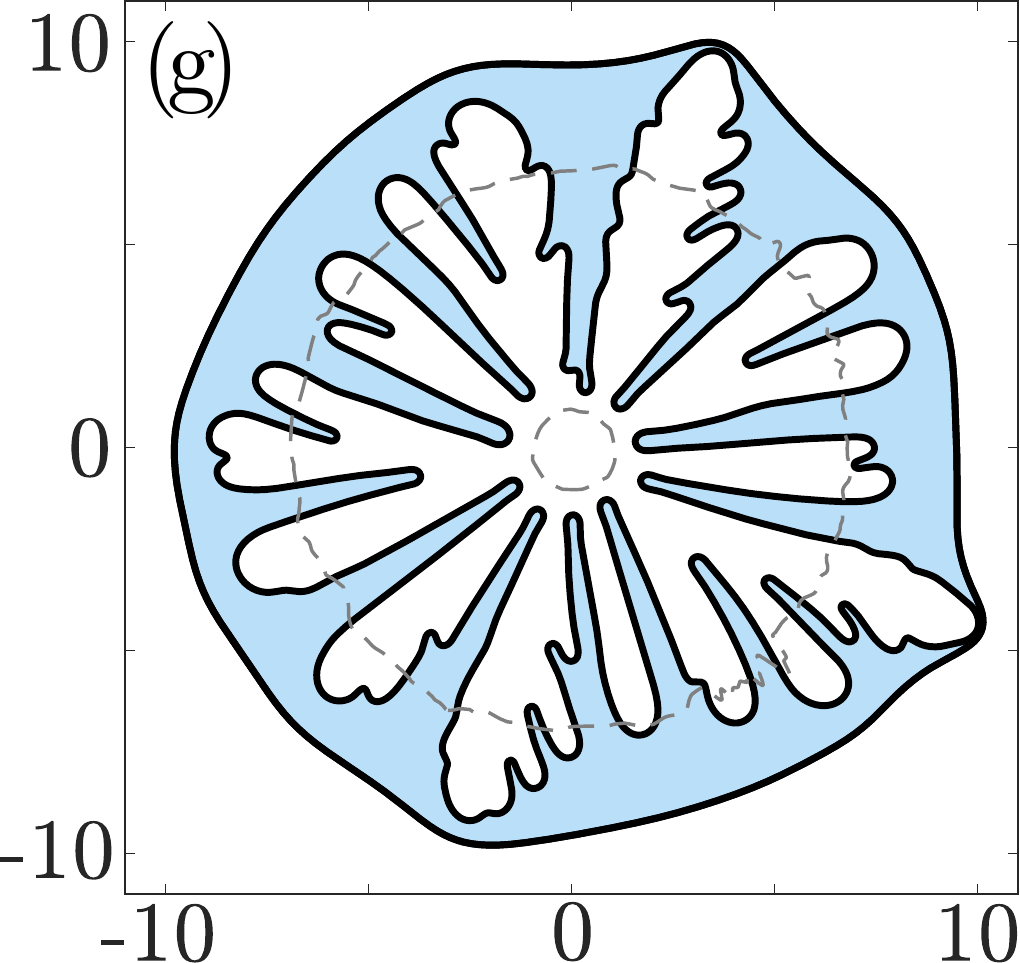}
	\includegraphics[width=0.15\linewidth]{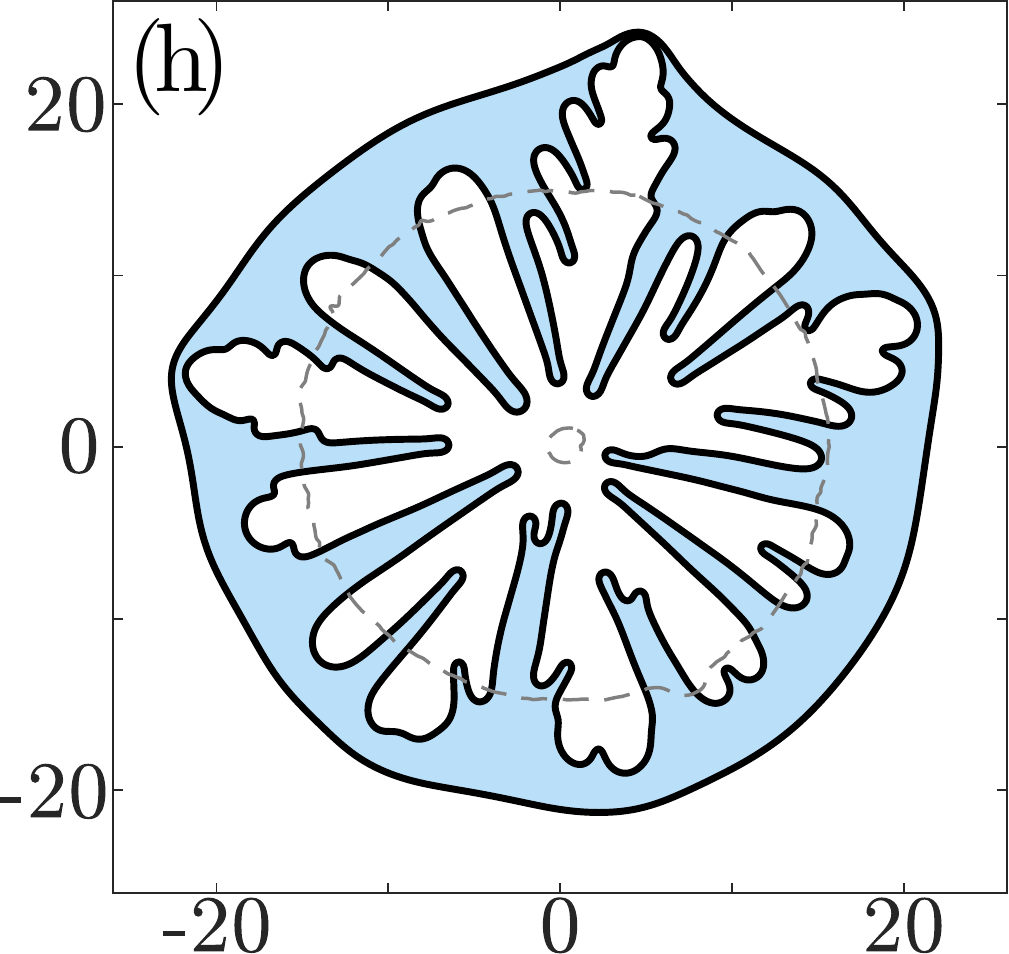}
	\includegraphics[width=0.15\linewidth]{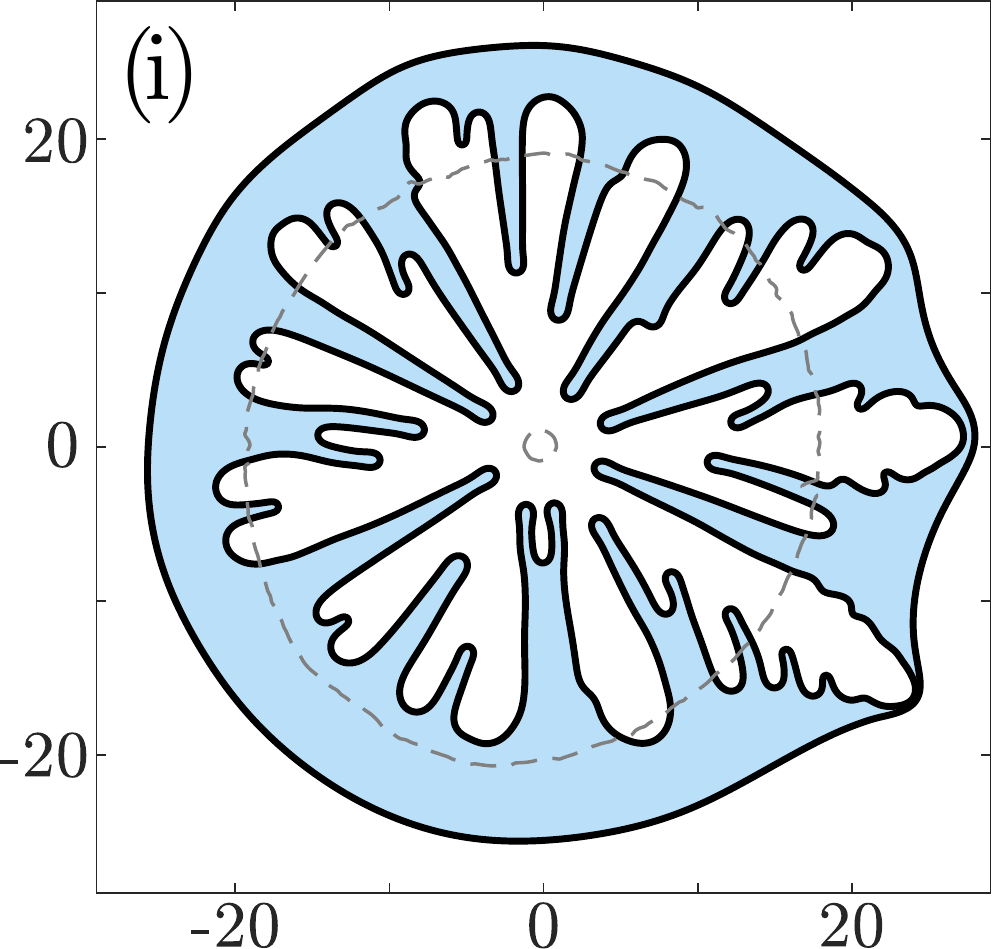}
	\includegraphics[width=0.15\linewidth]{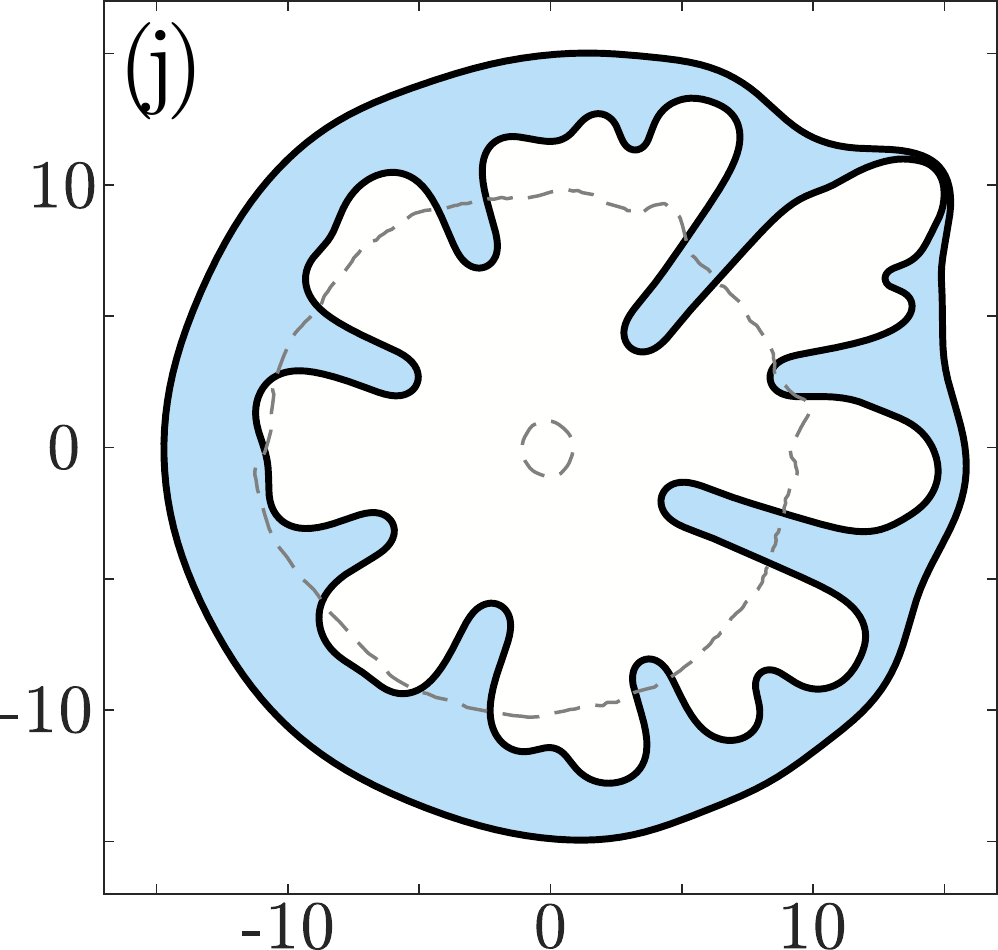}
	\includegraphics[width=0.15\linewidth]{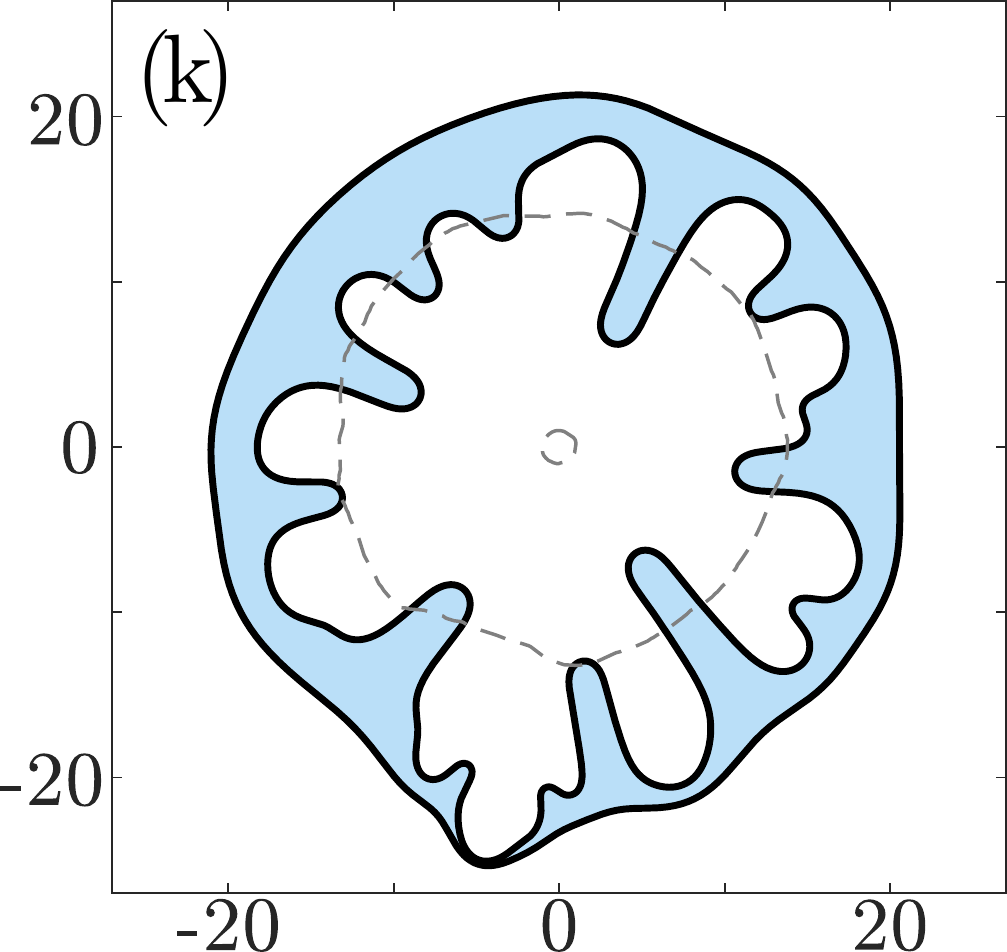}
	\includegraphics[width=0.15\linewidth]{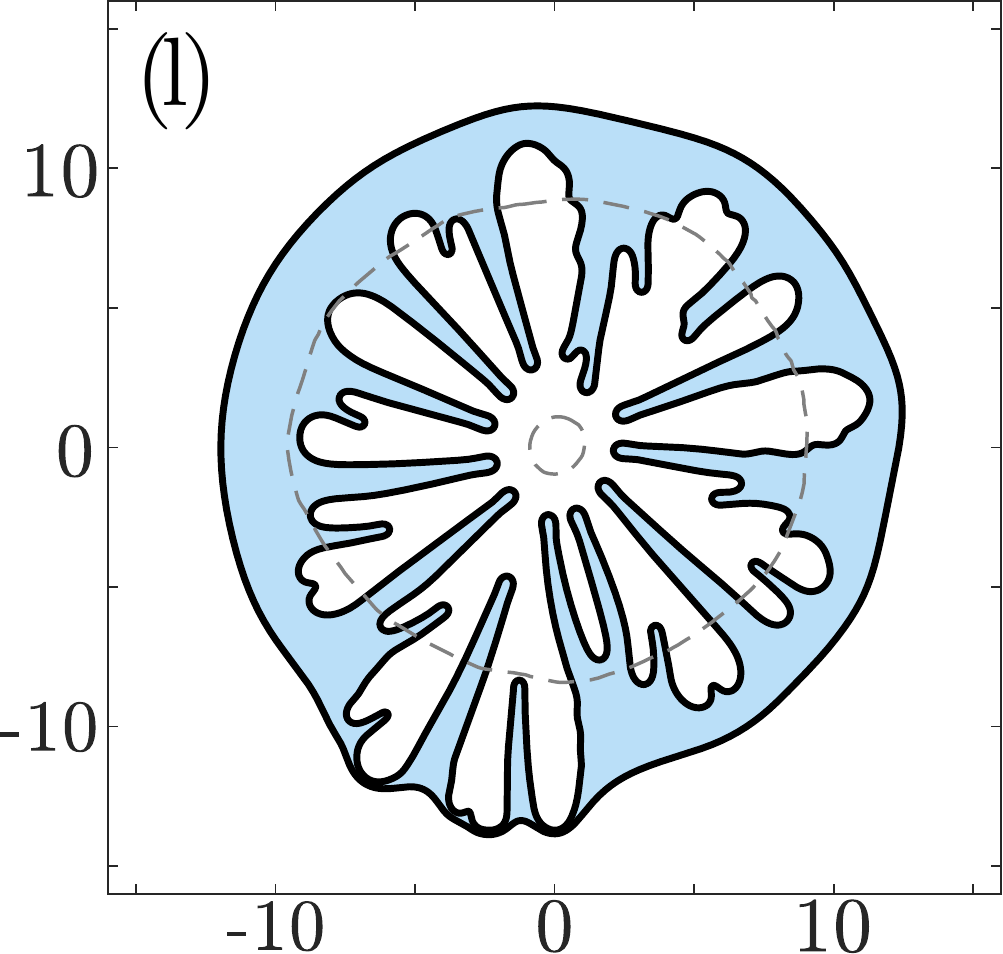}
    \caption{(a-f) Images from experiments with pressure differential $\bar{p}_I - \bar{p}_O$ (kPa) $(a)$ 16.2, $(b)$ 16.2, $(c)$ 16.2, $(d)$ 1.72, $(e)$ 1.72, and $(f)$ 16.2.  The average initial radius of the inner and outer interfaces are (mm) $(a)$ 6.3 and 43.6, $(b)$ 2.2 and 33.7, $(c)$ 1.6 and 30.5, $(d)$ 2.6 and 26, $(e)$ 1.7 and 21.94, and $(f)$ 2.71 and 24.02.
   	The black bars denote a length of 25 mm.
   	(g-l) Numerical simulations using initial conditions from (a-f). The non-dimensional pressure differential here (computed from \eqref{eq:deltap}) is $\Delta p=$ $(a)$ 1424 $(b)$ 497 $(c)$ 351 $(d)$ 62 $(e)$ and $(f)$ 610, and initial radius ratio $R_O = $ $(a)$ 6.9 $(b)$ 15.2 $(c)$ 19.6 $(d)$ 10 $(e)$ 13.4 and $(f)$ 8.9. The corresponding (non-dimensional) bursting times, at which the simulations are shown, are $t_{\textrm{burst}} = 0.008$, 0.125, 0.253, 0.541, and 1.682, 0.0317. Dashed lines denote the initial condition.}
\label{fig:FigureX}
\end{figure}

To support our numerical simulations, we conducted a small number of laboratory experiments with a Hele--Shaw cell whose two plates were made of plexiglass (see Fig.~\ref{fig:FigureX}(a-f)).  The viscous fluid used was water with surface tension $0.072$ Nm$^{-1}$, while the inviscid fluid was air.  A pressure differential $\Delta\bar{p}$ was applied between the inner and outer interfaces. Initial shapes of fluid interface were recorded using image analysis with Matlab, and average radius of inner and outer interfaces, $\bar{R}_i$ and $\bar{R}_o$, were estimated.  Subsequently, the dimensionless values $\Delta p$ and $R_O$ were calculated using (\ref{eq:deltap}) and (\ref{eq:FreeParameter2}), respectively.  Finally, to compare with the dimensionless time-scales via (\ref{eq:Scaling}), we note the two plates in the Hele--Shaw cell were separated by a distance of $b=56$ $\mu$m.

For each of the experimental images in Fig.~\ref{fig:FigureX}(a-f), we have provided an analogous simulation in Fig.~\ref{fig:FigureX}(g-l), so that (a) matches (g), (b) matches (h), and so on. From the experiments, using the MATLAB image processing toolbox, the initial positions of the inner and outer interfaces were extracted and used as the initial conditions for the simulations. For all cases, both experimental and numerical, we have shown images for times just before the inner interface bursts through the outer interface.  One observation is that it is clear that the fingering pattern is much more severe for the high pressure difference 16.2 kPa when compared to the low pressure difference 1.72 kPa. Further, while the system is highly unstable and sensitive to the choice of initial conditions, our simulations appear to compare well with the experimental results.

\subsubsection{Comparison with \citet{Ward2011}} \label{sec:ExpansionRate}

\begin{figure}
	\centering
	\includegraphics[width=0.5\linewidth]{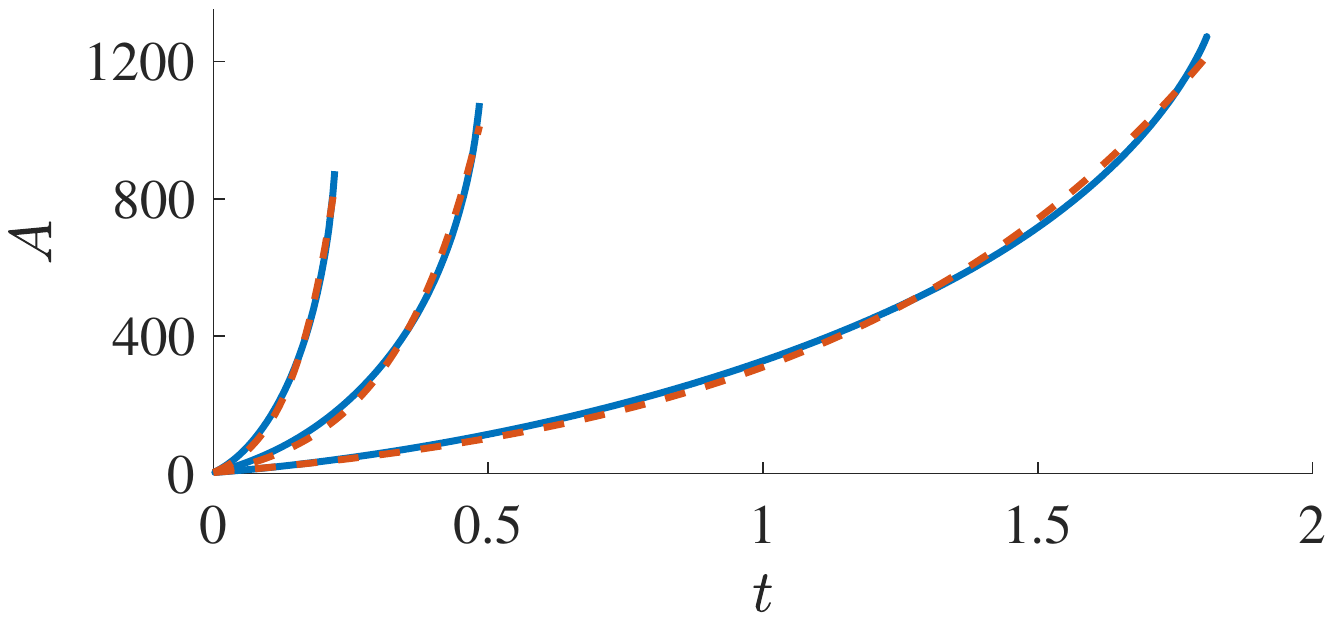}
	\caption{The area of the inner bubble computed from the numerical solution to \eqref{eq:Model1}-\eqref{eq:Model4} (blue) with $R_O = 15$ and (left to right) $\Delta p = 50$, 100, and 300. Simulations are performed with the initial conditions \eqref{eq:RadialIC} where $N = 12$ and $\varepsilon = 5 \times 10^{-3}$. A curve of the form of \eqref{eq:ExpansionRate1} is fitted to this data (red curves). For each choice of parameters, we find that the R$^2$ value is approximately 0.995.}
	\label{fig:expanding}
\end{figure}

We compare  numerical solutions of \eqref{eq:Model1}-\eqref{eq:Model4} with the some of the experimental results of \citet{Ward2011}, who reported on a more thorough investigation on how both the pressure differential between interfaces and the amount of viscous fluid influences the behaviour of the inner bubble.  As mentioned in Sec.~\ref{sec:NumericalSimulations}, when $\Delta p$ is constant there is no simple relationship between the model's parameters and the rate of expansion of the inner bubble. \citet{Ward2011} postulated that the area of the inner bubble increases exponentially in time according to
\begin{align} \label{eq:ExpansionRate1}
	A(t) = A(0) + C (\textrm{e}^{\beta t} - 1),
\end{align}
where $C$ is a constant and $\beta$ is the gas expansion parameter. The results of these experiments suggested a power law relationship between $\Delta p$ and $\beta$
\begin{align} \label{eq:ExpansionRate2}
	\beta \propto \Delta p^m,
\end{align}
where $m \approx 1.06$. We explore whether the expansion rate of $\partial \Omega_i$ is consistent with \eqref{eq:ExpansionRate1} and \eqref{eq:ExpansionRate2} by performing a series of numerical experiments for different values of $\Delta p$ and $R_O$. We approximate the area enclosed by the inner bubble using the level set function $\phi_i$ each time step via the volume integral
\begin{align} \label{eq:Area1}
	A = \int H(\phi_i) \,\textrm{d} \bmth{x},
\end{align}
where $H$ is the Heaviside function. Following \citet{Osher2003}, we use a first-order approximation of $H$ with the smeared-out Heaviside function
\begin{align} \label{eq:Area2}
	\hat{H}(\phi) =
	\begin{cases} 0 &\mbox{if } \phi < -\delta \\
		\dfrac{1}{2} + \dfrac{\phi}{2 \delta} + \dfrac{1}{2 \pi}\sin \dfrac{\pi \phi}{\delta} & \mbox{if } -\delta \le \phi \le \delta \\
		1 &\mbox{if } \phi > \delta
	\end{cases},
\end{align}
with the smoothing parameter $\delta = 1.5 \Delta x$. Finally, we determine $\beta$ by fitting a curve of the form \eqref{eq:ExpansionRate1} to the area of the bubble. To test whether our model predicts \eqref{eq:ExpansionRate1} is a reasonable approximation of the area of the inner bubble, we compute numerical solutions of \eqref{eq:Model1}-\eqref{eq:Model4} with different values of $\Delta p$, and fit a curve to this data of the form  of \eqref{eq:ExpansionRate1}, shown in Fig.~\ref{fig:expanding}. We find that for each parameter combination considered, we find a R$^2$ value of approximately 0.995, suggesting that \eqref{eq:ExpansionRate1} is a good approximation for how the area of the bubble evolves in time.

\begin{figure}
	\centering
	\includegraphics[width=0.8\linewidth]{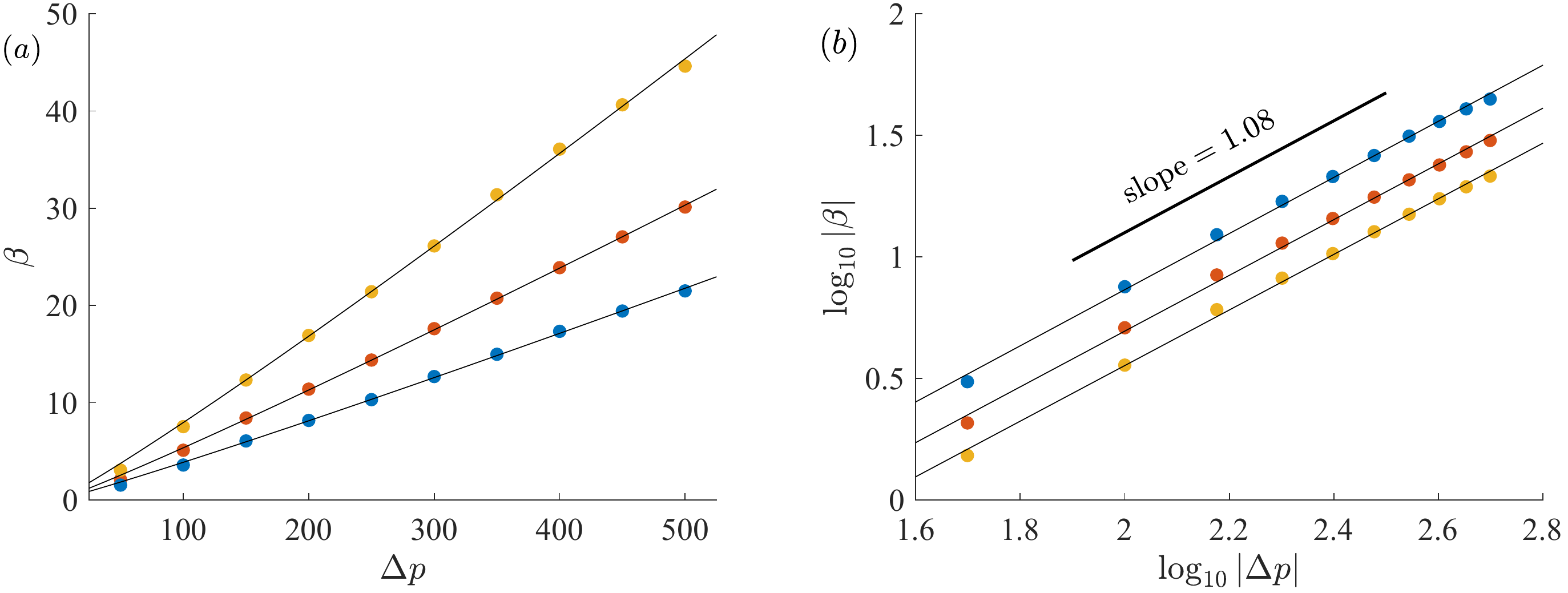}
	\caption{$(a)$ The gas expansion parameter $\beta$ as a function of $\Delta p$ for $R_O$ (top to bottom) 10, 12.5, and 15. We compute $\beta$ by fitting a curve of the form of \eqref{eq:ExpansionRate1}, where $A$ is determined from the numerical solution to \eqref{eq:Model1}-\eqref{eq:Model4}. $(b)$ The corresponding $\log$-$\log$ plot, where black line is line of best fit.}
	\label{fig:ExpansionRate}
\end{figure}

Figure \ref{fig:ExpansionRate}$(a)$ shows our estimate of $\beta$ as a function of $\Delta p$ for values of $R_O$ (top to bottom) 10, 12.5, and 15, and Fig.~\ref{fig:ExpansionRate}$(b)$ shows this data on a $\log$-$\log$ scale. For each data point, five simulations are performed, and $\beta$ is averaged over each of the simulations.  We find that either increasing $\Delta p$ or decreasing $R_O$ corresponds in an increase in $\beta$, resulting in the bubble expanding at a faster rate. This is to be expected, as \eqref{eq:Model2} indicates that either increasing $\Delta p$ or decreasing $R_O$ increases the normal speed of the interface. By taking a line of best fit (denoted by the black lines in Fig.~\ref{fig:ExpansionRate}$(b)$), we find $m = 1.08$ to two decimal places for each value of $R_O$ considered, which is in reasonable agreement with the experimental results of \citet{Ward2011}.

\begin{figure}
	\centering
	\includegraphics[width=0.8\linewidth]{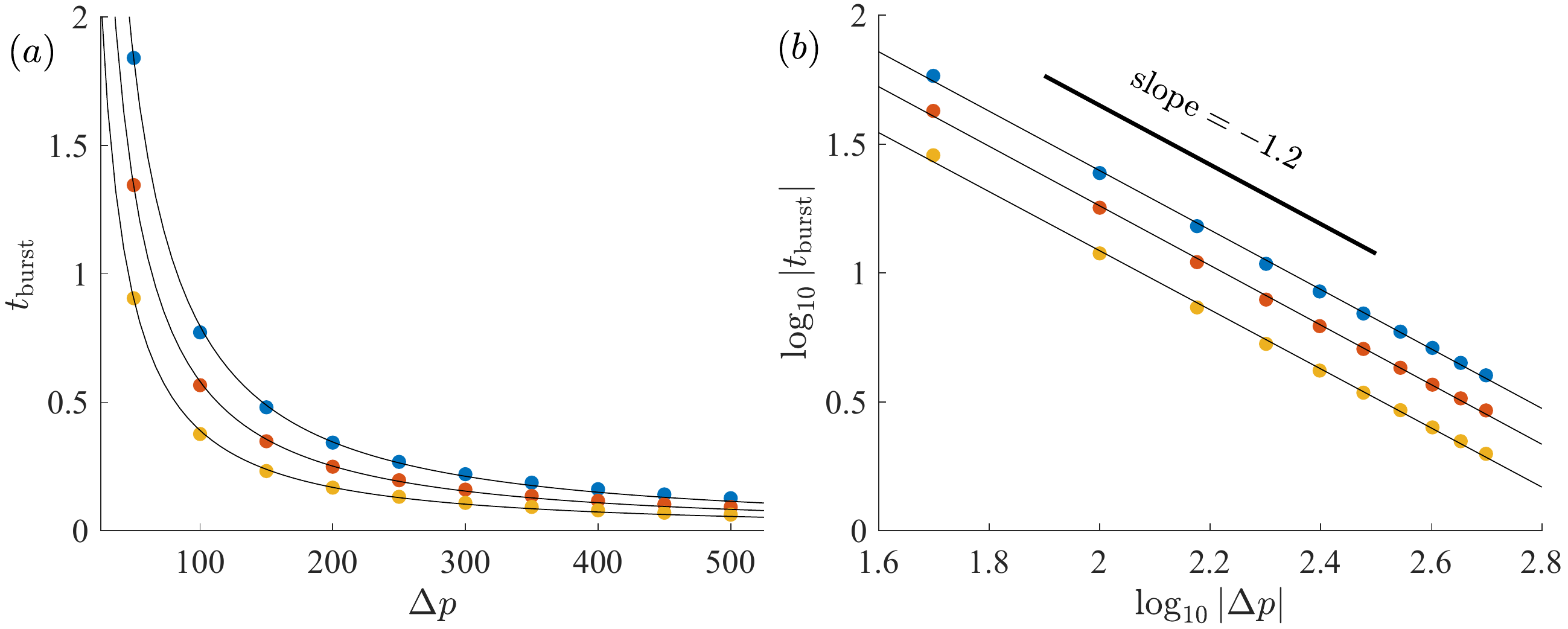}
	\caption{$(a)$ The time at which bursting occurs, $t_{\mathrm{burst}}$, as a function of $\Delta p$ for (bottom to top) $R_O = 10$, 12.5, and 15. $(b)$ The corresponding log-log plot, where black lines are a line of best fit. }
	\label{fig:tburstpsii}
\end{figure}

As discussed above, it was observed experimentally by \citet{Ward2011} that as the bubble is injected into the viscous blob, it becomes unstable and the fingers that develop ``burst'' through the outer interface, at which point the experiment is concluded.  These authors found that this bursting time follows the power-law relationship
\begin{align} \label{eq:BurstingTime}
	t _{\mathrm{burst}} \propto \Delta p^{\alpha},
\end{align}
where it was determined experimentally that $\alpha \approx -1.21$. We wish to determine whether numerical solutions of \eqref{eq:Model1}-\eqref{eq:Model4} are able to reproduce this power-law relationship. As our numerical scheme is stopped when the distance between the two interfaces is sufficiently small, we approximate $t_{\mathrm{burst}}$ by computing the minimum distance between the two interfaces via \eqref{eq:MinDist}, and linearly extrapolating to determine when this distance is zero. Figure \ref{fig:tburstpsii}$(a)$ compares $t_{\mathrm{burst}}$ as a function of $\Delta p$ with (bottom to top) $R_O = 10$, $12.5$, and $15$, while Fig.~\ref{fig:tburstpsii}$(b)$ shows this data on a $\log$-$\log$ scale. For each data point, five simulations are performed, and $t_{\mathrm{burst}}$ is averaged over each of the simulations. By taking a line of best fit for each value of $R_O$ considered, we find $\alpha = -1.20$ to two decimal places, which is in agreement with the experimental results of \citet{Ward2011}.

\subsection{Results for contracting bubble} \label{sec:Contracting}

We now briefly turn our attention to the case where $\Delta p < 0$ such that the inner bubble contracts.  We perform a series of simulations for different values of $R_O$ and $\Delta p$ to demonstrate how a fingering pattern develops on the outer interface.  In this scenario, the geometry remains doubly connected until either one of two outcomes occurs, depending on the parameters and choice of initial condition.  First, the outer interface can burst through the inner interface in a similar way to that described in Sec.~\ref{sec:Expanding}.  Or second, the leading interface may contract to a point before bursting occurs. In the latter case, simulations are halted if the minimum radius of the inner bubble is less than 0.05.

\begin{figure}
	\centering
	\begin{tabular}{c|ccc}
		\hline
		& $\Delta p = -500$ & $\Delta p = -1000$ & $\Delta p = -2000$ \\
		\hline
		\rotatebox{90}{\hspace{1.25cm} $R_O = 1.25$} &
		\includegraphics[width=0.25\linewidth]{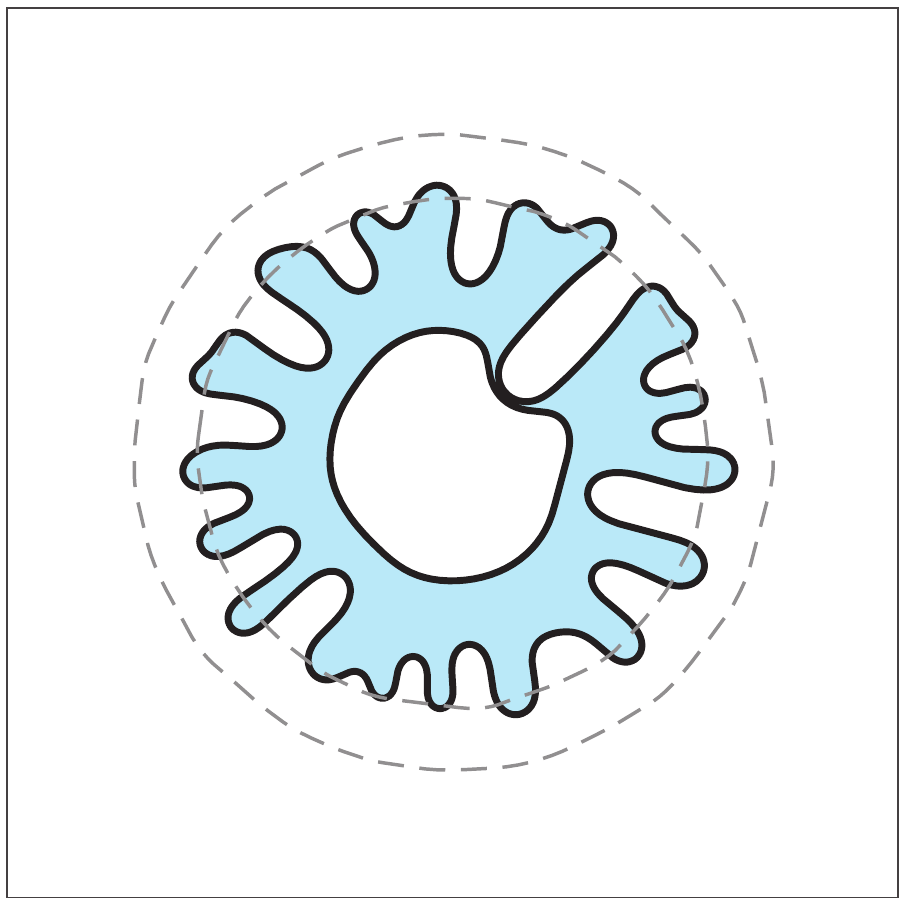} &
		\includegraphics[width=0.25\linewidth]{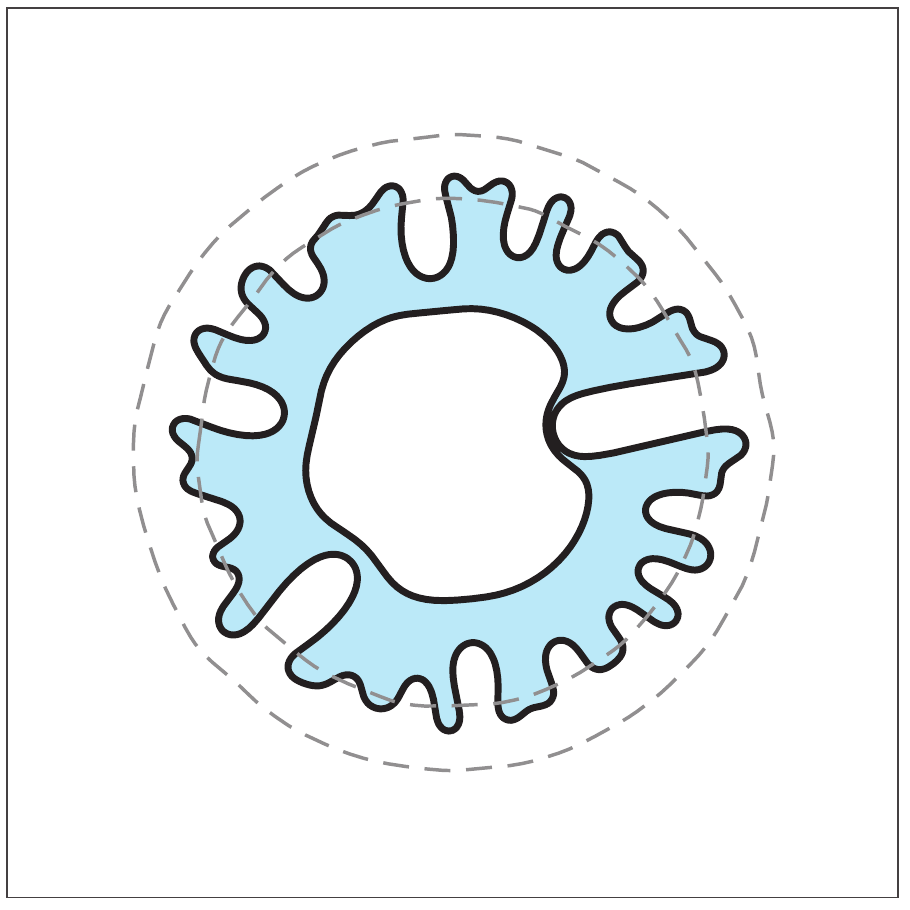} &
		\includegraphics[width=0.25\linewidth]{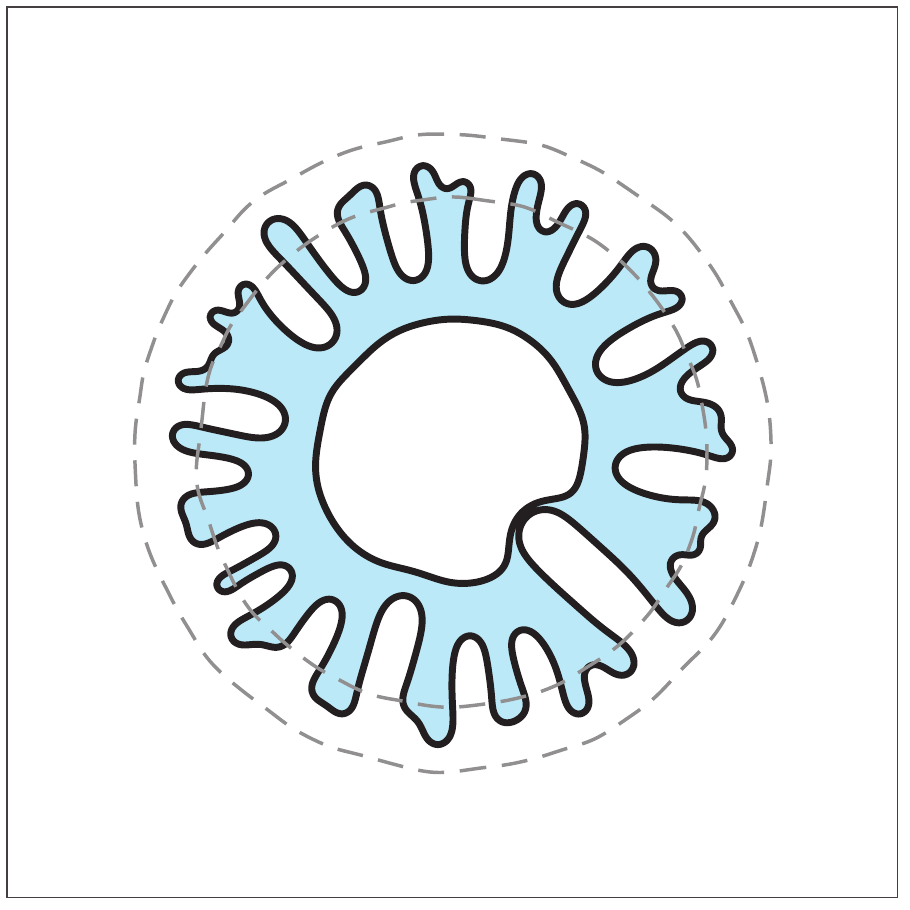}  \\
		\rotatebox{90}{\hspace{1.3cm} $R_O = 1.375$} &
		\includegraphics[width=0.25\linewidth]{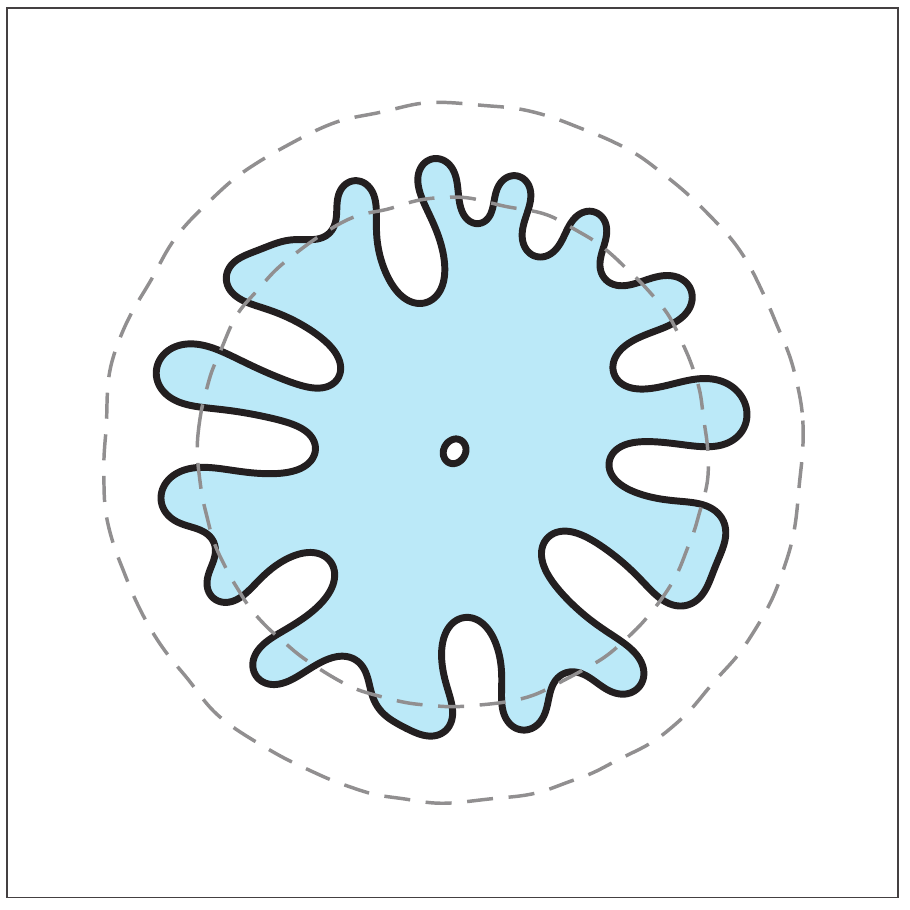} &
		\includegraphics[width=0.25\linewidth]{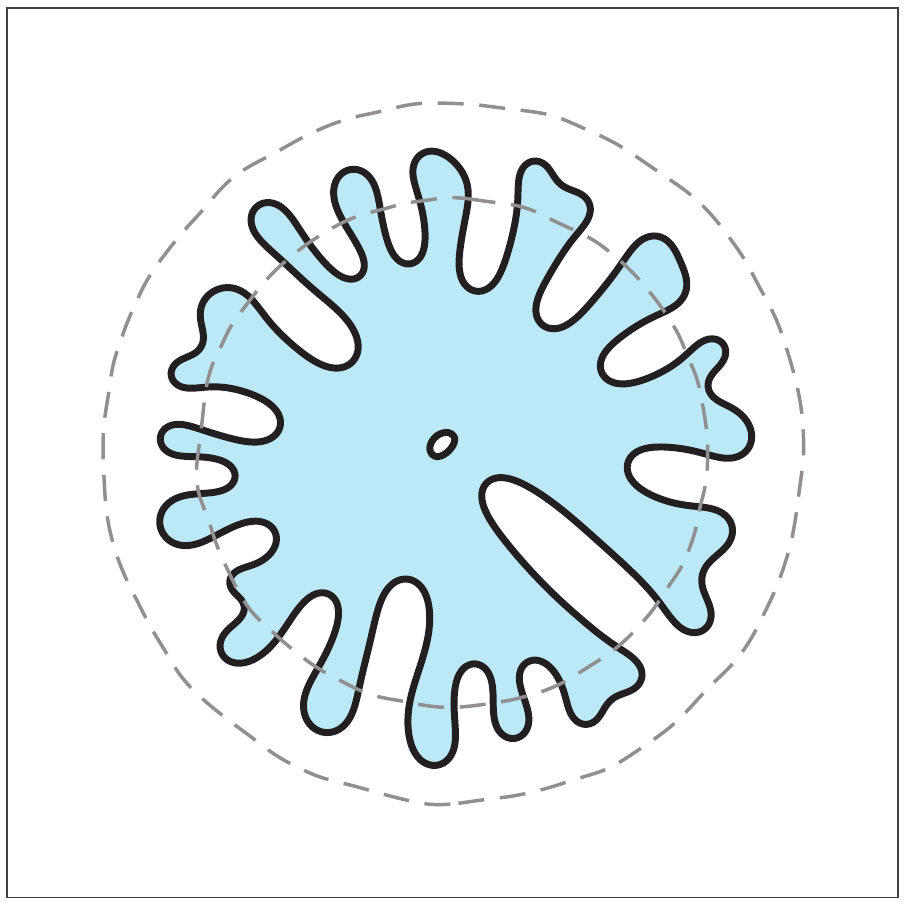} &
		\includegraphics[width=0.25\linewidth]{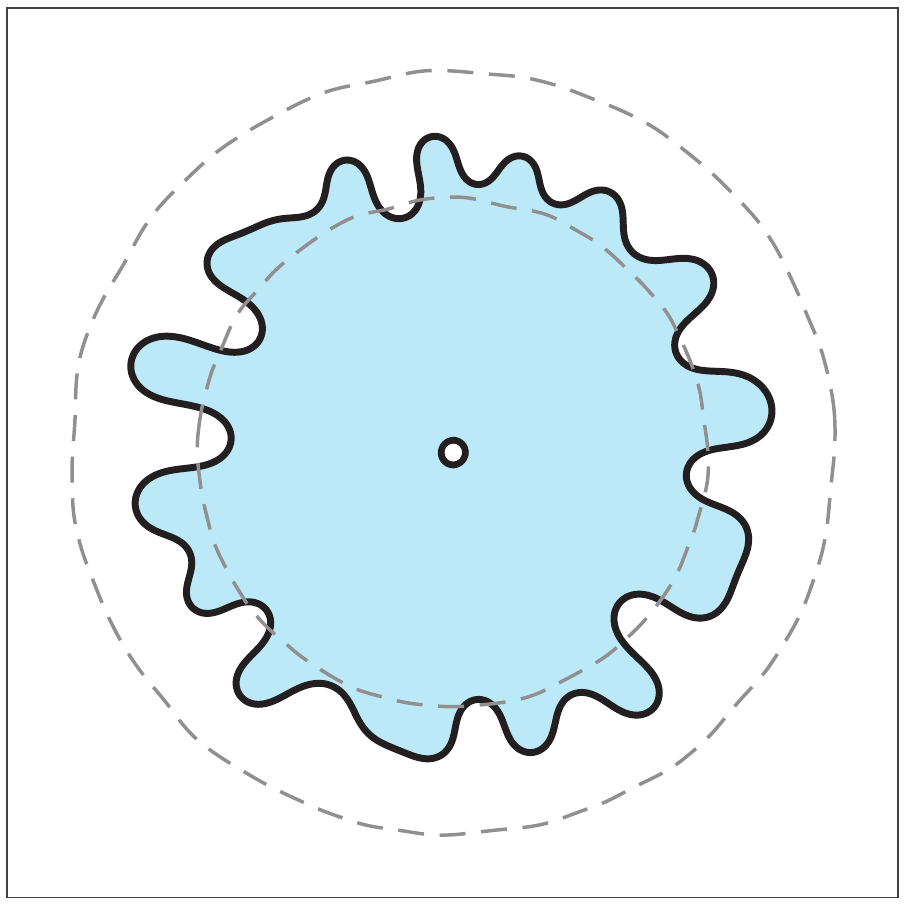}  \\
		\rotatebox{90}{\hspace{1.4cm} $R_O = 1.5$} &
		\includegraphics[width=0.25\linewidth]{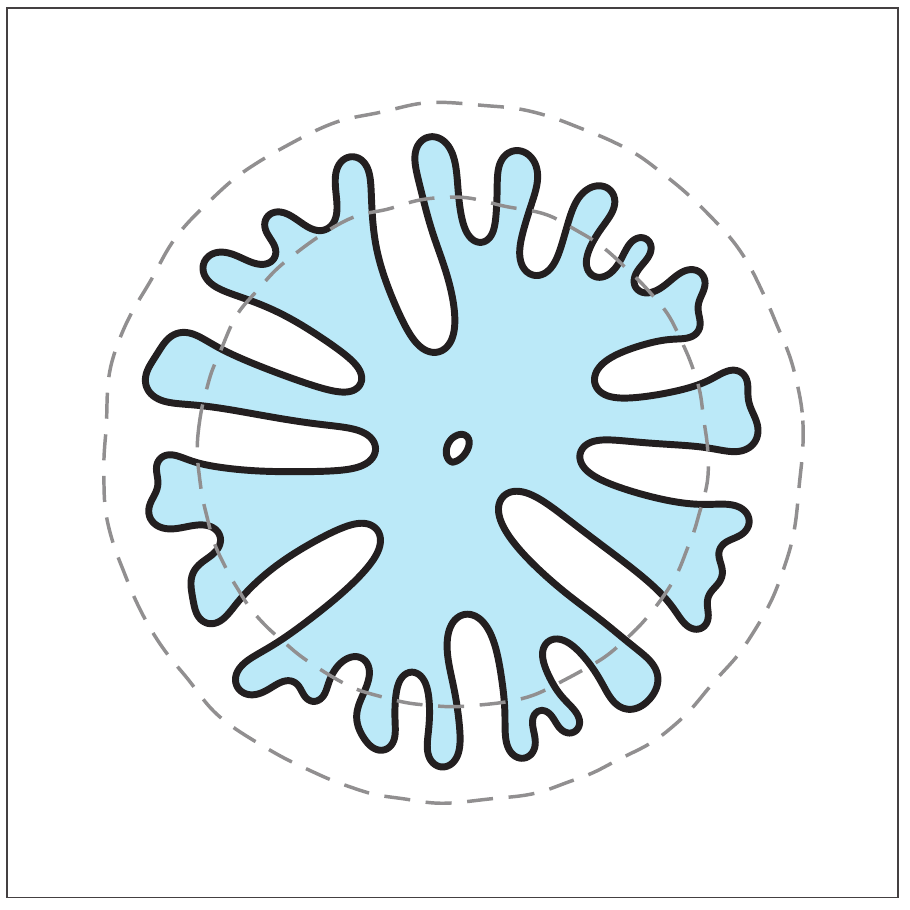} &
		\includegraphics[width=0.25\linewidth]{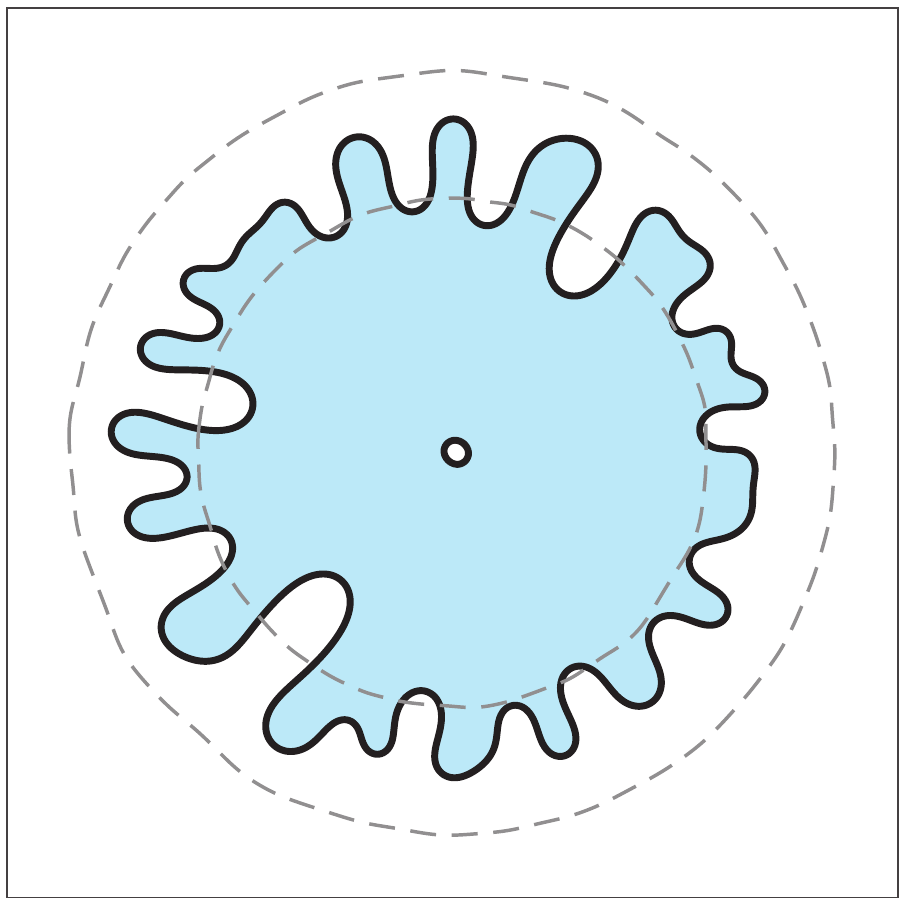} &
		\includegraphics[width=0.25\linewidth]{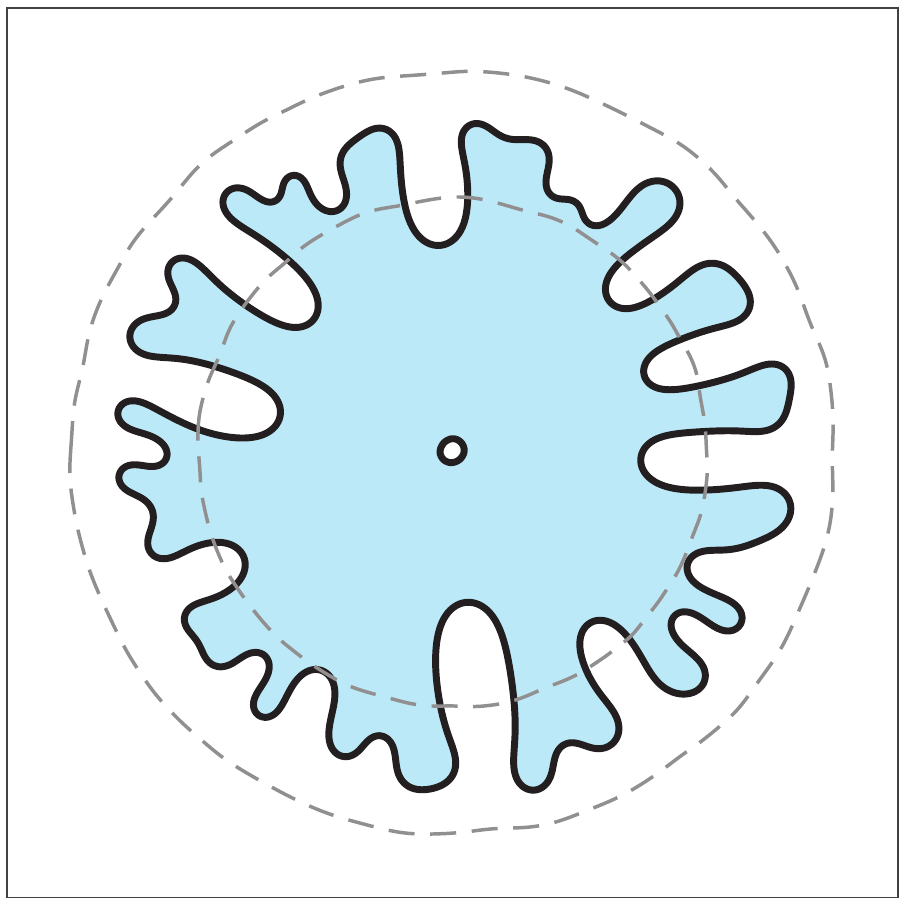}  \\
	\end{tabular}
	\caption{A selection of numerical solution to \eqref{eq:Model1}-\eqref{eq:Model4} for different values of $\Delta p$ and $R_O$. Simulations are performed with the initial conditions \eqref{eq:RadialIC}, where $N = 30$ and $\varepsilon = 5 \times 10^{-4}$. The dotted (grey) curves represent the initial condition for the inner and outer boundaries. Simulations are performed on the domain $-2 \le x \le 2$ and $-2 \le y \le 2$ using $600 \times 600$ equally spaced grid points.}
	\label{fig:Contracting}
\end{figure}

Figure \ref{fig:Contracting} shows numerical solutions of \eqref{eq:Model1}-\eqref{eq:Model4} for values of $R_O = 1.25$, $1.375$, and $1.5$ (rows one to three), and $\Delta p = -500$, $-1000$, and $-2000$ (columns one to three) at the time each simulation is concluded (in the top row this is the largest computational time before bursting, while for the remaining simulations this is the computational time immediately after the bubble radius contracts below 0.05). For each of the parameter combinations considered, we find that the trailing interface is unstable, and fingers develop inward towards the leading interface. The morphology of these fingers appears similar to the simply connected case in which a viscous blob is withdrawn from a point at some rate (see experiments by \citet{Thome1989} for example). In particular, fingers do not appear to tip split as they do for the expanding bubble case (see Fig.~\ref{fig:DoublyConnected}). Instead, the pressure differential between the boundaries `pulls' the fingers inward until either one of the fingers bursts through the leading bubble, or the leading bubble contracts to a point. The number of fingers that the trailing interface develops appears to increase with either an increase in $\Delta p$ or a decrease in $R_O$. This can be explained by noting that according to \eqref{eq:DarcysLaw}, either increasing the pressure differential or decreasing the distance between the interfaces results in a larger velocity, which in turn has a destabilising effect.

Comparing rows one and three of Fig.~\ref{fig:Contracting}, we see that whether or not the leading interface contracts to a point, as opposed to bursting occurring, is dependent upon the choice of $R_O$. If $R_O$ is sufficiently small (row one), for the range of $\Delta p$ considered, we find that the trailing interface will always burst through the leading interface before it contracts to a point. However for larger values of $R_O$ (rows two and three), we find that the leading interface will always contract to a point. In this case, we expect that the bubble will contract to a circle in the limit \citep{Dallaston2013}.

\section{Flow driven by angular velocity} \label{sec:Rotating}

In this section, we turn our attention to the configuration for which the velocity of the viscous fluid is not driven by a prescribed pressure differential, but instead by a centrifugal force due to the Hele--Shaw cell being rotated. For the simply connected case, this configuration is primarily studied with a viscous blob surrounded by inviscid fluid. The centrifugal force acts to propel the dense fluid outward, resulting in the interface becoming unstable. Experimental results \citep{Carrillo1996} and numerical simulations \citep{Alvarez2004,Miranda2005,Morrow2021,Paiva2019,Barua2022} indicate that these fingers are distinct from traditional Saffman-Taylor fingers, in that they appear more stretched out, and the number of fingers that develops appears to remain constant in time. Further, this number increases with the angular velocity of the plates. The configuration in which the inviscid fluid is injected into the viscous fluid while the plates are rotating was recently considered by \citet{Morrow2019}, who studied how the rotation rate acts to stabilise the interface.

The doubly connected configuration we are interested in here was first considered by \citet{Carrillo1999,Carrillo2000}, who performed a series of experiments which showed that either one or both of the interfaces can be unstable depending on the angular velocity of the Hele--Shaw plates and the volume of viscous fluid present. A family of exact solutions were derived by \citet{Crowdy2002} using conformal mapping techniques, and these solutions were shown to quantitatively reproduce experimental results. However, these solutions are valid under the zero-surface-tension assumption and, as such, the interfaces between the fluids generally form unphysical cusps in this model.  For the remainder of this section, we extend the work of \citet{Carrillo1999,Carrillo2000} and \citet{Crowdy2002} by performing a numerical investigation into the nonlinear behaviour of an annular blob in a rotating Hele--Shaw cell when the effects of surface tension are included.

\subsection{Summary of mathematical model}

Following \citet{Carrillo1999,Carrillo1996}, we can modify Darcy's law \eqref{eq:DarcysLaw} to include rotational effects
\begin{align} \label{eq:DarcysLaw2}
	\bar{\bmth{v}}= -\frac{b^2}{12 \mu} \left( \bar{\grad} \bar{p} - \rho \bar{\omega}^2 \bar{r} \bar{\bmth{e}}_r \right),
\end{align}
where $\rho$ is the density of the viscous fluid and $\bar{\omega}$ is the angular velocity of the plates. We take $\bar{\omega}$ to be constant, although we note that previous studies for the simply connected case have considered the angular velocity to be time-dependent \citep{Anjos2017}. We re-define pressure as
\begin{align}
	\bar{P} = \left(  \bar{p} - \frac{\rho \bar{\omega}^2 \bar{r}^2}{2} \right),
\end{align}
and, recalling $\bar{\nabla} \cdot \bar{\bmth{v}} = 0$, our model becomes
\begin{subequations}
	\begin{alignat}{3}
		\nabla^2 \bar{P} &= 0 &\bar{\bmth{x}} &\in \bar{\Omega}, \label{eq:Rotating1} \\
		\bar{v}_n &= -\frac{b^2}{12 \mu}\bar{\grad} \bar{P} \cdot \bar{\bmth{n}} &\bar{\bmth{x}} &\in \partial \bar{\Omega}_i, \partial \bar{\Omega}_o, \label{eq:Rotating2} \\
		\bar{P} &= -\frac{\rho \bar{\omega}^2 \bar{r}^2}{2} - \gamma \bar{\kappa}
        \qquad \qquad  &
        \bar{\bmth{x}} &\in \partial \bar{\Omega}_i, \label{eq:Rotating3}\\
		\bar{P} &= -\frac{\rho \bar{\omega}^2 \bar{r}^2}{2} - \gamma \bar{\kappa}  &
        \bar{\bmth{x}} &\in \partial \bar{\Omega}_o,  \label{eq:Rotating4}
	\end{alignat}
\end{subequations}
where $\gamma$ is the surface tension.

We non-dimensionalise \eqref{eq:Rotating1}-\eqref{eq:Rotating4} via \eqref{eq:Scaling} (noting that $\bar{p}$ is now $\bar{P}$) giving
\begin{subequations}
	\begin{alignat}{3}
		\nabla^2 P &= 0  &\bmth{x} &\in \Omega, \label{eq:Model5} \\
		v_n &= -\grad P \cdot \hat{n} & \bmth{x}
        & \in \partial \Omega_I, \partial \Omega_O, \label{eq:Model6} \\
		P &= -\omega^2 r^2 - \kappa \qquad \qquad
        & \bmth{x} & \in \partial \Omega_I, \label{eq:Model7}\\
		P &= -\omega^2 r^2 - \kappa
        & \bmth{x} & \in \partial \Omega_O.  \label{eq:Model8}
	\end{alignat}
\end{subequations}
Our dimensionless model \eqref{eq:Model5}-\eqref{eq:Model8} has two free parameters, the initial radius of the outer interface, $R_0$, defined by \eqref{eq:FreeParameter2}, and
\begin{align}
	\omega^2 = \frac{\bar{\omega}^2 \rho \bar{R}_i^3}{2 \gamma },
\end{align}
whereby the dimensionless angular velocity $\omega$ is the key centrifugal parameter. We can see that the velocity of the fluid is driven by a pressure differential resulting from the rotational terms in the dynamic boundary conditions \eqref{eq:Model7} and \eqref{eq:Model8}. However, this differential is not prescribed, but instead depends on the location of each interface. As $\omega^2 \ge 0$, the viscous fluid will always be propelled outward, and thus there is no analogous contracting case that was considered in Sec.~\ref{sec:Contracting}. Numerically speaking, it is straightforward to include the centrifugal terms in the dynamic boundary conditions into our scheme, presented in Sec.~\ref{sec:NumericalScheme}. However, for this rotating flows, when incorporating \eqref{eq:Rotating2} and \eqref{eq:Rotating3} into our finite difference stencil, separate level set functions are not required for each interface, and the problem can be solved using a single level set function as illustrated in Fig.~\ref{fig:LevelSet}$(a)$. 

\subsection{Linear stability analysis}\label{sec:lsa2}

Before presenting our numerical results, it is worth briefly summarising the results from linear stability analysis \citep{Carrillo2000}, as this theory allows us to compare and contrast with the results from Sec.~\ref{sec:PressureDifferential}. By applying the standard expansion (\ref{eq:pLSA})-(\ref{eq:sLSA}), we find that, to leading order
\begin{align}
	\frac{\textrm{d} s_{I0}}{\textrm{d} t} = \frac{1}{s_{I0} \ln (s_{O0}/s_{I0})} \left( \omega^2 (R_O^2 - 1) - \left( \frac{1}{s_{I0}} + \frac{1}{s_{O0}} \right)  \right),
\label{eq:rotateSI0}
\\
	\frac{\textrm{d} s_{O0}}{\textrm{d} t} = \frac{1}{s_{O0} \ln (s_{O0}/s_{I0})} \left( \omega^2 (R_O^2 - 1) - \left( \frac{1}{s_{I0}} + \frac{1}{s_{O0}} \right)  \right),
\label{eq:rotateSO0}
\end{align}
where $s_{I0}(0) = 1$ and  $s_{O0}(0) = R_O$.  These nonlinear differential equations can be solved numerically if necessary \citep{Carrillo2000}, but even without those calculations we can make some elementary observations.  First, dividing (\ref{eq:rotateSI0}) by (\ref{eq:rotateSO0}) leads to conservation of mass (\ref{eq:mass}).  Further, the sign of
$$
\omega^2(R_O^2-1)-\left(\frac{1}{s_{I0}}+\frac{1}{s_{O0}}\right)
$$
determines whether the interface speeds are positive or negative.  Indeed, if $\omega^2>1/(R_O(R_O-1))$, the system is expanding, while if $\omega^2<1/(R_O(R_O-1))$ the system is contracting.  In this way, we see that the term $\omega^2(R_O^2-1)$ in (\ref{eq:rotateSI0})-(\ref{eq:rotateSO0}) plays the same role as $\Delta p$ does in (\ref{eq:Circle1})-(\ref{eq:Circle2}).

Moving to the next order, by writing out $P_1$ and $s_{I1}$ as in (\ref{eq:P1andsI1andsO1}), we arrive at the linear system
\begin{align}
\frac{\textrm{d}\delta_{In}}{\textrm{d}t}=&
\left[\frac{1}{s_{I0}}\left(\frac{n(s_{I0}^{2n}+s_{O0}^{2n})}{s_{O0}^{2n}-s_{I0}^{2n}}
-1\right)\frac{\mathrm{d}s_{I0}}{\mathrm{d}t}
-\frac{n(s_{I0}^{2n}+s_{O0}^{2n})}{s_{O0}^{2n}-s_{I0}^{2n}}\left(\frac{n^2-1}{s_{I0}^3}+2\omega^2
\right)\right]\delta_{In}
\nonumber \\
& -\frac{2ns_{I0}^ns_{O0}^n}{s_{O0}^{2n}-s_{I0}^{2n}}
\left(\frac{1}{s_{O0}}\frac{\mathrm{d}s_{I0}}{\mathrm{d}t}
+\frac{n^2-1}{s_{I0}s_{O0}^2}-\frac{2\omega^2 s_{O0}}{s_{I0}}\right)\delta_{On},
\label{eq:gamman2}
\\
\frac{\textrm{d}\delta_{On}}{\textrm{d}t}=&
\frac{2ns_{I0}^ns_{O0}^n}{s_{O0}^{2n}-s_{I0}^{2n}}
\left(\frac{1}{s_{I0}}\frac{\mathrm{d}s_{O0}}{\mathrm{d}t}
-\frac{n^2-1}{s_{I0}^2s_{O0}}-\frac{2\omega^2 s_{I0}}{s_{O0}}\right)\delta_{In}
\nonumber \\
&+\left[-\frac{1}{s_{O0}}\left(\frac{n(s_{I0}^{2n}+s_{O0}^{2n})}{s_{O0}^{2n}-s_{I0}^{2n}}
+1\right)\frac{\mathrm{d}s_{O0}}{\mathrm{d}t}
+\frac{n(s_{I0}^{2n}+s_{O0}^{2n})}{s_{O0}^{2n}-s_{I0}^{2n}}\left(-\frac{n^2-1}{s_{O0}^3}+2\omega^2
\right)\right]\delta_{On}.
\label{eq:deltan2}
\end{align}
Writing the equations this way (without substituting in the expressions for $\textrm{d}s_{I0}/{\textrm{d}t}$ and $\textrm{d}s_{O0}/{\textrm{d}t}$), we can draw some broad conclusions and relate directly with (\ref{eq:gamman})-(\ref{eq:deltan}).  For a start, terms involving $n^2-1$ are due to surface tension and help stabilise higher order modes.  More crucially, the key point is that the rotation rate $\omega$ appears to have two opposing effects.  First, provided $\omega^2>1/(R_O(R_O-1))$, from (\ref{eq:rotateSI0}) and (\ref{eq:rotateSO0}) we see that increasing $\omega$ increases the interface speed which from (\ref{eq:gamman2}) and (\ref{eq:deltan2}) has the effect of making the inner interface more unstable and the outer interface less unstable (via the appearance of $(\textrm{d}s_{I0}/{\textrm{d}t})\gamma_n$ in (\ref{eq:gamman2}) and $-(\textrm{d}s_{O0}/{\textrm{d}t})\delta_n$ in (\ref{eq:deltan2})).  This is essentially the traditional Saffman-Taylor instability, which appears in the same way in (\ref{eq:gamman})-(\ref{eq:deltan}).  On the other hand, increasing $\omega$ has the effect of stabilising the inner interface and destabilising the outer interface (via the terms $-\omega^2\gamma_n$ in (\ref{eq:gamman2}) and $\omega^2\delta_n$ in (\ref{eq:deltan2})).  This effect is due to the centrifugal force pushing out viscous fluid as it rotates.  In summary, while these equations from linear stability analysis appear complicated, we are able to extract a sense of the most important ingredients in determining whether fingering patterns tend to occur.  In particular, these opposing forces provide the opportunity for both interfaces to be highly unstable at the same time, giving rise to distinct fingering patterns that do not appear in annular configurations that are driven by pressure differences only.

\begin{figure}
	\centering
	\includegraphics[width=0.4\linewidth]{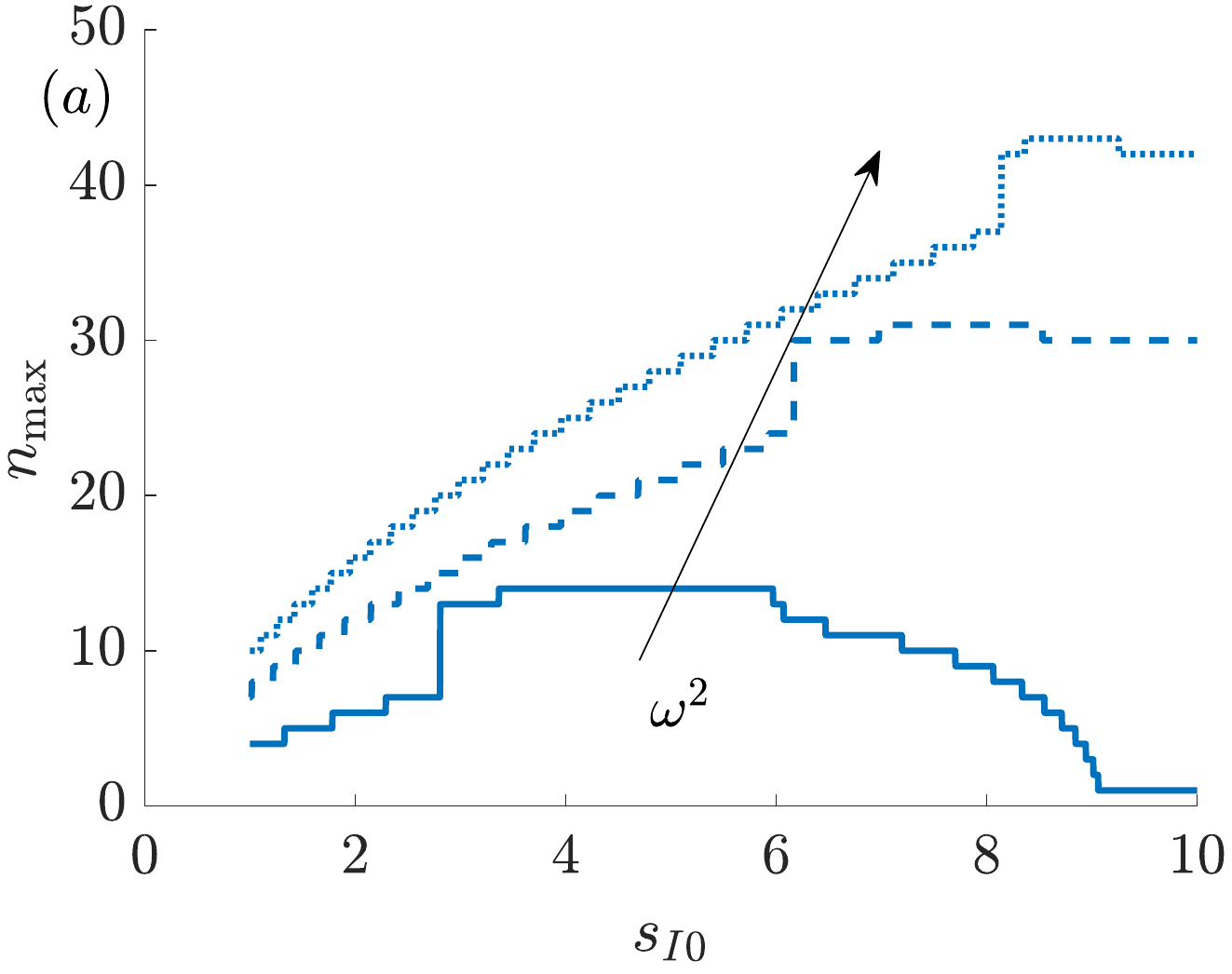}
	\includegraphics[width=0.4\linewidth]{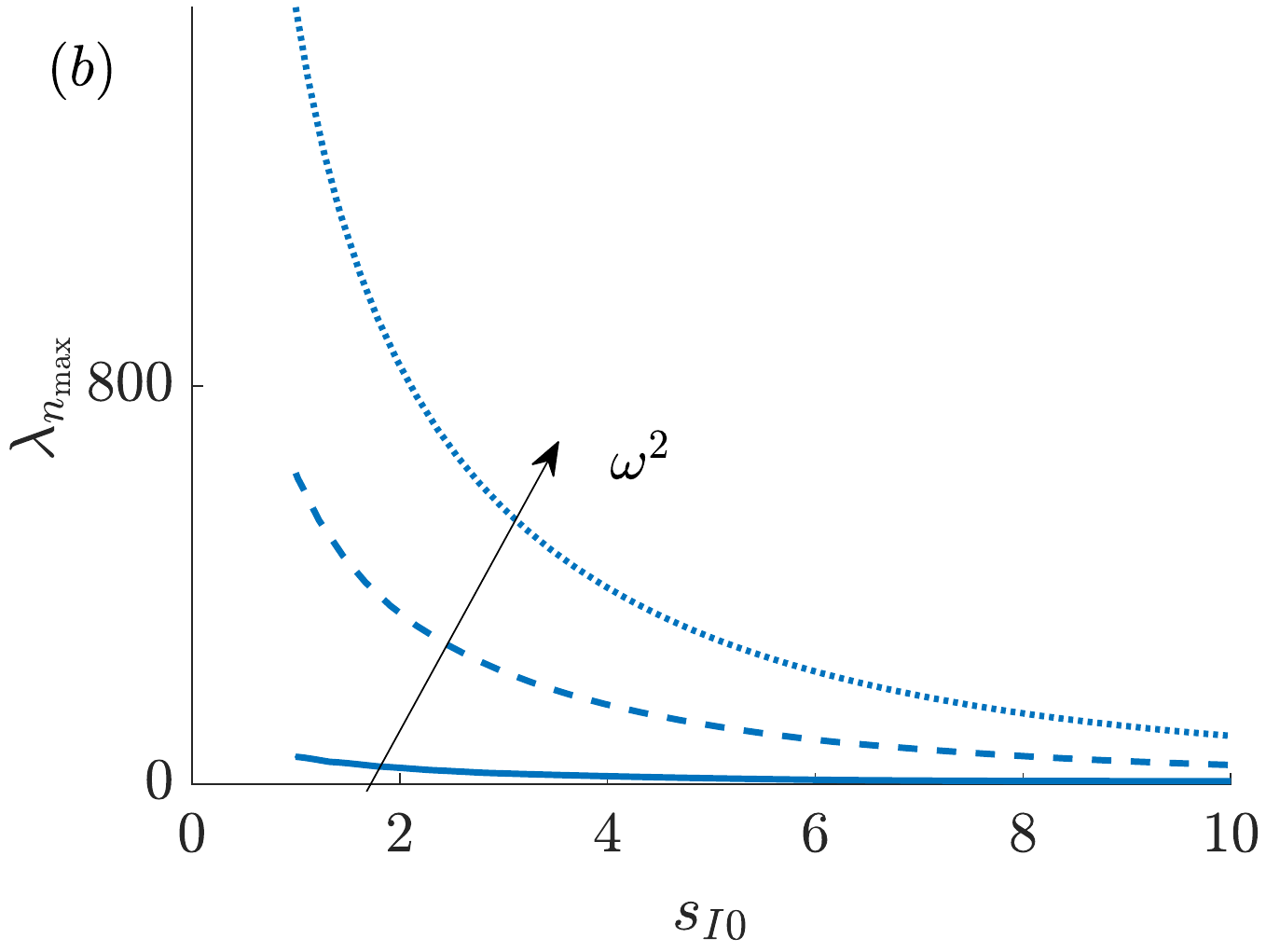}
	
	\includegraphics[width=0.4\linewidth]{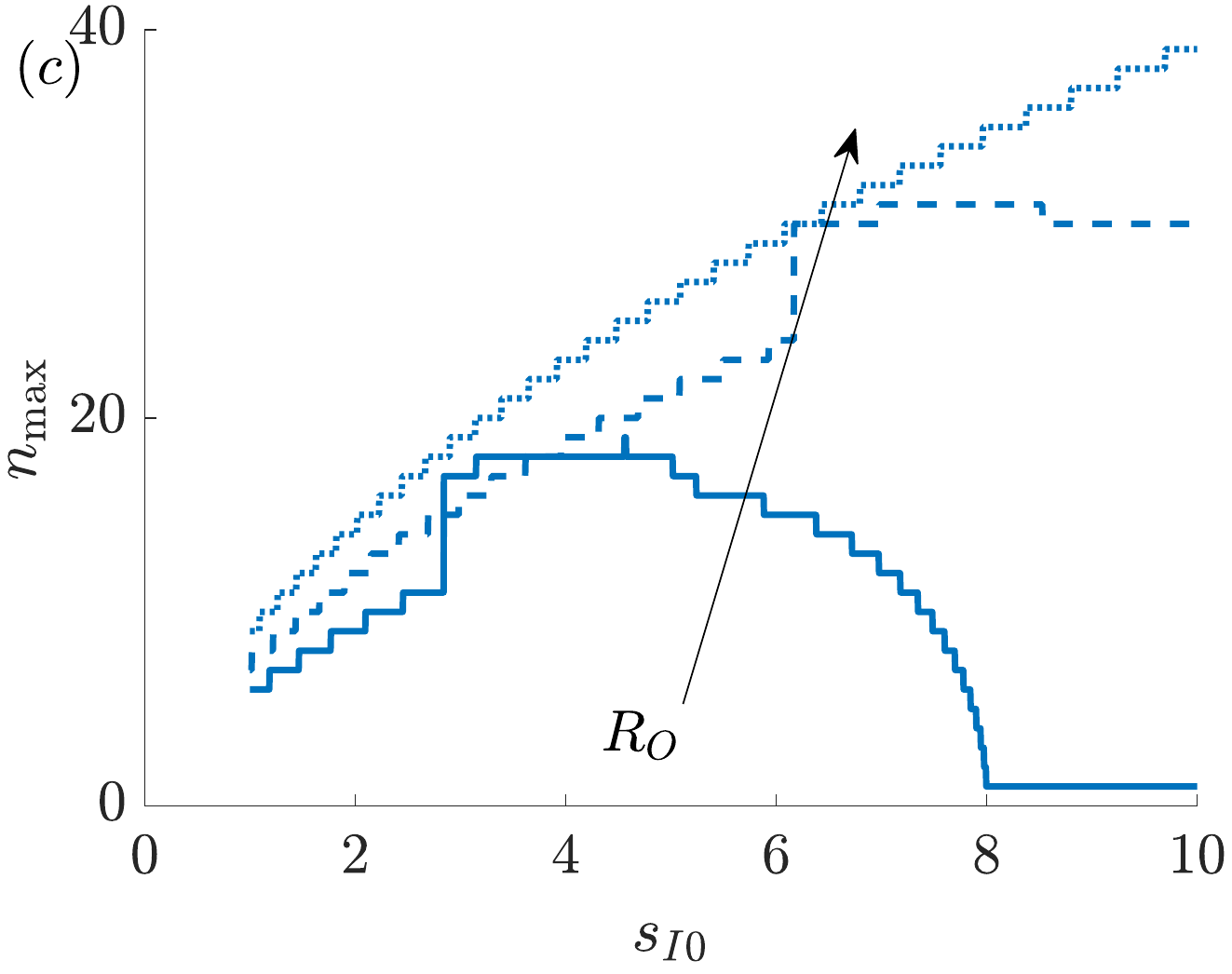}
	\includegraphics[width=0.4\linewidth]{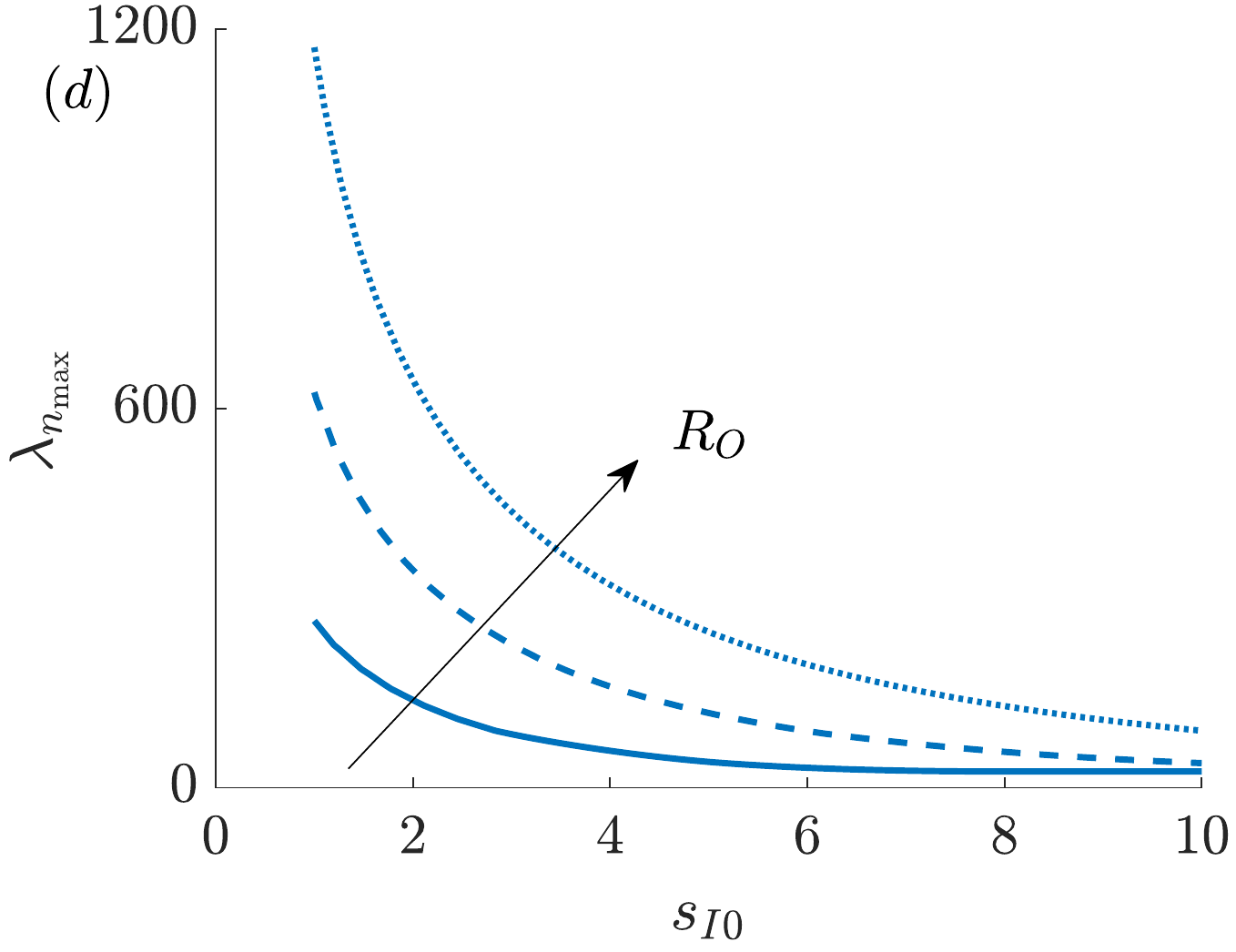}
	\caption{$(a)$ The most unstable mode of perturbation, $n_{\max}$, computed from \eqref{eq:gamman2} and \eqref{eq:deltan2} with $R_O = 5$ and (bottom to top) $\omega^2 = 3.2$, 12.8, and 22.4. $(b)$ Corresponding maximum eigenvalue at $n = n_{\max}$. $(c)$ $n_{\max}$ and $(d)$ $\lambda_{n_{\max}}$ with $\omega^2 = 12.8$ and (bottom to top) $R_O = 3.75$, 5, and 6.25.}
	\label{fig:LinearStabiliityAnalysis2}
\end{figure}

Before moving on to our numerical simulations of the fully nonlinear version of this problem, we provide some quantitative insight into (\ref{eq:gamman2})-(\ref{eq:deltan2}) by assuming quasi-steady-state conditions and thereby interpreting the two differential equations at each time as a constant-coefficient system (as we did for the system (\ref{eq:gamman})-(\ref{eq:deltan})).  We compute eigenvalues of the corresponding $2\times 2$ matrix and denote the maximum eigenvalue by $\lambda_n$ associated with $n$th mode of perturbation.  In Fig.~\ref{fig:LinearStabiliityAnalysis2} we plot the most unstable mode $n_{\max}$ and corresponding growth rate $\lambda_{n_{\max}}$ versus the unperturbed inner radius $s_{I0}$.  In (a)-(b) we fix $R_O=5$ and show results for three different rotation rates $\omega^2$, while in (c)-(d) we fix $\omega^2=12.8$ and vary $R_O$.  In all cases, we see the most unstable mode $n_{\max}$ first increases with $s_{I0}$ and then decreases, while the growth rate $\lambda_{n_{\max}}$ is monotonically decreasing with $s_{I0}$.  These results tentatively predict that this rotating Hele-Shaw configuration does not lead to ever increasing tip-splitting, which suggests the fingering patterns will be different from those that arise from the classical Saffman-Taylor instability.  It is interesting to note the overall trend of the system to stabilise for sufficiently large $s_{I0}$, although of course all of this theory is only relevant provided the perturbations are sufficiently small.

\subsection{Numerical simulations}\label{sec:numerical2}

In Sec.~\ref{sec:PressureDifferential}, we demonstrated that when the pressure differential between the inner and outer interfaces is prescribed, increasing the magnitude of $\Delta p$ increases the velocity of the viscous fluid, which in turn increases the complexity or severity of the viscous fingering patterns that develop on the trailing interface.  On the other hand, for the problem considered in this section, the influence of $\omega$ on the stability on each interface is not as straightforward. For example, we expect to find that increasing $\omega$ increases the pressure differential between the two interfaces, which in turn would destabilise and stabilise the interior and exterior interfaces, respectively. However, the angular velocity of the plates also acts to propel the dense fluid outward, which acts to stabilise and destabilise the interior and exterior interfaces, respectively \citep{Morrow2019}. In this section, we perform a series of numerical experiments to gain insight into how the rotation of the Hele--Shaw plates influences the evolution of both the leading and trailing boundaries.

\begin{figure}
	\centering
	\begin{tabular}{c|ccc}
		\hline
		& $\omega^2 = 3.2$ & $\omega^2 = 12.8$ & $\omega^2 = 22.4$ \\
		\hline
		\rotatebox{90}{\hspace{1.25cm} $R_O = 3.75$} &
		\includegraphics[width=0.25\linewidth]{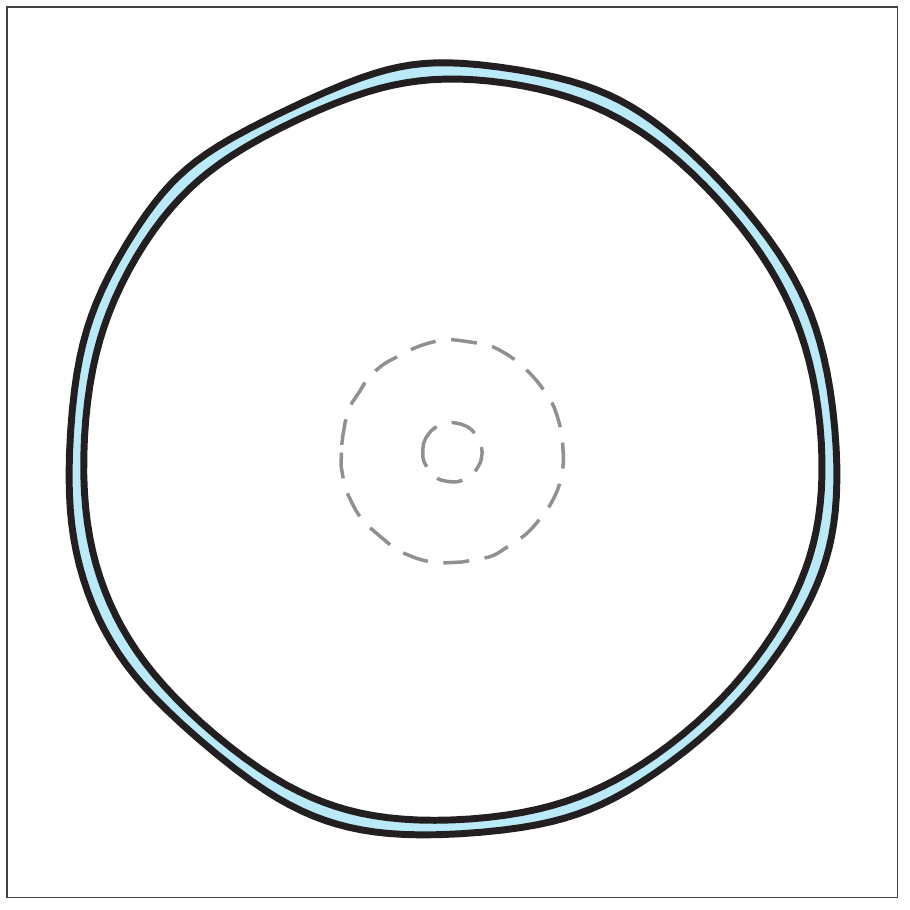} & \includegraphics[width=0.25\linewidth]{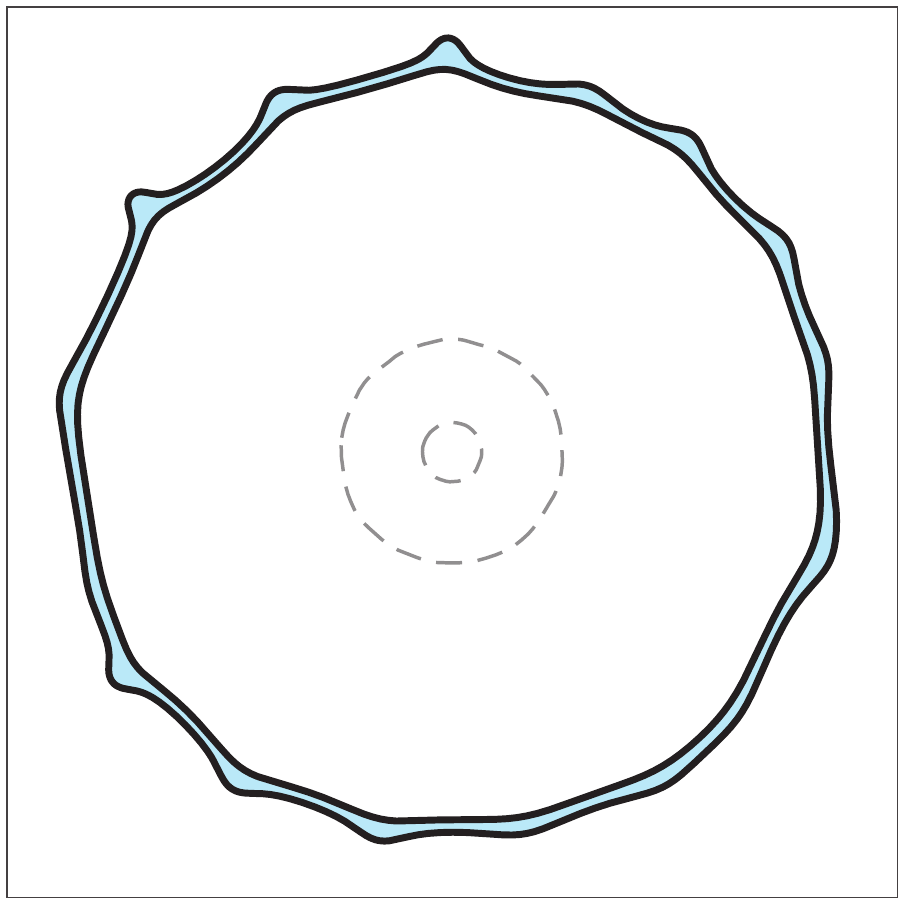}  & \includegraphics[width=0.25\linewidth]{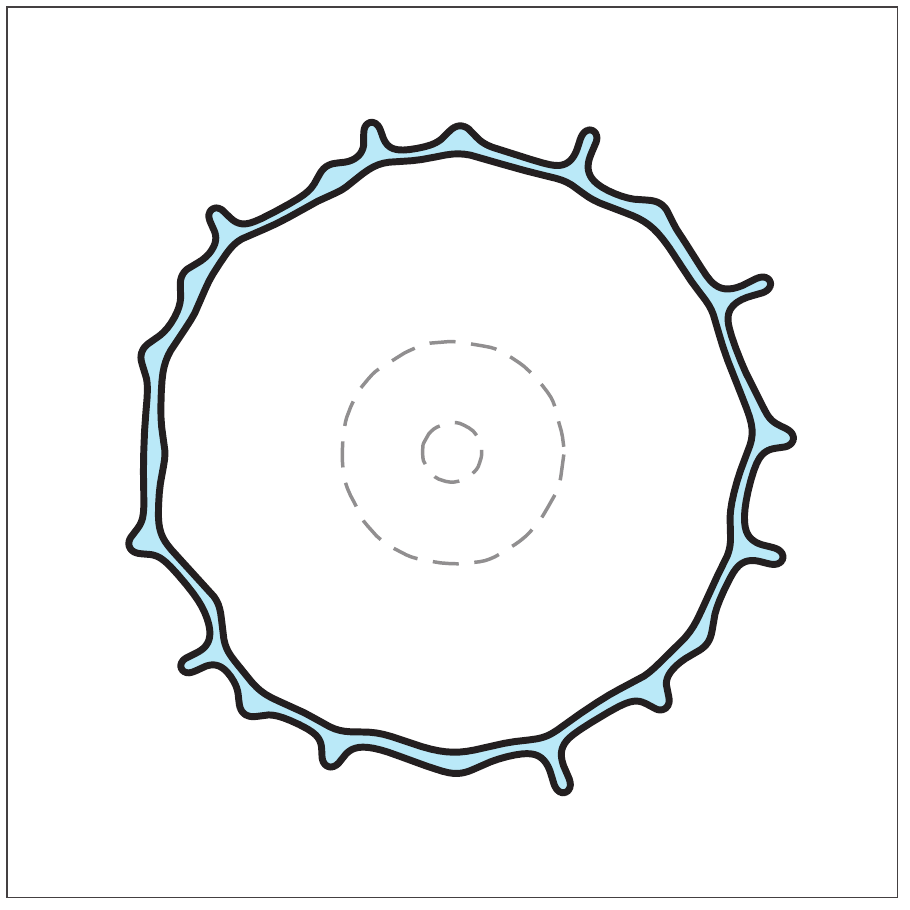}  \\
		\rotatebox{90}{\hspace{1.5cm} $R_O = 5$} &
		\includegraphics[width=0.25\linewidth]{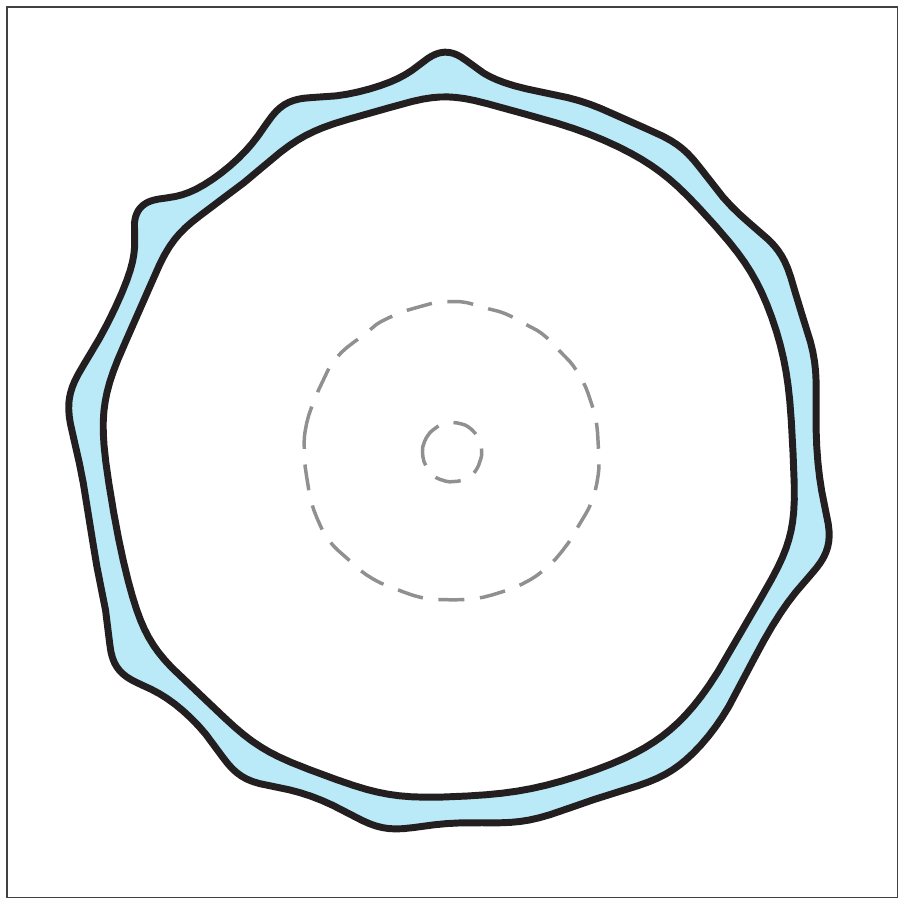} & \includegraphics[width=0.25\linewidth]{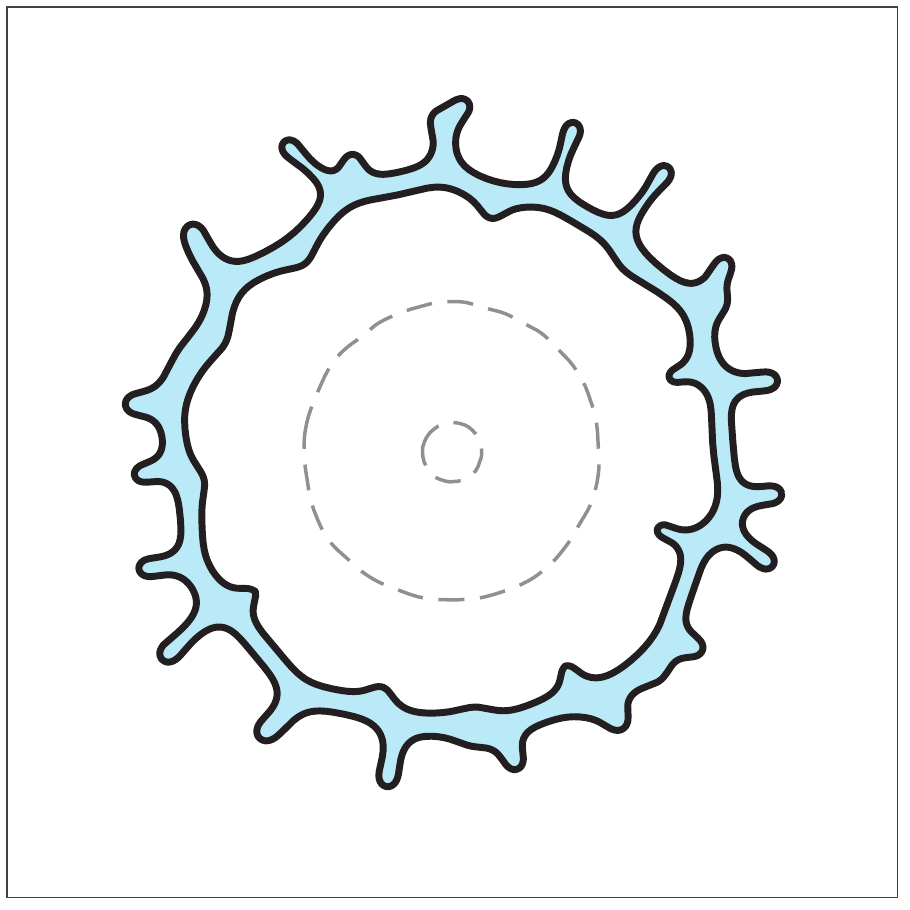}  & \includegraphics[width=0.25\linewidth]{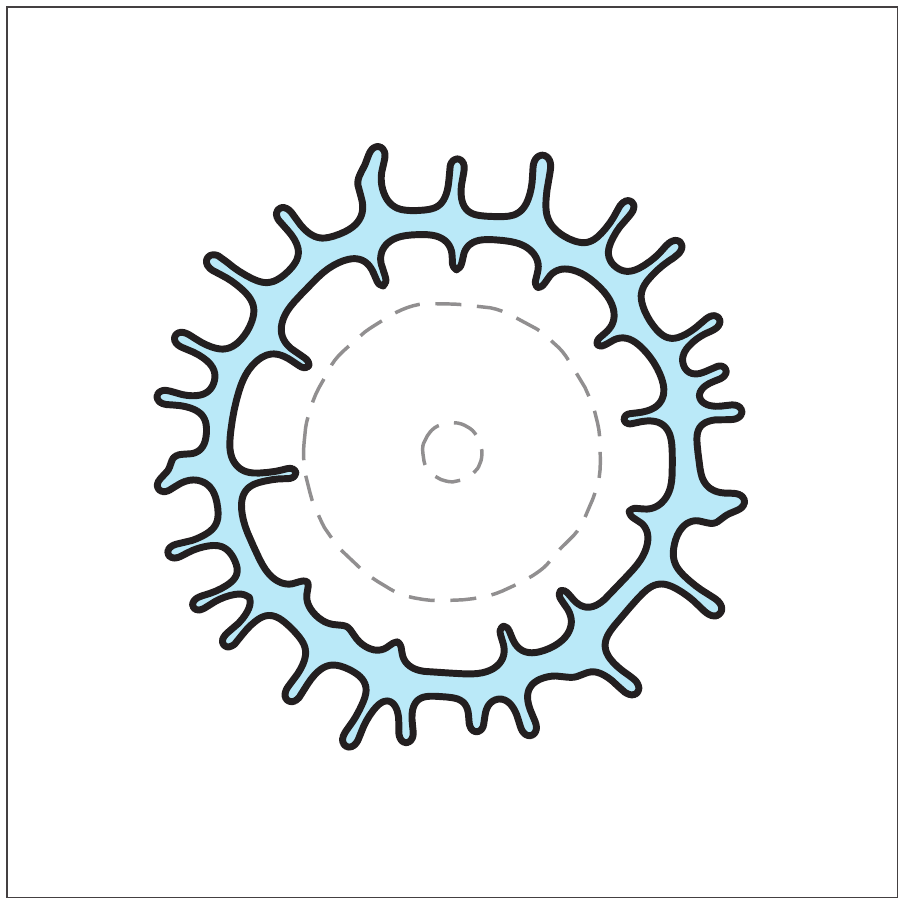}  \\
		\rotatebox{90}{\hspace{1.25cm} $R_O = 6.25$} &
		\includegraphics[width=0.25\linewidth]{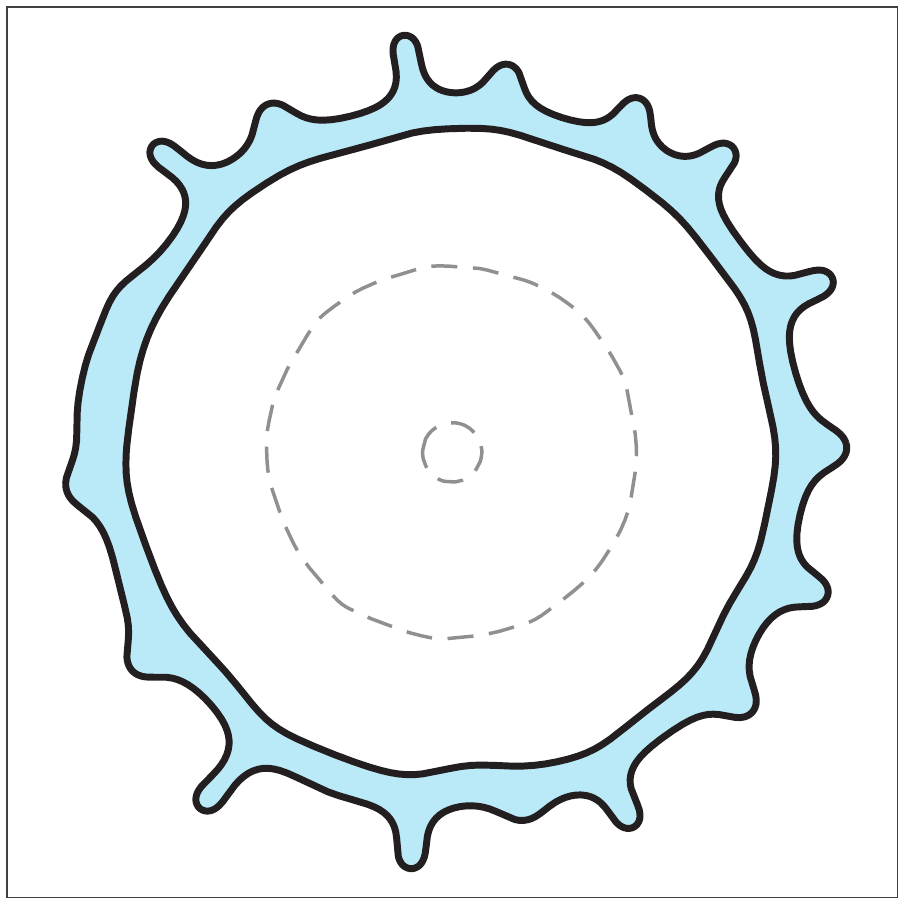} & \includegraphics[width=0.25\linewidth]{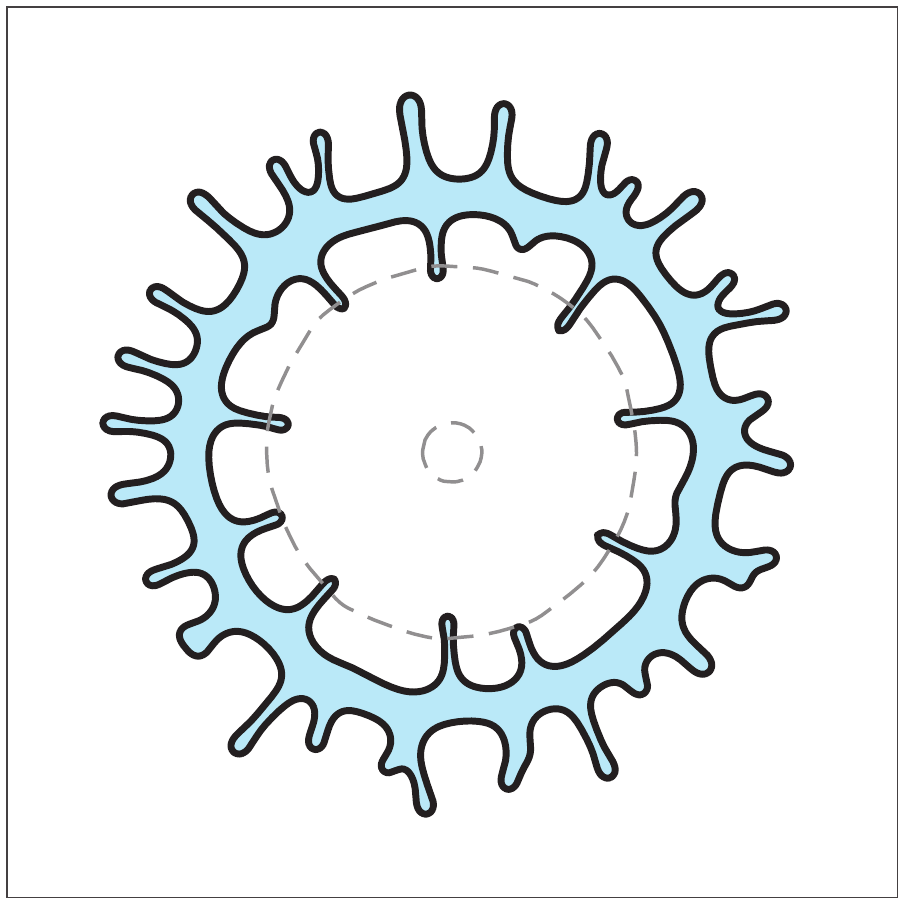}  & \includegraphics[width=0.25\linewidth]{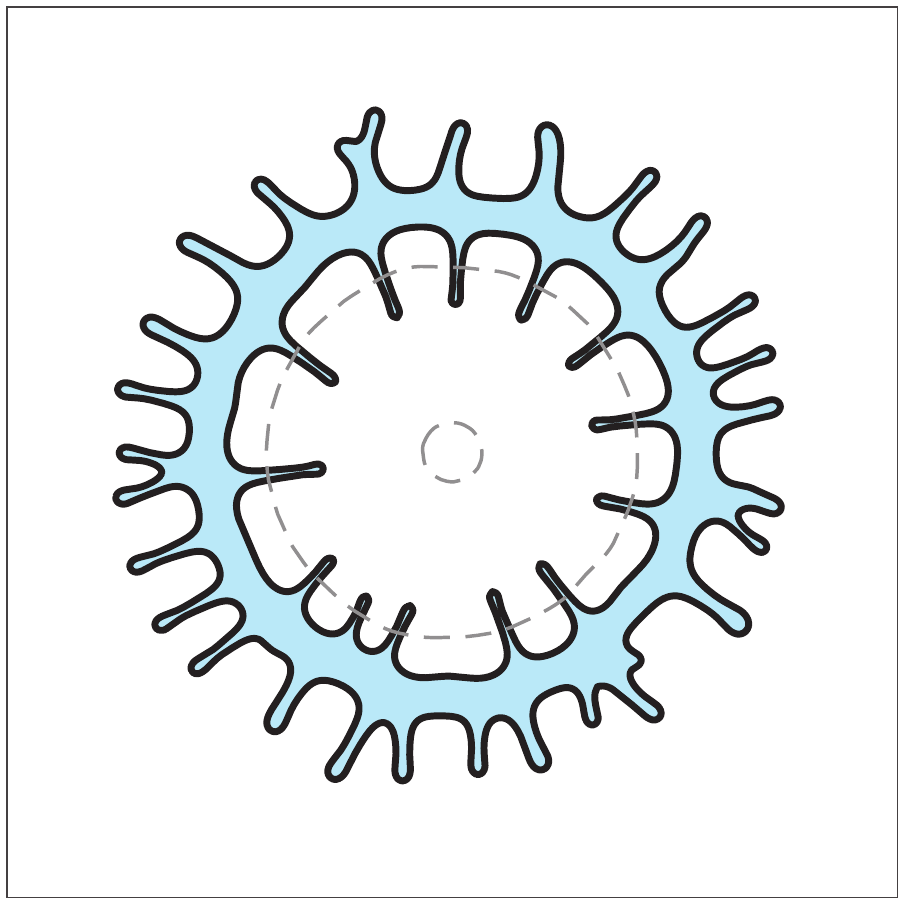}  \\
	\end{tabular}
	\caption{Numerical solutions of \eqref{eq:Rotating1}-\eqref{eq:Rotating4} for different values of $R_O$ and $\omega$. Simulations are performed with the initial conditions \eqref{eq:RadialIC}, where $N = 30$ and $\varepsilon = 5 \times 10^{-4}$. The dotted (grey) curves represent the initial condition for the inner and outer boundaries. Simulations are performed on the domain $-15 \le x \le 15$ and $-15 \le y \le 15$ using $900 \times 900$ equally spaced grid points.}
	\label{fig:RotatingExamples}
\end{figure}

We perform a series of simulations over a range of values of $R_O$ and $\omega$, shown in Fig.~\ref{fig:RotatingExamples}. For the smallest parameter combination of $\omega$ and $R_O$ considered (row one column one), both the leading and trailing interfaces remain near circular over the entire simulation, and form a continually thinning annulus. As $R_O$ is increased (top to bottom), the leading interface destabilises, with the number of fingers developing increasing with $R_O$. For larger values of the centrifugal parameter (rows two and three), both the leading and trailing interfaces develop fingers, where both the number and length of these fingers increases with either $\omega$ or $R_O$. The simulations indicate that increasing $R_O$ or $\omega$ appears to have a de-stabilising effect on both interfaces (as suggested by the results in Fig.~\ref{fig:LinearStabiliityAnalysis2}). It is interesting to note the different morphological features of the fingers that each interface develops. For the leading interface, fingers are short and thin, projecting outwards.  On the other hand, for the trailing interface, fingers appear wider and flatter with minimal tip splitting, and the gap between neighbouring fingers is narrow, particularly compared to traditional Saffman-Taylor fingers (see Fig.~\ref{fig:DoublyConnected} for example).  With this geometry of fingering, we speculate that the inner interface does not eventually burst through the outer interface.  We note that the patterns observed from the numerical simulations are consistent with experimental results  (see figures 7 and 8 in \citet{Carrillo2000} for example).  Finally, while it is difficult to precisely determine the number of fingers in each image in Fig~\ref{fig:RotatingExamples}, a rough count suggests that the predictions from linear stability theory (via the most unstable mode $n_{\max}$ plotted in Fig.~\ref{fig:LinearStabiliityAnalysis2}) are slightly higher than the number of fingers present on the outer interface in Fig~\ref{fig:RotatingExamples}, while the inner interface has roughly half as many fingers.  It is clear that the nonlinear shape of these fingers is far from sinusoidal and that nonlinearity is overwhelming any linear effects that dominate for very small amplitude perturbations.

These are only visual observations. To better quantify the influence of $R_O$ and $\omega$ on each interface, we introduce two metrics for measuring the complexity or severity of the viscous finger patterns. The first is the isoperimetric ratio defined by
\begin{align} \label{eq:Isoperimetric}
	\mathcal{I} = \frac{L^2}{4 \pi A},
\end{align}
where $L$ is the arc length of the closed curve that describes the interface and $A$ is
the area enclosed by the interface. The area is computed from \eqref{eq:Area1} and \eqref{eq:Area2}, while the arc length is determined via an analogous formula
\begin{align}
	L = \int \delta(\phi) |\nabla \phi| \, \textrm{d} \bmth{x},
\end{align}
where $\delta = \textrm{d} H / \textrm{d} \phi$ \citep{Osher2003}. The second metric is a measure of the length of the fingers that develop, which we refer to as the circularity ratio
\begin{align} \label{eq:Circularity}
	\mathcal{C} = \frac{R_{\mathrm{outer}}}{R_{\mathrm{inner}}},
\end{align}
where $R_{\mathrm{outer}}$ and $R_{\mathrm{inner}}$ are the maximum and minimum radii of the interface (sweeping across all angles), respectively. Note that for a circular interface, $\mathcal{I} = 1$ and $\mathcal{C} = 1$; otherwise, both of these metrics increase from unity as the shape of the interface becomes less circular.

\begin{figure}
	\centering
	Isoperimetric ratio \\
	\includegraphics[width=0.4\linewidth]{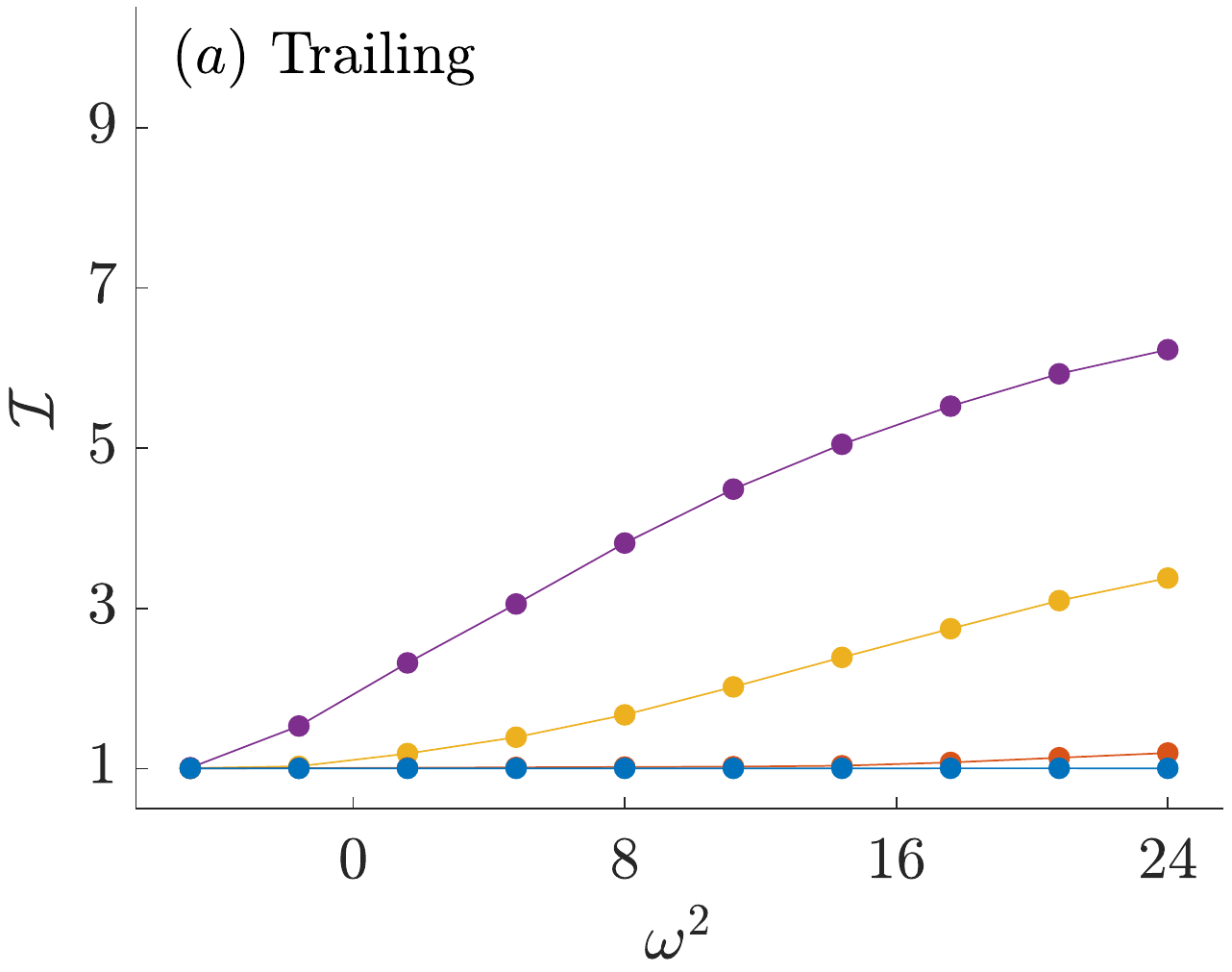}
	\includegraphics[width=0.4\linewidth]{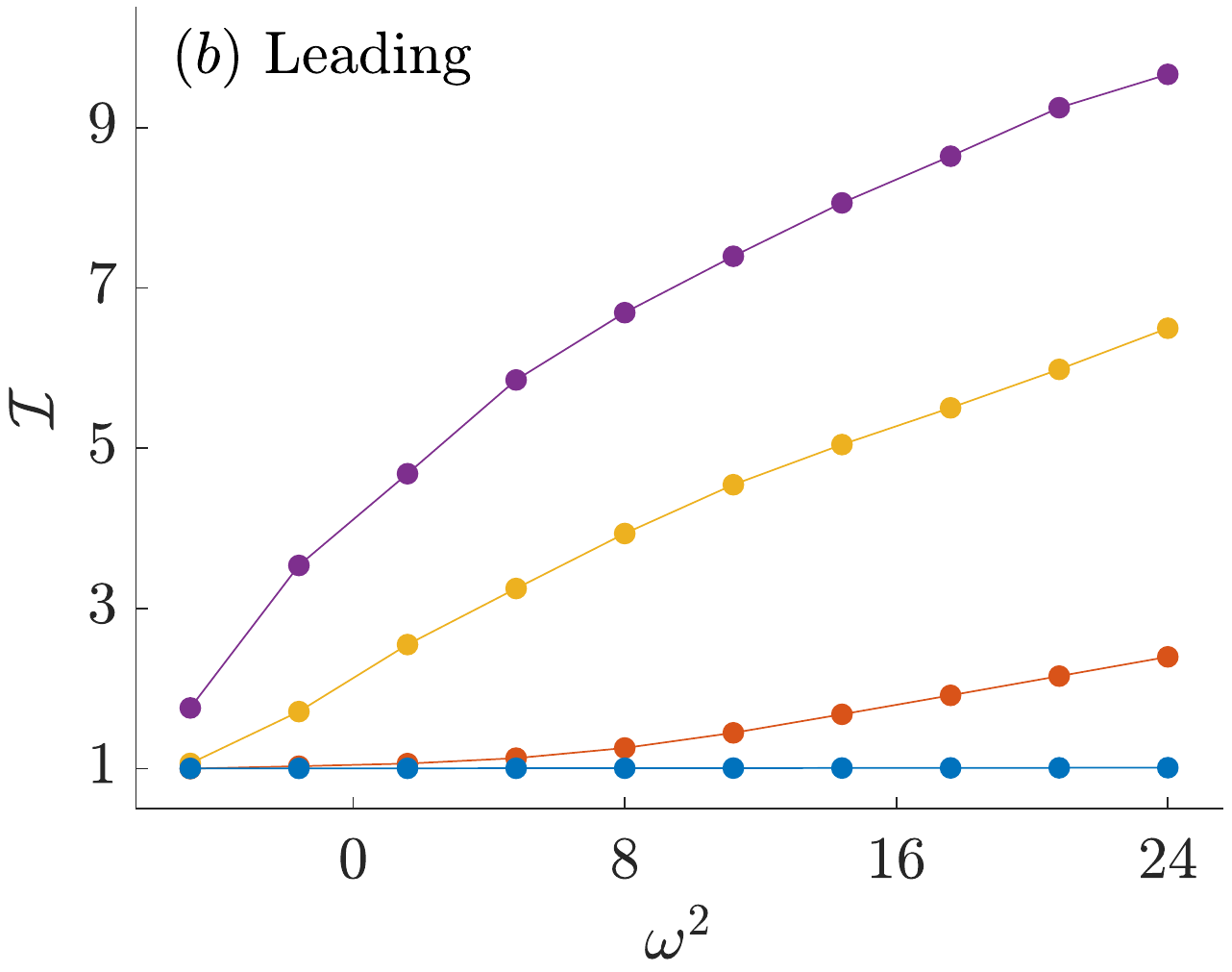}	 \\
	Circularity ratio \\
	\includegraphics[width=0.4\linewidth]{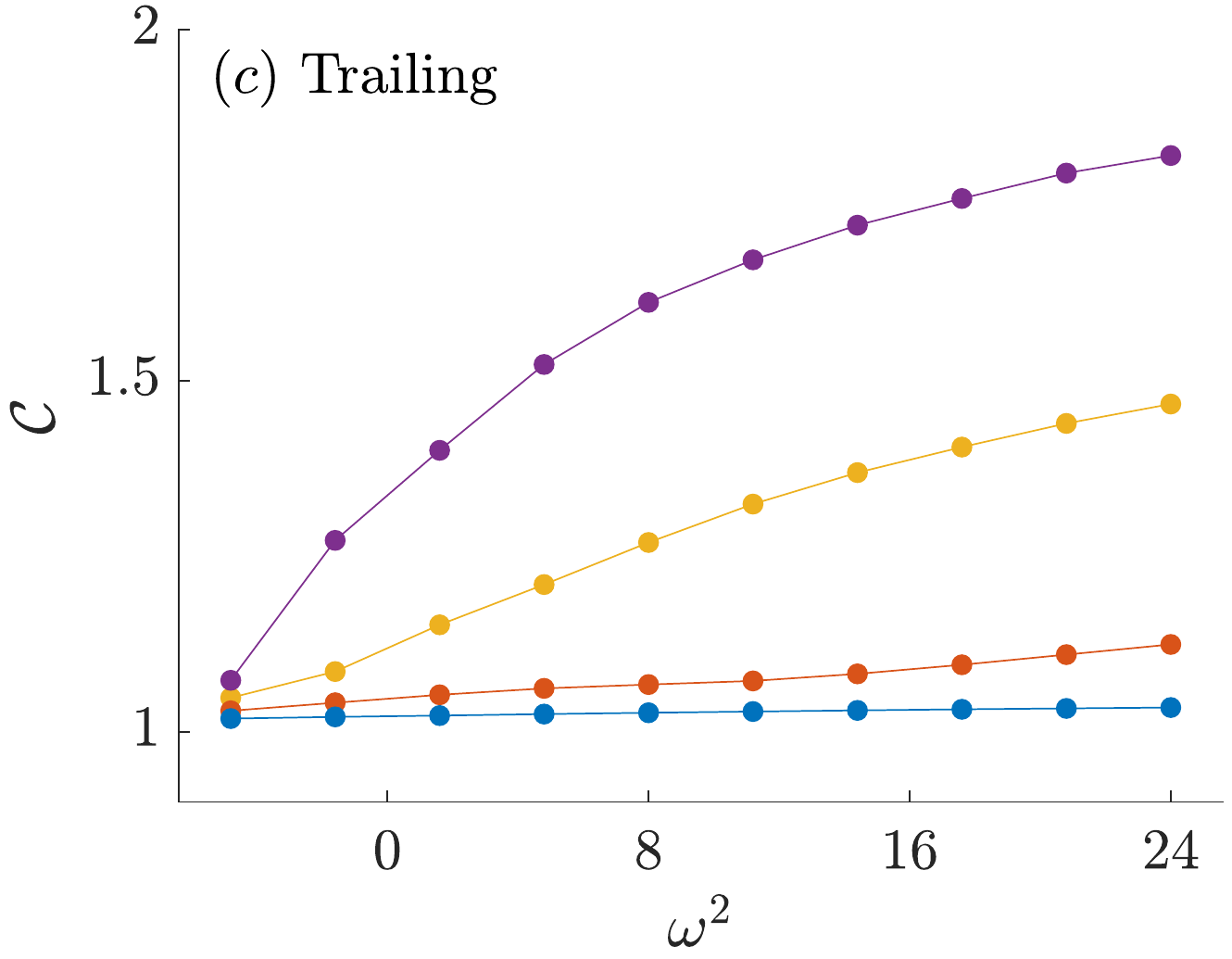}
	\includegraphics[width=0.4\linewidth]{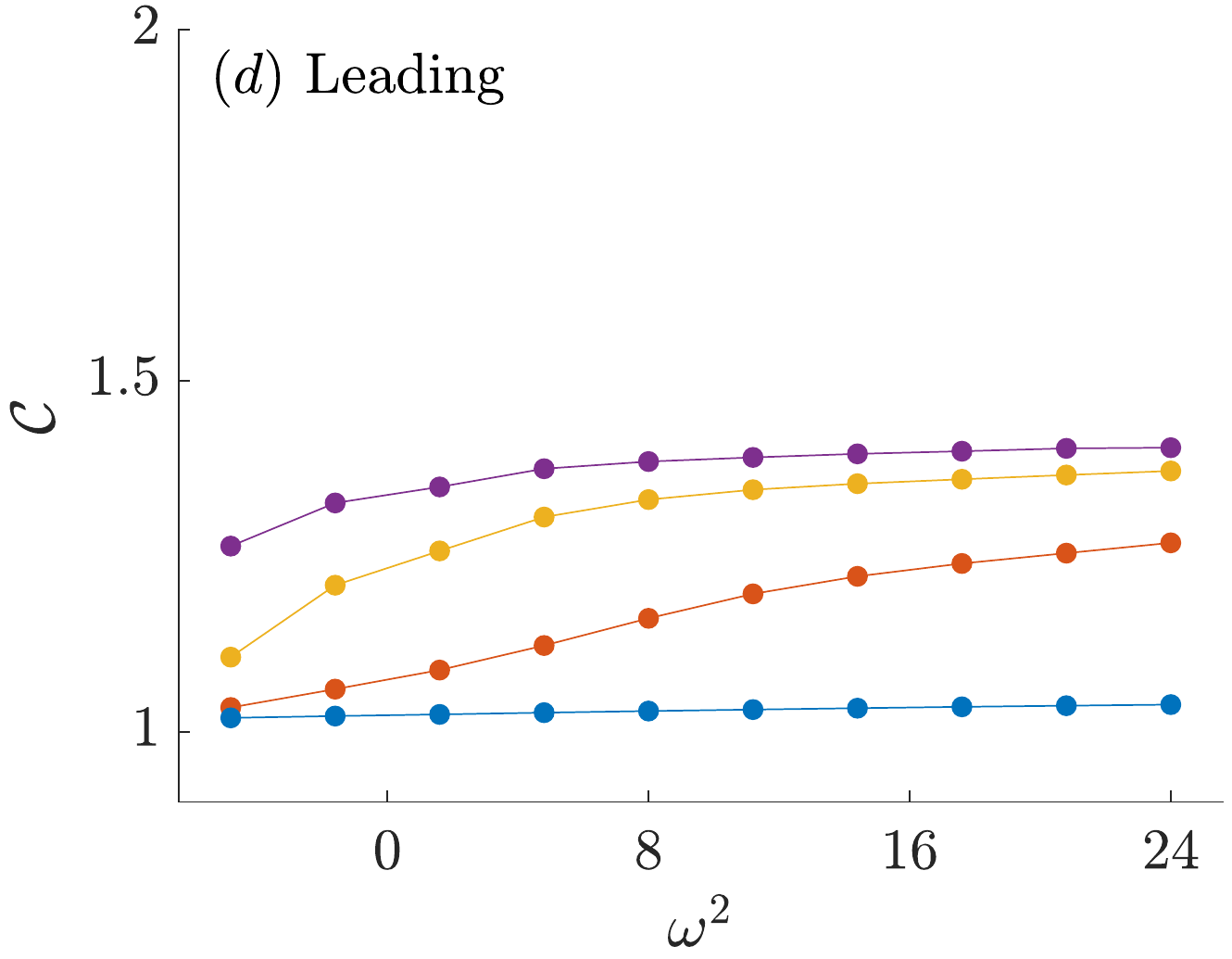}	
	\caption{$(a)$-$(b)$ The isoperimetric \eqref{eq:Isoperimetric} and $(c)$-$(d)$ circularity \eqref{eq:Circularity} ratios of the trailing (inner) and leading (outer) interfaces as a function of the angular velocity $\omega$. Simulations are performed with the initial conditions \eqref{eq:RadialIC}, where $N = 30$, $\varepsilon = 5 \times 10^{-4}$ and (bottom to top) $R_O = 2.5$, 3.75, 5, and 6.25. For each pair of $\omega$ and $R_O$, five simulations are performed, and $\mathcal{I}$ and $\mathcal{C}$ are averaged over each simulation.}
	\label{fig:FingerServeity}
\end{figure}

Figure \ref{fig:FingerServeity} shows the isoperimetric and circularity ratios for both the trailing and leading interfaces as a function of $\omega$ with $R_O=2.5$, $3$, $5$, and $6.25$ (bottom to top). For each pair of $R_O$ and $\omega$, five simulations are performed (each with $\theta_m$ and $\theta_n$ in (\ref{eq:RadialIC}) chosen randomly), and the isoperimetric and circularity ratios are averaged over each simulation. Across the range of parameters considered we find a general trend that both $\mathcal{I}$ and $\mathcal{C}$ increase with $R_O$ and $\omega$ for both the leading and trailing interfaces. This trend suggests that increasing either the amount of viscous fluid or the angular velocity results in both interfaces becoming less circular. Comparing figure~\ref{fig:FingerServeity}$(a)$ and $(b)$, the isoperimetric ratio of the leading interface is larger than that for the trailing interface across all parameter combinations considered. This behaviour is slightly different from that for the circularity ratio. For small values of $R_O$ and $\omega$, the circularity ratio of the leading interface (Fig.~\ref{fig:FingerServeity}$(d)$) is larger than that for the trailing interface (Fig.~\ref{fig:FingerServeity}$(c)$). However, for larger values of $R_O$ and $\omega$, $\mathcal{C}$ becomes smaller for the leading interface. This behaviour can be explained by noting that for large values of $R_O$ and $\omega$, while the leading interface produces a larger number of fingers, the length of the fingers of the trailing interface are larger relative to the average radius of the inner bubble (see row three, column three in Fig.~\ref{fig:RotatingExamples} for example).

\section{Summary} \label{sec:Discussion}

We have conducted a numerical investigation into two different but related models of doubly connected Hele--Shaw flow. By using a scheme based on the level set method, we have been able to compute solutions when the velocity of the viscous annular blob is driven by either a prescribed pressure differential between the interfaces or a centrifugal force due to the Hele--Shaw plates being rotated. While exact solutions of both of these models have previously been derived in the zero-surface-tension case \citep{Crowdy2002,Dallaston2012}, we extend this work by including the regularising effects of surface tension which prevents solutions developing unphysical cusps. Our numerical scheme is able to capture the complex interfacial patterns on a uniform computational grid, and simulations have been shown to compare well with existing experimental results.

In Sec.~\ref{sec:PressureDifferential}, we considered a model for doubly connected Hele--Shaw flow in which the pressure differential between the air on the inside and outside of the viscous fluid is prescribed, leading to cases in which the inner bubble either expands or contracts. Linear stability analysis (Sec.~\ref{sec:lsa1}) together with these simulations reveal that the trailing interface develops the expected viscous fingering patterns due to the Saffman-Taylor instability, while the leading interface is also (mildly) unstable, even though it involves a viscous fluid displacing an inviscid fluid (in isolation, the leading interface would be stable).  The fingers that develop on the trailing interface are morphologically distinct depending on whether the interior bubble is expanding or contracting.

For the expanding case (Sec.~\ref{sec:Expanding}), we observe tip-splitting and branching behaviour typical of classic Saffman-Taylor fingers, becoming more pronounced when the pressure differential, $\Delta p$, is increased.  We also demonstrate how sensitive this highly unstable system is to initial conditions with perturbed circular interfaces whose centres are slightly offset.  Here, nonlinear competition favour fingers with a slight advantage in pressure gradients, leading to a break in radial symmetry.  These simulations compare well with a small number of experiments we performed, together with more detailed experiments reported by Ward \& White~\citep{Ward2011}.  Regarding the latter, we were able to show that both the expansion rate of the interior bubble and the time at which bursting occurs computed from our numerical solutions compares well with these experimental results~\citep{Ward2011}.  On the other hand (Sec.~\ref{sec:Contracting}), for $\Delta p<0$, the interior bubble contracts and the outer interface is shown to develop fingers whose tips tend to not split (as they do in the expanding case) but instead appear to be `pulled' inward. These inward pointing fingers can either burst through the inner boundary or, for a large enough value of $R_O$, the interior bubble can contract to a point before bursting can take place.

In Sec.~\ref{sec:Rotating}, we study the evolution of a fluid annulus in a rotating Hele--Shaw cell.  The linear stability analysis summarised in Sec.~\ref{sec:lsa2} is complicated and difficult to interpret; however, this theory demonstrates that while surface tension tends to stabilise both interfaces, the angular velocity promotes a traditional Saffman-Taylor instability on the inner interface (due to a tendency for the inviscid fluid to displace the viscous fluid) and a centrifugal instability on the outer interface (due to viscous fluid being pushed outwards).  Our numerical simulations (Sec.~\ref{sec:numerical2}) revealed that either one or both interfaces can develop fingering patterns depending on the angular velocity of the plates or the amount of viscous fluid present.  It was found that increasing either the angular velocity or the amount of viscous fluid leads to more pronounced finger growth for both the leading and trailing interfaces, either visually or via the isoperimetric ratio or circularity ratio. Morphologically speaking, the fingering patterns each interface develops are distinct from each other, whereby the outer boundary typically develops a larger number of short outward pointing fingers, while the inner boundary grows fewer fingers that are longer, flatter and wider in appearance.

Looking ahead, the study of Hele-Shaw flows is currently an active research area, especially for various non-standard scenarios that involve different configurations of the Hele-Shaw apparatus (tapered plates, lifting plates, elastic plates, for example), non-Newtonian fluids (or fluids with suspended particles), ferrofluids and flows with applied electric fields.  There is scope for developing new schemes for computing fully nonlinear numerical simulations in these scenarios for doubly connected or multiply connected domains, either by adapting the level set scheme outlined here or via a boundary integral formulation.  In terms of mathematical modelling, there are interesting questions about how the model presented in Sec.~\ref{sec:PressureDifferential} behaves when the annular viscous domain is very thin and also in the limit that the trailing interface bursts through, or ruptures, the leading interface.  Progress on this problem will be reported elsewhere.

\section{Acknowledgements}

LCM and SWM acknowledge the support of the Australian Research Council via the Discovery Project DP140100933.  They thanks Michael Dallaston for help with the numerical verification described in Sec.~\ref{fig:verification} and Michael Jackson for many discussions about linear stability analysis of doubly connected Hele--Shaw flows, including the radial geometry considered in this study.  NDC and SWM are very grateful for the French National Research Institute for Agriculture, Food and Environment in Montpellier  (formally IRSTEA) for their technical support and generous hospitality while the experiments were being performed.

\bibliography{references}

\appendix
\section{Further details of numerical scheme}\label{sec:furthernumerical}
\subsection{General algorithm} \label{sec:GeneralAlgorithm}

A summary of our numerical algorithm for solving \eqref{eq:Model1}-\eqref{eq:Model4} is as follows:
\begin{itemize}[leftmargin=15.0mm]
	\item[\textit{Step 1}] Given initial conditions for the inner and outer interfaces, $\partial \Omega_i(0)$ and $\partial \Omega_o(0)$, construct two level set functions, $\phi_i$ and $\phi_o$, such that $\phi_i < 0$ in the inner bubble region and $\phi_i>0$ otherwise, while $\phi_o < 0$ in the outer bubble region and $\phi_o > 0$ otherwise. Each level set function is initialised as a signed distance function.
	\item[\textit{Step 2}] Solve \eqref{eq:Model1}, \eqref{eq:Model3}, and \eqref{eq:Model4} for the pressure in the region $\bmth{x} \in \Omega$ using a modified finite difference stencil as described in Sec.~\ref{sec:SolvingVelocityPotential}
	\item[\textit{Step 3}] Compute $F_i$ and $F_o$ according to \eqref{eq:SpeedFunction}, where the spatial derivatives are approximated using central differences. Both $F_i$ and $F_o$ are extended into the region $\bmth{x} \in \mathbb{R}^2 \backslash \Omega$ by solving the biharmonic equation as described in Section \ref{sec:VelocityExtension}.
	\item[\textit{Step 4}] Update $\phi_i$ and $\phi_o$ by solving the level set equations \eqref{eq:DClevelseteqn}. The spatial derivatives are approximated using a second order essentially non-oscillatory scheme, and we integrate in time using second order total variation diminishing Runge-Kutta, where $\Delta t = 0.25 \times \Delta x / \max(|F_i|,|F_o|)$.
	\item[\textit{Step 5}] Both $\phi_i$ and $\phi_o$ are reinitialised by solving
	\begin{align} \label{eq:reinitialised}
		\frac{\partial \phi}{\partial \tau} + S(\phi) (|\grad \phi| - 1) = 0,
	\end{align}
	where $\tau$ is a pseudo-time variable and
	\begin{align}
		S = \frac{\phi}{\sqrt{\phi^2 + (|\grad \phi \Delta x|)^2)}}.
	\end{align}
	We perform five iterations of \eqref{eq:reinitialised} using $\Delta \tau = \Delta x / 5$ every four time steps.
	\item[\textit{Step 6}] Determine the minimum distance between the inner and outer interfaces. We approximate this distance with
	\begin{align} \label{eq:MinDist}
		D(t) \approx \min(|\phi_i| + |\phi_o|),
	\end{align}
	and simulations are stopped if $D < 4 \Delta x$.
\end{itemize}

\subsection{Verification}\label{fig:verification}

\begin{figure}
	\centering
	\includegraphics[width=0.35\linewidth]{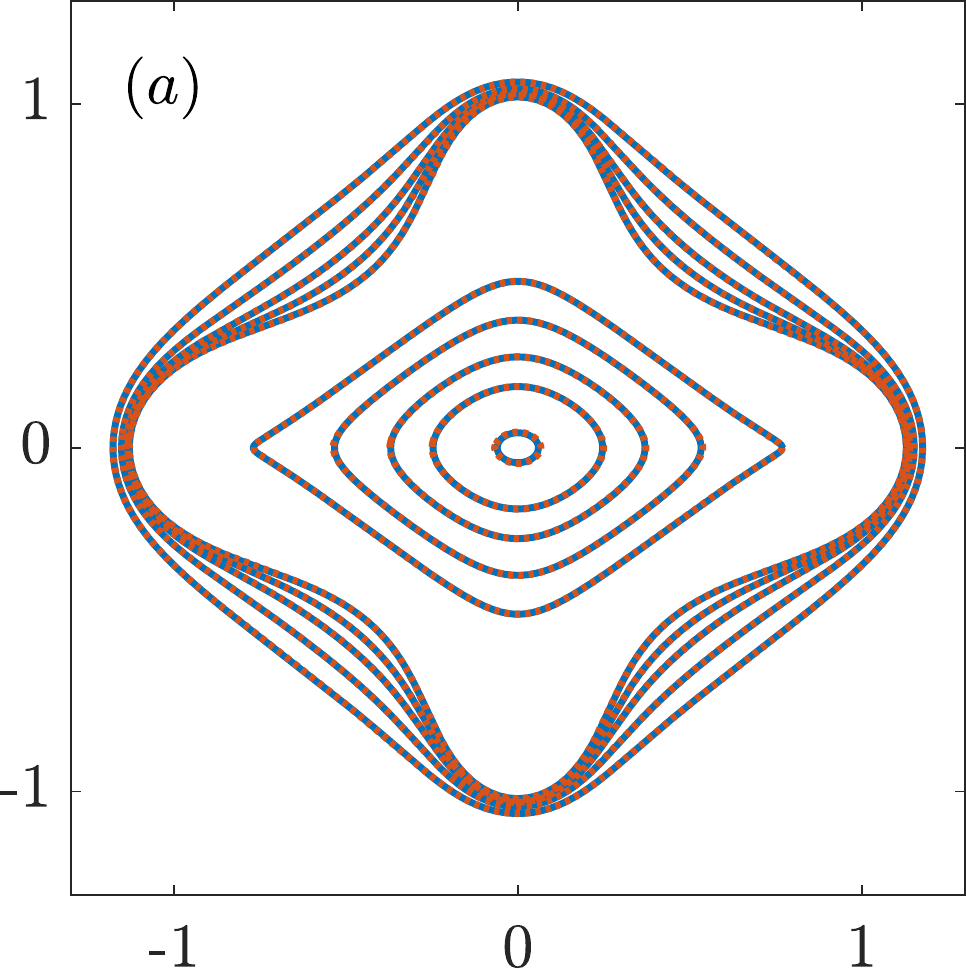}	
	\includegraphics[width=0.35\linewidth]{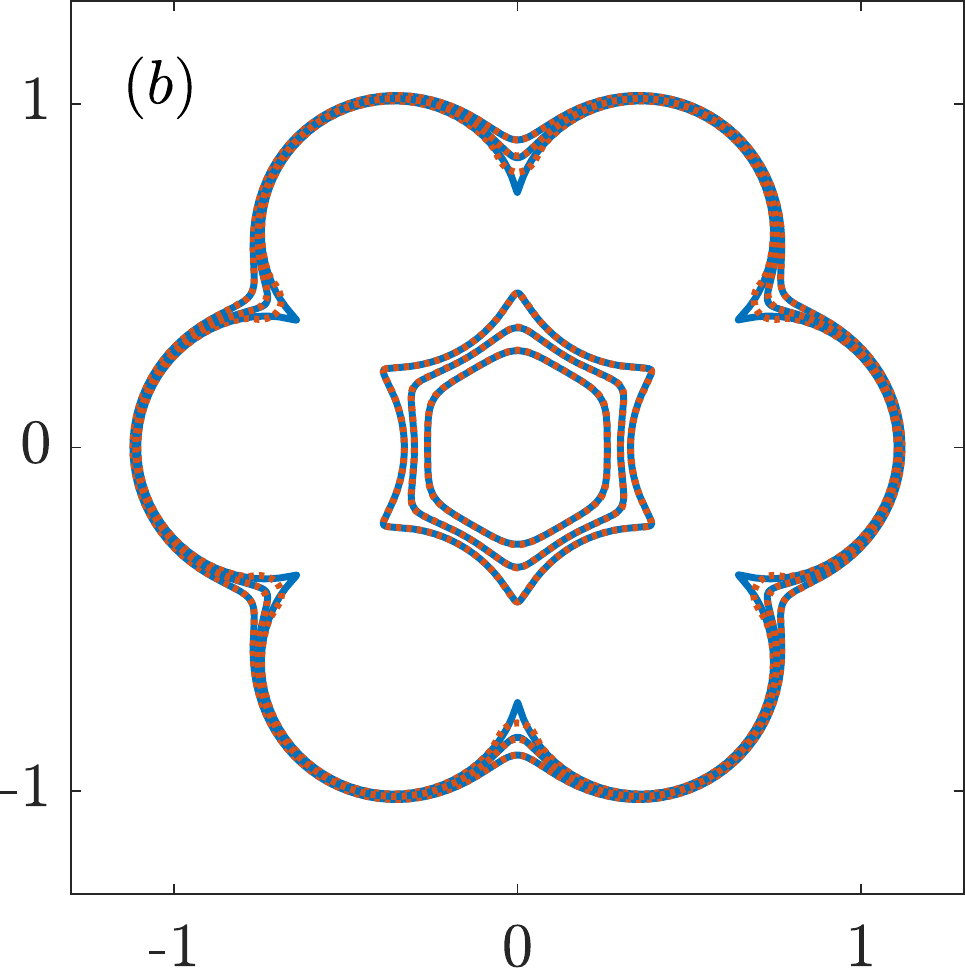}	
	\caption{A comparison of an exact solution derived by Ref.~\cite{Dallaston2012} (solid blue) with numerical solution to \eqref{eq:Model1}-\eqref{eq:Model4} (dashed red) where $\Delta p = -1$ and $\gamma = 0$. For $(a)$, the initial condition is given by \eqref{eq:mapping1} and \eqref{eq:special} where $p_1(0) = 1.2$, $q_1(0) = 1.7$, $p_2(0) = 1.1$, $q_2(0) = 1.4$, and $R(0) = 1$ while for $(b)$, the initial condition is given by \eqref{eq:mapping2} and \eqref{eq:special} where $N = 6$, $p(0) = 1.4$, $p(0) = 1.06$, $\rho(0) = 0.44$, and $R(0) = 1$. Both simulations are performed on the computational domain $-1.25 \le x \le 1.25$ and $-1.25 \le x \le 1.25$ using $300 \times 300$ equally spaced nodes. The direction of increasing time is denoted by the arrows.}
	\label{fig:ZeroSurfaceTension}
\end{figure}

We perform a test of our scheme by comparing the numerical solution to \eqref{eq:Model1}-\eqref{eq:Model4} with exact solutions derived in Ref.~\cite{Dallaston2012} when the effects of surface tension are ignored ($\gamma = 0$). We consider two different initial conditions. Introducing the time-dependent mapping function $z = f(\zeta, t)$ where $z \in \mathbb{C}$ with 4 poles and zeros at $\pm p_1$, $\pm i p_2$ and $\pm q_1$, $\pm i q_2$, where $\rho(t)$ is the conformal modulus, and $R(t)$ as a scaling factor, the first initial condition is given by
\begin{equation}
f_\zeta(\zeta, 0) = R\frac{P(\zeta^2/q_1^2, \rho^2) P(-\zeta^2/q_2^2, \rho^2)}{P(\zeta^2/p_1^2, \rho^2) P(-\zeta^2/p_2^2, \rho^2)},
\label{eq:mapping1}
\end{equation}
where
\begin{equation}
P(\zeta, \rho) = (1 - \zeta) \prod_{j=1}^{\infty} (1 - \rho^{2j} \zeta) (1 - \rho^{2j} / \zeta),
\label{eq:special0}
\end{equation}
\begin{equation}
\frac{\dot{R}}{R} = \zeta_R I(\zeta_R) \frac{f_\zeta}{f} - \frac{\dot{\zeta}_R}{\zeta_R},
\quad
f(\zeta_R) = R\zeta_R,
\quad
\zeta_R = \rho^{1 - 1/N} p,
\end{equation}
\begin{equation}
\frac{\dot{\rho}}{\rho} = \frac{1}{2 \pi \ln \rho} \int_{0}^{2 \pi} \frac{1}{|f(e^{i s})|} - \frac{1}{\rho^2 |f(\rho e^{i s})|^2} \textrm{ d} s,
\end{equation}
\begin{equation}
\frac{\dot{p}_k}{p_k} = -I(p_k),
\quad
\frac{\dot{q}_k}{q_k} = -I(q_k) - \frac{2}{\ln \rho \bar{f}_\zeta(1/q_k)},
\quad
I(\zeta) = \frac{f_t}{\zeta f_\zeta}.
\label{eq:special}
\end{equation}
We refer to Ref.~\cite{Dallaston2012} for further details. The outer and inner boundaries are parametrised by setting $\zeta = e^{i \theta}$ and $\zeta = \rho e^{i \theta}$, respectively, where $0 \le \theta < 2 \pi$. We compare the exact (solid blue) and numerical (dashed red) solutions in Fig.~\ref{fig:ZeroSurfaceTension}$(a)$, and find the two are indistinguishable (at this scale) when using  $300 \times 300$ equally spaced nodes. Furthermore, the interior boundary appears to contract to an ellipse, which is what should happen in this zero-surface-tension case \cite{Dallaston2012}.

For the second initial condition, Fig.~\ref{fig:ZeroSurfaceTension}$(b)$, the initial mapping function is
\begin{align} \label{eq:mapping2}
	f(\zeta, 0) &= R \zeta \frac{P(\rho^N \zeta^N / p^N, \rho^N)}{P(\zeta^N/p^N, \rho^N)},
\end{align}
along with \eqref{eq:special0}-\eqref{eq:special}. In this case, as the interior bubble contracts, the exact solution develops a sharp cusp on the exterior interface due to \eqref{eq:Model1}-\eqref{eq:Model4} being ill-posed when surface tension is absent. The level set method acts to provide a form of regularisation, and as such, the sharp cusp in not observed in the numerical solution. However, we still see agreement between the numerical and exact solutions for the interior interface, as well as the exterior interface in the times before the cusp develops.

Finally, we make a brief comment on mass conservation. Our numerical scheme is based on the level set method, which can suffer from mass loss (or gain). Across the simulations presented in this work, we monitor the mass of the viscous fluid via \eqref{eq:Area1} and \eqref{eq:Area2}. By using a sufficiently refined grid and performing reinitialisation (discussed in Sec.~\ref{sec:GeneralAlgorithm}) frequently enough, we find that mass loss (compared to the initial amount of fluid in the system) is around $0.1\%$.

\end{document}